\documentclass[prl,twocolumn,amsmath,amssymb,superscriptaddress,floatfix]{revtex4-2}

\usepackage[english]{babel}
\usepackage{amsfonts}
\usepackage{graphicx}
\usepackage{times}
\usepackage{color}
\usepackage[normalem]{ulem}
\usepackage{bm}
\usepackage{hyperref}
\usepackage{dcolumn}
\usepackage{xcolor}

\begin{document}

\preprint{APS/123-QED}


\title{Universal thermal breakdown of polaron coherence in one, two, and three dimensions}

\author{Jeet Shannigrahi}
\affiliation{Department of Physics and Astronomy, University of British Columbia, Vancouver, British Columbia, Canada, V6T 1Z1}
\affiliation{Quantum Matter Institute, University of British Columbia, Vancouver, British Columbia, Canada, V6T 1Z4}

\author{Janez Bon\v{c}a}
\affiliation{Faculty of Mathematics and Physics, University of Ljubljana, 1000 Ljubljana, Slovenia}
\affiliation{Josef Stefan Institute, 1000 Ljubljana, Slovenia}

\author{Mona Berciu}
\affiliation{Department of Physics and Astronomy, University of British Columbia, Vancouver, British Columbia, Canada, V6T 1Z1}
\affiliation{Quantum Matter Institute, University of British Columbia, Vancouver, British Columbia, Canada, V6T 1Z4}

\begin{abstract}

How a polaron loses its quasiparticle coherence with increasing temperature is a long-standing open problem. Holstein addressed it in 1959 only in the extreme antiadiabatic, strong-coupling limit, while more recent numerically exact approaches are largely restricted to one dimension. Here we solve this problem on square and simple cubic lattices across the weak-, intermediate-, and strong-coupling regimes. We show that the polaron effective mass $m^\ast$ and inverse lifetime $1/\tau$ increase monotonically with temperature until, at a $T \sim 0.5 \,\Omega$, the quasiparticle peak dissolves into a broad incoherent thermal continuum. The phonon frequency $\Omega$ therefore defines a universal coherence scale, independent of dimensionality and coupling strength, validating Holstein's prediction far beyond the regime in which it was derived. These results follow from a finite-temperature generalization of the Momentum Average (MA) approximation, yielding a closed-form, diagrammatically derived self-energy that is asymptotically exact in the strong-coupling limit at all temperatures. Benchmark comparisons demonstrate excellent quantitative agreement of the resulting 1D spectral functions with the numerically exact Variational Exact Diagonalization--Finite-Temperature Lanczos Method (VED-FTLM) and finite-$T$ Density Matrix Renormalization Group (DMRG).

\end{abstract}

\maketitle


\textit{Introduction:} Polarons -- charge carriers dressed by the phonon cloud generated through their interaction with the lattice -- govern the low-temperature thermodynamic and transport properties of low-carrier-density materials of broad current interest, including transition-metal oxides, organic semiconductors, and halide perovskites~\cite{vanMechelen2008, Herz2017, Wang2016, Franchini2021}. At finite temperature $T$, the polaron effective mass $m^\ast$, quasiparticle (\textit{qp}) weight $Z$, and lifetime $\tau$ evolve through scattering from thermally excited phonons, which also reshape the surrounding phonon cloud. These quantities are directly probed by angle-resolved photoemission spectroscopy (ARPES)~\cite{Damascelli2003} and determine transport coefficients. Understanding their temperature dependence, and how the dressed quasiparticle ultimately loses its coherence, remains a central open problem in the theory of strongly coupled carrier--boson systems.

By contrast, single polaron behaviour at zero temperature is well understood. As one of the paradigmatic quasiparticles of condensed-matter physics~\cite{Landau1933,Pekar1946,Pekar1948}, they have been studied extensively in models such as Holstein~\cite{Holstein1959,Holstein1959_2}, Fr\"{o}hlich~\cite{Frohlich1950}, and Peierls/SSH~\cite{SSH,SSH2}, using a wide variety of analytical and numerical methods~\cite{Marsiglio1993,Ranninger1992,Alexandrov1994,Marsiglio1995,Fehske1997,Fehske2000,Kornilovitch1998,Prokofev1998,Cataudella2007,Linden2005,Sous2021,Nery2018}. In all these models, the single polaron forms a coherent quasiparticle with an infinite lifetime $\tau\rightarrow\infty$, while its \textit{qp} weight $Z$ and effective mass $m^\ast$ can be strongly renormalized, particularly at strong couplings.

Much less is known at finite temperature. Early work was limited to two-site systems~\cite{Ranninger1992,Ranninger1997,Ciuchi2006,Berciu2007}. Subsequent progress came primarily from numerically exact studies of finite 1D systems: Bon\v{c}a {\it et al.} computed the single Holstein polaron (HP) spectral function using the Variational Exact Diagonalization--Finite-Temperature Lanczos Method (VED-FTLM) on rings of up to 12 sites~\cite{Bonca2019}, while finite-$T$ Density Matrix Renormalization Group (DMRG) extended these calculations to chains of 101 sites, yielding accurate results for $T \lesssim 0.4\,\Omega$~\cite{Meisner2020,Meisner2022}. Other approaches include Hierarchical Equations of Motion in momentum space~\cite{Jankovic2022}, mixed quantum-classical methods~\cite{Reichman2025}, cumulant-expansion techniques~\cite{ReichmannCEI2022,ReichmannCEII2022,Mitric2023}, and Dynamical Mean-Field Theory (DMFT)~\cite{Ciuchi1997,Mitric2022}. Nevertheless, numerically exact finite-temperature methods remain essentially restricted to one dimension, as their computational cost grows prohibitively with both system size and dimensionality. As a result, finite-temperature polaron properties in higher dimensions have remained inaccessible, and, to our knowledge, no such finite-$T$ result exist for $D \geq 2$.

\begin{figure*}
  \includegraphics[width=0.95\textwidth]{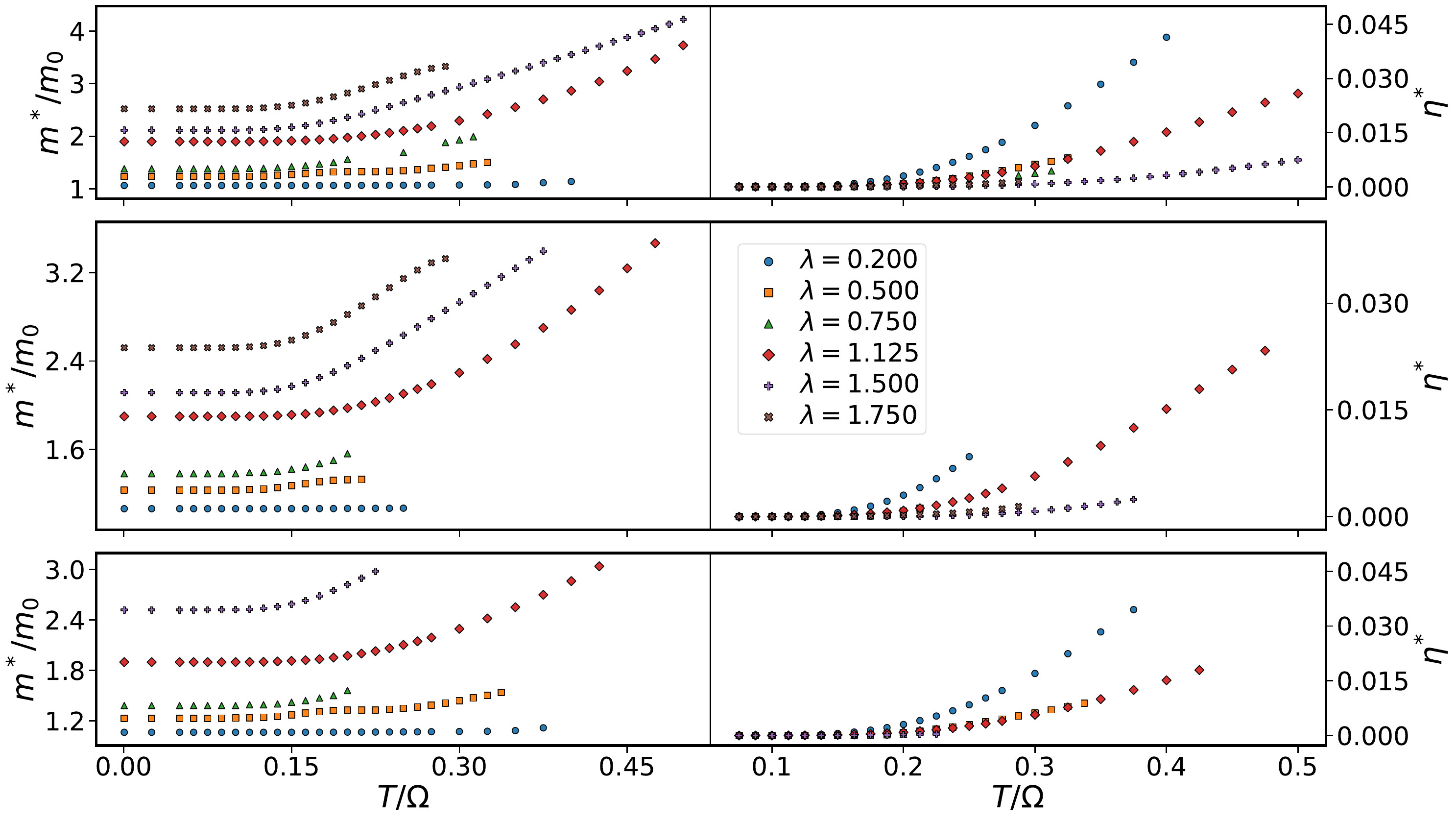}
  \caption{Temperature dependence up to $T \sim 0.5\,\Omega$ of the effective polaron mass $m^\ast$ and the thermal broadening $\eta^\ast\sim 1/\tau$ of the 1D, 2D and 3D  HP (top to bottom) across different coupling regimes, according to MA$^{(1)}$.  Here $t=1, \Omega=2$ and $\eta_0=0.001$ and $0.01$ for $D=1$ and $D\ge2$, respectively. The worst-case  fitting standard error was below $10^{-2}$ for the \textit{qp}  weight and below  $10^{-3}$ for $\eta^\ast$. Remarkably, $m^\ast$ and $\eta^\ast$ values are nearly independent of $D$, unlike in the extreme strong-coupling limit where $m^\ast/m_0 \to e^{2D\lambda t/\Omega}$ grows exponentially with dimensionality; even our largest couplings are thus far from that asymptotic regime considered by Holstein~\cite{Holstein1959_2}.  See text for more details.}
   \label{fig:eff_mass_lifetime}
\end{figure*}

In this Letter we solve this problem for the single Holstein polaron on hypercubic lattices with $D=1$, 2, and 3, in the thermodynamic limit. Our central result, summarized in Fig.~\ref{fig:eff_mass_lifetime}, is that across all coupling regimes and dimensions, the effective mass $m^\ast$ and inverse lifetime $1/\tau$ increase monotonically with temperature up to $T \sim 0.5\,\Omega$, above which the quasiparticle (\textit{qp}) peak dissolves into an incoherent thermal continuum and a coherent quasiparticle ceases to exist. The phonon frequency $\Omega$ therefore defines the universal coherence scale, independent of dimensionality and coupling strength, confirming Holstein's prediction~\cite{Holstein1959_2} far beyond the extreme antiadiabatic, strong-coupling limit in which it was derived.

These results are made possible by the finite-temperature generalization of the Momentum Average (MA) approximation~\cite{Mona2006,Mona2007,Glenn2006}, derived in the End Matter through a new diagrammatic resummation. It yields a closed-form expression for the finite-$T$ self-energy that is exact in the atomic limit ($t=0$) at any coupling strength and temperature, and goes smoothly to the known zero-temperature MA result as $T\rightarrow0$; the finite-$T$ theory thus interpolates between two analytically controlled limits. We are not aware of  any other explicit, closed-form finite-temperature self-energy that becomes asymptotically exact in a physical limit, for any other method. Instead, self-energies are typically accessible only indirectly or through analytic continuation. The lattice type enters solely through the free-particle propagator, making the evaluation numerically trivial in any dimension. The accuracy of MA at $T=0$ is well documented and arises from its variational nature; we show that MA remains similarly accurate at finite $T$, in excellent agreement with available 1D results~\cite{Bonca2019,Meisner2020,Mitric2022}. Moreover, since MA accuracy improves with increasing $D$~\cite{Glenn2006}, the 2D and 3D results presented here are not just qualitatively, but also quantitatively accurate. Although we focus here on the Holstein model on hyper-cubic lattices, the MA framework applies with comparable efficiency and accuracy to a broad class of electron--boson coupling models and lattice geometries, where finite-$T$ properties are even less explored. The present work thus establishes a general framework for investigating finite-temperature polaron physics well beyond the Holstein model.

\textit{Model:}  We study a single polaron described by the Holstein Hamiltonian $\hat{H} = \hat{H}_\mathrm{e} + \hat{H}_\mathrm{ph} + \hat{H}_\mathrm{ep}$.  The nearest-neighbor hopping $ \hat{H}_\mathrm{e} = -t\sum_{\langle i,j\rangle}(c^\dagger_i c_j + \mathrm{h.c.}) = \sum_{\bf k} \epsilon_{\bf k} c^\dagger_{\bf k} c_{\bf k}, $ where $c^\dagger_i$ creates an electron at site $i$ of a hyper-cubic lattice (lattice constant $a=1$, number of sites $N\rightarrow \infty$), $c^\dagger_{\bf k}$ is its Fourier transform, and $\epsilon_{\bf k}= -2t\sum_{\gamma=x,y,z}\cos(k_\gamma )$ in 3D. The electron spin is trivial and suppressed. The branch of Einstein phonons is described by $\hat{H}_\mathrm{ph} = \Omega\sum_i b^\dagger_i b_i$ (we set $\hbar = k_B=1$) where   $b^\dagger_i$ creates a phonon at site $i$. The  Holstein electron-phonon coupling $\hat{H}_\mathrm{ep} = g\sum_i c^\dagger_i c_i(b^\dagger_i + b_i)$ is characterized by the dimensionless coupling $\lambda = g^2/(2Dt\Omega)$.  All sums over momenta are over the first Brillouin zone and  the phonon frequency $\Omega$ is defined in units of $t$.

The finite-$T$ retarded single-electron propagator is~\cite{Mahan2000,ReichmannCEI2022,Bonca2019}:
\begin{equation}
  G({\bf k},\omega) = \Bigl\langle c_{\bf k}\,\hat{G}\!\Bigl(\omega + \hat{H}_\mathrm{ph}\Bigr)c^\dagger_{\bf k}\Bigr\rangle_\mathrm{th},
\label{eqn:freq_space_G}
\end{equation}
where $\hat{G}(\omega) = [\omega + i\eta_0 - \hat{H}]^{-1}$ with  $\eta_0 \to 0^+$ an artificial broadening. Here $ \langle\cdots\rangle_\mathrm{th} = \sum_{\{n\}}\frac{e^{-\beta\Omega\sum_i n_i}}{\mathcal{Z}}\langle \{n \}| \cdots |\{n\}\rangle$ denotes the thermal average over the empty lattice, with $|\{n\}\rangle = \prod_i \,(b^\dagger_i)^{n_i}/\sqrt{n_i!}\,|0\rangle$ a product of phonon Fock eigenstates of $\hat{H}_\mathrm{ph}$, the partition function $\mathcal{Z} = [1 - e^{-\beta\Omega}]^{-N}$ and $\beta = 1/T$.
At $T = 0$, the thermal average reduces to the vacuum  expectation value ($n_{i}=0,  \forall i$), and the lowest-energy feature of $A(\omega,{\bf k}) = -\frac{1}{\pi}\,\mathrm{Im}\,G({\bf k},\omega)$ is a Lorentzian of width $\eta_0$ centered at the polaron energy $E_P({\bf k})$. At finite $T$, the polaron peak is no longer the lowest-energy feature, nor is it  Lorentzian: lower-energy features arise from  absorption of thermal phonons~\cite{Bonca2019}, see Sec.~IV of the Supplemental Material~\cite{SM}.
As detailed in the End Matter, this significantly complicates the extraction of $Z, m^\ast$ and $\tau$ from  $A(\omega,{\bf k})$.

\textit{Momentum Average Approximation:} The simplest MA$^{(0)}$  variational ansatz confines the phonon cloud to a single site, keeping only configurations of the form $c^\dagger_i (b^\dagger_j)^m|\{n\}\rangle$. In the End Matter we present a new diagrammatic resummation formalism that allows us to derive  the MA$^{(0)}$ self-energy $\Sigma^{(0)}_T(\omega)=  [ G_0({\bf k},\omega)]^{-1} - [ G({\bf k},\omega)]^{-1}$, where  $ G_0({\bf k},\omega) = (\omega + i\eta_0 - \epsilon_{\bf k})^{-1}$ is the bare propagator. The result is:
\begin{equation}
  \Sigma^{(0)}_T(\omega) = \frac{\epsilon(\omega)}{1 +
  g_0(\omega)\,\epsilon(\omega)},
\label{eqn:SE_MA0}
\end{equation}
where $\epsilon(\omega)$ (defined in Eq.~\ref{p3} of End Matter) is somewhat akin to a  retarded, $T$-dependent on-site potential, and $g_0(\omega) = \frac{1}{N} \sum_{\bf k}  G_0({\bf k},\omega)$ is the on-site free propagator.

\begin{figure}[t]
\includegraphics[width=\columnwidth]{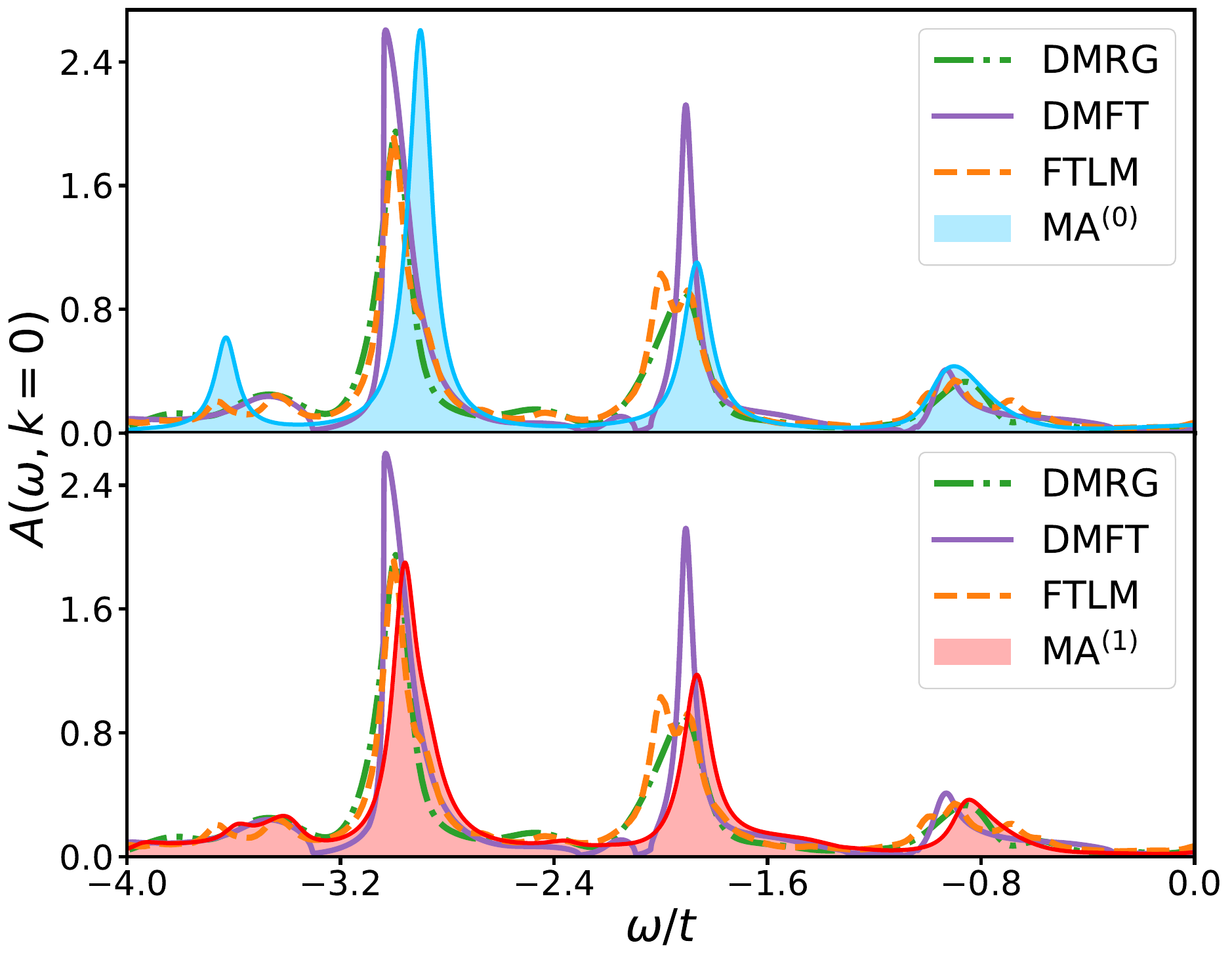}
  \caption{Top:  MA$^{(0)}$ (blue)  and Bottom:  MA$^{(1)}$ (red) 1D spectral functions $A(\omega,k=0)$ at $T = 0.4\,\Omega$, together with VED-FTLM \cite{Bonca2019} (orange), DMRG \cite{Meisner2020} (green) and  DMFT \cite{Mitric2022} (purple) results.  Parameters are $t= \Omega = 1, \lambda=1, \eta_0 = 0.05$.}
   \label{fig:Overlap}
\end{figure}

\begin{figure*}[!th]
    \centering  \includegraphics[width=\textwidth]{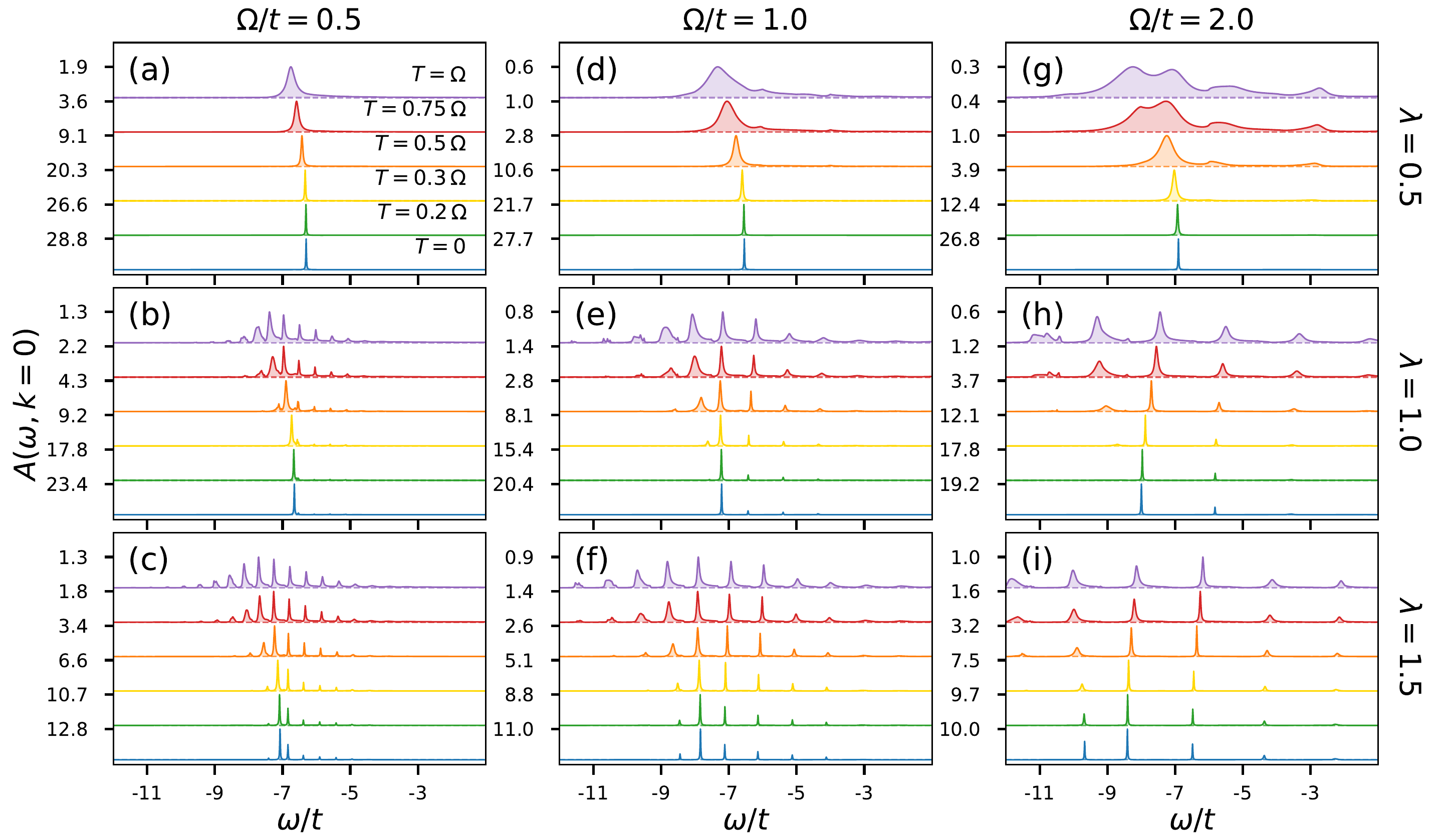} 
 \caption{3D MA$^{(1)}$ spectral functions  $A(\omega,{\bf k}=0)$ of the HP. Columns correspond to phonon energies $\Omega/t=0.5$ (left), $\Omega/t=1.0$ (middle),  $\Omega/t=2.0$ (right). Rows correspond to effective couplings  $\lambda = 0.5$ (top), $\lambda = 1.0$ (middle), $\lambda = 1.5$ (bottom). For each ($\Omega,\lambda$) set, spectral plots are shown for the same set of temperatures $T/\Omega =  0.0, 0.2, 0.3, 0.5, 0.75, 1.0$ from bottom to top; Other parameters are $t=1, \eta_0=0.01$.} 
\label{fig:3D_Akw}
\end{figure*}

We  emphasize that despite the complexity of the problem,  the self-energy admits the closed form solution of Eq.~\ref{eqn:SE_MA0}, making it numerically trivial to implement and evaluate at low numerical cost for any lattice in any dimension. Our solution is exact in the atomic limit $t = 0$ ($\lambda\rightarrow \infty$), therefore the MA results are expected to be very accurate for strong couplings $\lambda\gg 1$. For $t \neq 0$, taking $T \to 0$ recovers the known MA$^{(0)}$ self-energy, which accurately describes polaron  ground-state properties for $\Omega/t \geq 0.5$ in any $D$~\cite{Glenn2006}. The single-site phonon cloud ansatz is most accurate in the anti-adiabatic regime, which is why the results below focus on $\Omega \geq 0.5t$. Extension to multi-site clouds \cite{Dominic2017} will improve accuracy in the adiabatic limit, and is left for future work.

The MA$^{(0)}$ polaron bandwidth is overestimated for weak-to-medium couplings~\cite{Glenn2006}. We have therefore also carried out the finite-$T$ generalization of the one-site MA$^{(1)}$, which includes configurations with an additional phonon unbound to the cloud~\cite{Mona2007}, correcting the polaron bandwidth for all $\lambda$ and improving the placement of the first-order thermal and non-thermal continua. The MA$^{(1)}$ self-energy has the same form as Eq.~(\ref{eqn:SE_MA0})   however with a renormalized $\epsilon(\omega) \to \tilde{\epsilon}(\omega)$, see  Sec.~III of the Supplemental Material \cite{SM}.

\textit{Results:} Below we show representative results; additional ones are included in the  Supplemental Material \cite{SM}.

\textit{Benchmarking against exact 1D methods:} The MA$^{(0)}$ and MA$^{(1)}$ spectral functions $A(\omega,k=0)$ at $T = 0.4\,\Omega$  are compared with DMFT~\cite{Mitric2022}, VED-FTLM~\cite{Bonca2019}, and finite-$T$ DMRG~\cite{Meisner2020} in Fig.~\ref{fig:Overlap}, at crossover coupling $\lambda = 1$ in the extreme quantum regime $\Omega = t = 1$. The agreement between both MA variants and these methods is excellent, especially considering the trivial computational cost of MA. As expected, MA$^{(1)}$ has the quantitative edge. Comparisons at other $\lambda$ and at the Brillouin-zone edge are shown in  Supplemental Material, Sec.~VI~\cite{SM}. We emphasize that MA becomes even more accurate in higher D because the real-space bare propagator decays faster with distance than in 1D \cite{Economou2006}.

\textit{Spectral evolution in 3D:}  Figure~\ref{fig:3D_Akw} shows the evolution with $T$ of the 3D HP  MA$^{(1)}$ spectral functions at  ${\bf k}=(0,0,0)$ for weak ($\lambda = 0.5$, top), intermediate ($\lambda = 1$, middle), and strong ($\lambda = 1.5$, bottom) couplings, at  $\Omega/t = 0.5$ (left), $\Omega/t= 1$ (middle) and $\Omega/t=2$ (right panels). The polaron peaks at $T > 0$ are no longer pure Lorentzians, instead they are resonances inside thermal continua shaped by the polaron density of states \cite{Mirko2013}. We note that for lower-$T$, these continua are too small to be seen on the scale of Figure~\ref{fig:3D_Akw}; this is further discussed in End Matter. Spectral weight transfers from the \textit{qp} peak into these continua as $T$ increases, reflecting more scattering off of thermal phonons. At lowest order perturbation in the coupling strength, the thermal continuum is due to single thermal-phonon absorption and spans [$\min (E_P({\bf k})) - \Omega,\, \max (E_P({\bf k})) - \Omega]$, see Sec. IVB of Supp. Mat \cite{SM}. Higher-order thermal continua involve complex multi thermal-phonon absorption processes, starting from either the polaron band or from higher-energy $T=0$ features such as the second bound state (where it exists) and the polaron+one-phonon continuum, shifted appropriately by multiples of $\Omega$ to lower energies~\cite{Mirko2013}.

\textit{Role of phonon frequency:} For the larger phonon frequencies,  Figure~\ref{fig:3D_Akw} shows that the \textit{qp} peak is robust up to higher-$T$ for intermediate and strong couplings, because here the small-polaron bandwidth is strongly renormalized.  The thermal continua  are thus narrower, resulting in less overlap with the polaron peak.
The $\lambda = 1.5$ results begin to resemble the Lang--Firsov limit, with features spaced by $\sim\Omega$.  This parallels 1D HP behavior~\cite{Bonca2019}, however, in 3D the \textit{qp} weight in the small-polaron regime falls more sharply with increasing $\lambda$, making it increasingly difficult to disentangle the polaron peak from the growing thermal background.

Based on systematic fitting of a large number of data sets generated with very small $\eta_0$ values (see End Matter and Supplemental Material~\cite{SM}),
we find that generically, a \textit{qp} peak can no longer be separated with confidence from  the thermal background above $T \sim 0.5-0.7\,\Omega$; in some regimes, this range is narrower.

\textit{Effective mass and lifetime across dimensions:}  Figure~\ref{fig:eff_mass_lifetime} presents the temperature dependence of $m^\ast$ and the intrinsic thermal broadening (proportional to $\tau^{-1}$) $\eta^\ast(T) = \eta_T - \eta_{T=0}$  for the 1D, 2D, and 3D HP across all coupling regimes, computed via MA$^{(1)}$ at $\Omega=2t$. Results for $\Omega = t$ and $\Omega = 0.75t$  are in  Sec.~V of the Supp. Mat.~\cite{SM}.
   The \textit{qp} weight $Z_T$ and broadening $\eta_T$ are extracted from a Lorentzian fit of the ${\bf k=0}$  \textit{qp} peak, while that remains a good approximation (see End Matter). Because the MA$^{(1)}$ self-energy is local, ${m^\ast}/{m_0} = \left[1 - \left.\frac{d\Sigma_T(\omega)}{d\omega}  \right|_{E_P({\bf k}=0,T)}\right]^{-1}={1}/{Z_{T}},$ where $m_0 = 1/(2t)$ is the bare Bloch electron mass for hyper-cubic lattices. We checked that this agrees with $m^\ast$ extracted from the curvature of $E_P({\bf k},T)$. We also verified that $\eta^\ast(T)$ is independent of the value of $\eta_{T=0}=\eta_0$ when $\eta_0\rightarrow 0$.

Figure~\ref{fig:eff_mass_lifetime} thus maps the $T$-dependence of $m^\ast$ and $\tau\sim 1/\eta^\ast$ across the full parameter space in all dimensions. We note that at stronger coupling, $m^\ast$ renormalizes more strongly with temperature, as expected because  the electron interacts with thermal phonons through the electron--phonon coupling, which is insufficient to strongly renormalize $m^\ast$ at weak coupling. Conversely, at any $T$ the polaron peak broadens more rapidly at weak coupling, because here the thermal continuum widths are of order $\Omega$ and therefore multiple orders of thermal continua overlap with the \textit{qp} peak. As a result, here reliable fitting is limited to $T/\Omega \lesssim 0.3$--$0.5$.

Across all coupling regimes and dimensions, both $m^\ast$ and $\eta^\ast$ increase monotonically with $T$ while the \textit{qp} peak is well defined. This conclusively validates Holstein's 1959 prediction~\cite{Holstein1959_2} that small polarons have a monotonically increasing effective mass and growing scattering rate up to a crossover temperature $T_t \propto {0.5 \Omega}$ above which a coherent quasiparticle cannot be defined,  and extends it well beyond the extreme anti-adiabatic, strong-coupling limit where it was derived. It also corroborates the three-regime picture of Fratini and Ciuchi~\cite{Ciuchi2003} and extends the 1D findings of Bon\v{c}a et al.~\cite{Bonca2019} to a  multi-dimensional and multi-coupling setting.

\textit{Conclusion: } We have shown how single Holstein polarons lose their quasiparticle coherence with increasing temperature, in 1D, 2D, and 3D across the full parameter space. The key physical results are as follows: (i) Both the effective mass and the inverse quasiparticle lifetime increase monotonically with temperature up to {$T\!\sim\!0.5-0.7\Omega$}, beyond which the quasiparticle gradually dissolves into an incoherent thermal continuum. (ii) The characteristic temperature scale governing polaron coherence is set by the phonon frequency $\Omega$, independent of dimensionality and electron--phonon coupling, validating Holstein's original prediction across the entire parameter space and across dimensions. (iii) In 1D, the calculated spectral functions are in excellent quantitative agreement with VED-FTLM, finite-$T$ DMRG, and DMFT; in 2D and 3D, where no other finite-$T$ results are available, this work provides comprehensive predictions for the temperature dependence of the effective mass and quasiparticle lifetime.

These results rest on the finite-$T$ generalization of the Momentum Average approximation, whose closed-form, diagrammatically derived self-energy is asymptotically exact in the strong-coupling limit at arbitrary temperatures, and is evaluated at negligible computational cost in any dimension. We believe this combination to be without analog among finite-temperature methods. Since the derivation of the MA self-energy is not restricted to the Holstein coupling, this framework extends naturally to Fr\"ohlich, Peierls/SSH, and other electron--phonon models, whose finite-$T$ spectral properties are largely unexplored. Furthermore,  extensions of the phonon cloud beyond a single site \cite{Dominic2017} are expected to improve the quantitative accuracy  in the adiabatic regime at a moderate increase in computational cost. These directions provide a natural route toward a broadly applicable analytical framework for finite-temperature properties of polarons.

\textit{Note: } During the completion of
this work, one of us (M. Berciu) and other collaborators developed a complementary method~\cite{TGCEarXiv} for calculating finite-temperature
Green’s functions,  using  the thermofield double formalism extension of the GGCE cluster expansion method \cite{Sous2021}.  The
main qualitative features of our results agree where they
intersect. Specifically, the two works affirm that the polaron effective mass and inverse lifetime grow monotonically with temperature, and that above a temperature
scale set by a fraction of the phonon frequency the quasiparticle peak can no longer be cleanly separated from the
thermal background, signaling the breakdown of a simple
Lorentzian quasiparticle description.

\textit{Acknowledgements: } We thank P. Mitrić, V. Janković, N. Vukmirović, D. Tanasković for sharing with us the DMFT data from Ref.~\cite{Mitric2022}, and D. Jansen and F. Heidrich-Meisner for sending us the finite-$T$ DMRG data from  Ref.~\cite{Meisner2020}. J.S thanks D. Reichman, S. Fomichev, M.Y. Chu, O. Tong and P.  Mitrić for useful discussions. This project was supported in part by the Max Planck--UBC--UTokyo Center for Quantum Materials. J.S. acknowledges funding from the Natural Sciences and Engineering Research Council of Canada in Quantum Computing Program, grant number 543245 and from Mitacs Globalink Research Award IT46276.  J.B. acknowledges the support by the program No. P1-0044 of the Slovenian Research and Innovation Agency (ARIS) and   VIP project KTTK21 under contract no. SN-ZRD/22-27/510. M.B. acknowledges funding from the Natural Sciences and Engineering Research Council of Canada. The finite-$T$ Lorentzian fitting used library functions from LMFIT \cite{LMFIT}.

\textit{Data availability:} The data plotted in this article and its Supplemental Material  is available at Ref.~\cite{HPdata2026}. The data shown for comparison that were produced by other methods (VED-FTLM, finite-$T$ DMRG, and DMFT) were provided by their respective authors and are available from the corresponding original publications~\cite{Bonca2019,Meisner2020,Mitric2022}.

\bibliographystyle{apsrev4-2}
\bibliography{ref}

@article{Franchini2021,
  author  = {Franchini, Cesare and Reticcioli, Michele and Setvin, Martin and Diebold, Ulrike},
  title   = {Polarons in materials},
  journal = {Nat. Rev. Mater.},
  year    = {2021},
  volume  = {6},
  number  = {7},
  pages   = {560--586},
  doi     = {10.1038/s41578-021-00289-w},
  url     = {https://doi.org/10.1038/s41578-021-00289-w}
}

@article{Mirko2013,
  title = {Low-temperature evolution of the spectral weight of a spin-up carrier moving in a ferromagnetic background},
  author = {M\"oller, Mirko and Berciu, Mona},
  journal = {Phys. Rev. B},
  volume = {88},
  issue = {19},
  pages = {195111},
  numpages = {12},
  year = {2013},
  month = {Nov},
  publisher = {American Physical Society},
  doi = {10.1103/PhysRevB.88.195111},
  url = {https://link.aps.org/doi/10.1103/PhysRevB.88.195111}
}

@article{Landau1933,
  author  = {Landau, L. D.},
  title   = {{\"U}ber die Bewegung der Elektronen in Kristallgittern},
  journal = {Phys. Z. Sowjetunion},
  year    = {1933},
  volume  = {3},
  pages   = {664--665}
}

@article{Pekar1946,
  author  = {Pekar, S. I.},
  title   = {Local quantum states of electrons in an ideal ion crystal},
  journal = {Zh. Eksp. Teor. Fiz.},
  year    = {1946},
  volume  = {16},
  pages   = {341--348}
}

@article{Pekar1948,
  author  = {Landau, L. D. and Pekar, S. I.},
  title   = {Effective mass of a polaron},
  journal = {Zh. Eksp. Teor. Fiz.},
  year    = {1948},
  volume  = {18},
  pages   = {419--423}
}

@article{Berciu2007,
  title = {Exact Green's functions for the two-site Hubbard-Holstein Hamiltonian},
  author = {Berciu, Mona},
  journal = {Phys. Rev. B},
  volume = {75},
  issue = {8},
  pages = {081101(R)},
  numpages = {4},
  year = {2007},
  month = {Feb},
  publisher = {American Physical Society},
  doi = {10.1103/PhysRevB.75.081101},
  url = {https://link.aps.org/doi/10.1103/PhysRevB.75.081101}
}

@article{Holstein1959,
  author       = {Holstein, T.},
  title        = {Studies of Polaron Motion. Part I. The Molecular-Crystal Model},
  journal      = {Ann. Phys.},
  volume       = {8},
  number       = {3},
  pages        = {325--342},
  year         = {1959},
  doi          = {10.1016/0003-4916(59)90002-8}
}

@article{Holstein1959_2,
  author       = {Holstein, T.},
  title        = {Studies of polaron motion: Part II. The “small” polaron},
  journal      = {Ann. Phys.},
  volume       = {8},
  number       = {3},
  pages        = {343--389},
  year         = {1959},
  doi          = {https://doi.org/10.1016/0003-4916(59)90003-X}
}

@article{SSH,
  author       = {Su, W. P. and Schrieffer, J. R. and Heeger, A. J.},
  title        = {Solitons in Polyacetylene},
  journal      = {Phys. Rev. Lett.},
  volume       = {42},
  number       = {25},
  pages        = {1698--1701},
  year         = {1979},
  doi          = {10.1103/PhysRevLett.42.1698}
}

@article{SSH2,
  author       = {Bari{\v{s}}i{\'c}, S. and Labb{\'e}, J. and Friedel, J.},
  title        = {Tight Binding and Transition-Metal Superconductivity},
  journal      = {Phys. Rev. Lett.},
  volume       = {25},
  number       = {14},
  pages        = {919--922},
  year         = {1970},
  doi          = {10.1103/PhysRevLett.25.919}
}

@article{Marsiglio1995,
  author  = {Marsiglio, F.},
  title   = {Pairing in the Holstein model in the dilute limit},
  journal = {Physica C},
  year    = {1995},
  volume  = {244},
  number  = {1-2},
  pages   = {21--34},
  doi     = {10.1016/0921-4534(95)00046-1}
}

@article{Marsiglio1993,
  title   = {The spectral function of a one-dimensional Holstein polaron},
  author  = {Marsiglio, F.},
  journal = {Phys. Lett. A},
  volume  = {180},
  number  = {3},
  pages   = {280--284},
  year    = {1993},
  doi     = {10.1016/0375-9601(93)90711-8}
}

@article{Prokofev1998,
  author  = {Prokof'ev, N. V. and Svistunov, B. V.},
  title   = {Polaron problem by diagrammatic quantum Monte Carlo},
  journal = {Phys. Rev. Lett.},
  volume  = {81},
  pages   = {2514--2517},
  year    = {1998},
  doi     = {10.1103/PhysRevLett.81.2514}
}

@article{Cataudella2007,
  author  = {Cataudella, V. and De~Filippis, G. and Mishchenko, A. S. and Nagaosa, N.},
  title   = {Temperature dependence of the angle resolved photoemission spectra in the undoped cuprates: Self-consistent approach to the $t$-$J$-{Holstein} model},
  journal = {Phys. Rev. Lett.},
  volume  = {99},
  pages   = {226402},
  year    = {2007},
  doi     = {10.1103/PhysRevLett.99.226402}
}

@article{Ciuchi1997,
  author  = {Ciuchi, S. and de Pasquale, F. and Fratini, S. and Feinberg, D.},
  title   = {Dynamical mean-field theory of the small polaron},
  journal = {Phys. Rev. B},
  year    = {1997},
  volume  = {56},
  number  = {8},
  pages   = {4494--4512},
  doi     = {10.1103/PhysRevB.56.4494}
}

@article{Kornilovitch1998,
  author  = {Kornilovitch, P. E.},
  title   = {Continuous-Time Quantum Monte Carlo Algorithm for the Lattice Polaron},
  journal = {Phys. Rev. Lett.},
  year    = {1998},
  volume  = {81},
  number  = {24},
  pages   = {5382--5385},
  doi     = {10.1103/PhysRevLett.81.5382}
}

@article{Nery2018,
  author  = {Nery, Jean Paul and Allen, Philip B. and Antonius, Gabriel and Reining, Lucia and Miglio, Anna and Gonze, Xavier},
  title   = {Quasiparticles and phonon satellites in spectral functions of semiconductors and insulators: Cumulants applied to the full first-principles theory and the Fr{\"o}hlich polaron},
  journal = {Phys. Rev. B},
  volume  = {97},
  number  = {11},
  pages   = {115145},
  year    = {2018},
  doi     = {10.1103/PhysRevB.97.115145}
}

@article{Mona2006,
  author  = {Berciu, Mona},
  title   = {Green's Function of a Dressed Particle},
  journal = {Phys. Rev. Lett.},
  volume  = {97},
  number  = {3},
  pages   = {036402},
  year    = {2006},
  doi     = {10.1103/PhysRevLett.97.036402}
}

@article{Mona2007,
  author  = {Berciu, Mona and Goodvin, Glen L.},
  title   = {Systematic improvement of the momentum average approximation for the Green's function of a Holstein polaron},
  journal = {Phys. Rev. B},
  volume  = {76},
  number  = {16},
  pages   = {165109},
  year    = {2007},
  doi     = {10.1103/PhysRevB.76.165109}
}

@article{Glenn2006,
  author  = {Goodvin, Glen L. and Berciu, Mona and Sawatzky, George A.},
  title   = {Green's function of the Holstein polaron},
  journal = {Phys. Rev. B},
  volume  = {74},
  number  = {24},
  pages   = {245104},
  year    = {2006},
  doi     = {10.1103/PhysRevB.74.245104}
}

@book{Economou2006,
  author    = {Eleftherios N. Economou},
  title     = {Green's Functions in Quantum Physics},
  edition   = {3},
  series    = {Springer Series in Solid-State Sciences},
  volume    = {7},
  publisher = {Springer},
  address   = {Berlin, Heidelberg},
  year      = {2006},
  isbn      = {978-3-540-28838-1},
  doi       = {10.1007/3-540-28841-4}
}

@article{Dominic2017,
  title = {Dual coupling effective band model for polarons},
  author = {Marchand, Dominic J. J. and Stamp, Philip C. E. and Berciu, Mona},
  journal = {Phys. Rev. B},
  volume = {95},
  issue = {3},
  pages = {035117},
  numpages = {18},
  year = {2017},
  month = {Jan},
  publisher = {American Physical Society},
  doi = {10.1103/PhysRevB.95.035117},
  url = {https://link.aps.org/doi/10.1103/PhysRevB.95.035117}
}

@article{Sous2021,
  author  = {Carbone, Matthew R. and Reichman, David R. and Sous, John},
  title   = {Numerically exact generalized Green's function cluster expansions for electron-phonon problems},
  journal = {Phys. Rev. B},
  year    = {2021},
  volume  = {104},
  number  = {3},
  pages   = {035106},
  doi     = {10.1103/PhysRevB.104.035106}
}

@article{Ranninger1992,
  author  = {Ranninger, J. and Thibblin, U.},
  title   = {Two-site polaron problem: Electronic and vibrational properties},
  journal = {Phys. Rev. B},
  volume  = {45},
  pages   = {7730--7738},
  year    = {1992},
  doi     = {10.1103/PhysRevB.45.7730}
}

@article{Alexandrov1994,
  author  = {Alexandrov, A. S. and Kabanov, V. V. and Ray, D. K.},
  title   = {From electron to small polaron: An exact cluster solution},
  journal = {Phys. Rev. B},
  volume  = {49},
  pages   = {9915--9923},
  year    = {1994},
  doi     = {10.1103/PhysRevB.49.9915}
}

@article{Fehske1997,
  author  = {Fehske, H. and Loos, J. and Wellein, G.},
  title   = {Spectral properties of the 2D Holstein polaron},
  journal = {Z. Phys. B},
  volume  = {104},
  pages   = {619--627},
  year    = {1997},
  doi     = {10.1007/s002570050498}
}

@article{Fehske2000,
  author  = {Fehske, H. and Loos, J. and Wellein, G.},
  title   = {Lattice polaron formation: Effects of nonscreened electron-phonon interaction},
  journal = {Phys. Rev. B},
  volume  = {61},
  pages   = {8016--8025},
  year    = {2000},
  doi     = {10.1103/PhysRevB.61.8016}
}

@article{Mitric2022,
  author  = {Mitri{\'c}, Petar and Jankovi{\'c}, Veljko and Vukmirovi{\'c}, Nenad and Tanaskovi{\'c}, Darko},
  title   = {Spectral Functions of the Holstein Polaron: Exact and Approximate Solutions},
  journal = {Phys. Rev. Lett.},
  year    = {2022},
  volume  = {129},
  number  = {9},
  pages   = {096401},
  doi     = {10.1103/PhysRevLett.129.096401}
}

@article{Mitric2023,
  author  = {Mitri{\'c}, Petar and Jankovi{\'c}, Veljko and Vukmirovi{\'c}, Nenad and Tanaskovi{\'c}, Darko},
  title   = {Cumulant expansion in the Holstein model: Spectral functions and mobility},
  journal = {Phys. Rev. B},
  year    = {2023},
  volume  = {107},
  number  = {12},
  pages   = {125165},
  doi     = {10.1103/PhysRevB.107.125165}
}

@article{Jankovic2022,
  author  = {Jankovi{\'c}, Veljko and Vukmirovi{\'c}, Nenad},
  title   = {Spectral and thermodynamic properties of the Holstein polaron: Hierarchical equations of motion approach},
  journal = {Phys. Rev. B},
  year    = {2022},
  volume  = {105},
  number  = {5},
  pages   = {054311},
  doi     = {10.1103/PhysRevB.105.054311}
}

@article{Meisner2022,
  author  = {Jansen, David and Bon{\v c}a, Janez and Heidrich-Meisner, Fabian},
  title   = {Finite-temperature optical conductivity with density-matrix renormalization group methods for the Holstein polaron and bipolaron with dispersive phonons},
  journal = {Phys. Rev. B},
  year    = {2022},
  volume  = {106},
  pages   = {155129},
  doi     = {10.1103/PhysRevB.106.155129},
  note    = {Corrected 13 February 2023}
}

@article{Frohlich1950,
  author  = {Fr{\"o}hlich, Herbert and Pelzer, H. and Zienau, S.},
  title   = {Properties of Slow Electrons in Polar Materials},
  journal = {Philos. Mag.},
  volume  = {41},
  number  = {314},
  pages   = {221--242},
  year    = {1950},
  doi     = {10.1080/14786445008521794}
}

@misc{Reichman2025,
  author       = {Nguyen, Haimi and Mandal, Arkajit and Mahajan, Ankit and Reichman, David R.},
  title        = {Mixed Quantum-Classical Methods for Polaron Spectral Functions},
  year         = {2025},
  eprint       = {2505.13365},
  archivePrefix= {arXiv},
  primaryClass = {physics.chem-ph},
  doi          = {10.48550/arXiv.2505.13365},
  note         = {arXiv:2505.13365v2, revised Oct 23, 2025}
}

@misc{TGCEarXiv,
  author        = {Carbone, M. R. and Fomichev, S. and Kloss, B. and Millis, A. J. and Berciu, M. and Reichman, D. R. and Sous, J.},
  title         = {Finite-temperature Green's function cluster expansion from thermofield doubles: Breakdown of the polaron picture},
  year          = {2026},
  eprint        = {2608.18267},
  archivePrefix = {arXiv},
  primaryClass  = {cond-mat.str-el},
  doi           = {10.48550/arXiv.2608.18267}
}

@article{ReichmannCEI2022,
  author  = {Robinson, Paul J. and Dunn, Ian S. and Reichman, David R.},
  title   = {Cumulant methods for electron-phonon problems. I. Perturbative expansions},
  journal = {Phys. Rev. B},
  year    = {2022},
  volume  = {105},
  number  = {22},
  pages   = {224304},
  doi     = {10.1103/PhysRevB.105.224304}
}

@article{ReichmannCEII2022,
  author  = {Robinson, Paul J. and Dunn, Ian S. and Reichman, David R.},
  title   = {Cumulant methods for electron-phonon problems. II. The self-consistent cumulant expansion},
  journal = {Phys. Rev. B},
  year    = {2022},
  volume  = {105},
  number  = {22},
  pages   = {224305},
  doi     = {10.1103/PhysRevB.105.224305}
}

@article{Ciuchi2006,
  author  = {Paganelli, S. and Ciuchi, S.},
  title   = {Tunnelling system coupled to a harmonic oscillator: An analytical treatment},
  journal = {J. Phys.: Condens. Matter},
  year    = {2006},
  volume  = {18},
  number  = {32},
  pages   = {7669--7685},
  doi     = {10.1088/0953-8984/18/32/015}
}

@misc{LMFIT,
  author       = {Newville, Matthew and Otten, Renee and Nelson, Andrew and Stensitzki, Till and Ingargiola, Antonino and Allan, Daniel and Fox, Austin and Carter, Faustin and Rawlik, Michal},
  title        = {LMFIT: Non-Linear Least-Squares Minimization and Curve-Fitting for Python},
  howpublished = {Zenodo (software)},
  year         = {2025},
  month        = jul,
  note         = {Version 1.3.4. Published 2025-07-19},
  doi          = {10.5281/zenodo.16175987},
  url          = {https://doi.org/10.5281/zenodo.16175987}
}

@article{Barisic2007,
  author  = {Bari{\v{s}}i{\'c}, O. S.},
  title   = {Comment on ``Holstein polaron''},
  journal = {Phys. Rev. Lett.},
  year    = {2007},
  volume  = {98},
  number  = {20},
  pages   = {209701},
  doi     = {10.1103/PhysRevLett.98.209701}
}

@article{Glenn2008,
  author  = {Goodvin, G. L. and Berciu, M.},
  title   = {Momentum average approximation for models with electron-phonon coupling dependent on the phonon momentum},
  journal = {Phys. Rev. B},
  volume  = {78},
  number  = {23},
  pages   = {235120},
  year    = {2008},
  doi     = {10.1103/PhysRevB.78.235120}
}

@misc{SM,
  howpublished = {See Supplemental Material at [URL will be inserted by publisher] for detailed analysis and supporting data.},
  year         = {2026}
}

@article{vanMechelen2008,
  author  = {van Mechelen, J. L. M. and van der Marel, D. and
             Grimaldi, C. and Kuzmenko, A. B. and Armitage, N. P. and
             Reyren, N. and Hagemann, H. and Mazin, I. I.},
  title   = {Electron-Phonon Interaction and Charge Carrier Mass Enhancement
             in {SrTiO$_3$}},
  journal = {Physical Review Letters},
  volume  = {100},
  number  = {22},
  pages   = {226403},
  year    = {2008},
  doi     = {10.1103/PhysRevLett.100.226403}
}

@article{Wang2016,
  author  = {Wang, Z. and McKeown Walker, S. and Tamai, A. and Wang, Y. and
             Ristic, Z. and Bruno, F. Y. and de la Torre, A. and
             Ricc{\`o}, S. and Plumb, N. C. and Shi, M. and Hlawenka, P. and
             S{\'a}nchez-Barriga, J. and Varykhalov, A. and Kim, T. K. and
             Hoesch, M. and King, P. D. C. and Meevasana, W. and
             Diebold, U. and Mesot, J. and Moritz, B. and
             Devereaux, T. P. and Radovic, M. and Baumberger, F.},
  title   = {Tailoring the Nature and Strength of Electron--Phonon
             Interactions in the {SrTiO$_3$}(001) 2D Electron Liquid},
  journal = {Nature Materials},
  volume  = {15},
  number  = {8},
  pages   = {835--839},
  year    = {2016},
  doi     = {10.1038/nmat4623}
}

@article{Herz2017,
  author  = {Herz, L. M.},
  title   = {Charge-Carrier Mobilities in Metal Halide Perovskites:
             Fundamental Mechanisms and Limits},
  journal = {ACS Energy Letters},
  volume  = {2},
  number  = {7},
  pages   = {1539--1548},
  year    = {2017},
  doi     = {10.1021/acsenergylett.7b00276}
}

@misc{HPdata2026,
  author       = {Shannigrahi, Jeet},
  title        = {Data for Universal thermal breakdown of polaron coherence in one, two, and three dimensions},
  year         = {2026},
  version      = {1.0},
  publisher    = {Borealis},
  howpublished = {Dataset},
  doi          = {10.5683/SP4/LYHBP7},
  url          = {https://doi.org/10.5683/SP4/LYHBP7}
}

@article{Linden2005,
author = {Hohenadler, Martin and Evertz, Hans Gerd and von der Linden, Wolfgang},
title = {Quantum Monte Carlo approach to the Holstein polaron problem},
journal = {Phys. Status Solidi B},
volume = {242},
number = {7},
pages = {1406-1413},
doi = {https://doi.org/10.1002/pssb.200440019},
year = {2005}
}

@article{Hague2003,
  author  = {Hague, J. P.},
  title   = {Electron and phonon dispersions of the two-dimensional Holstein model: effects of vertex and non-local corrections},
  journal = {J. Phys.: Condens. Matter},
  volume  = {15},
  number  = {17},
  pages   = {2535--2550},
  year    = {2003},
  doi     = {10.1088/0953-8984/15/17/309}
}

@article{Ciuchi2003,
  author  = {Fratini, Simone and Ciuchi, Sergio},
  title   = {Dynamical Mean-Field Theory of Transport of Small Polarons},
  journal = {Phys. Rev. Lett.},
  volume  = {91},
  number  = {25},
  pages   = {256403},
  year    = {2003},
  doi     = {10.1103/PhysRevLett.91.256403},
  url     = {https://link.aps.org/doi/10.1103/PhysRevLett.91.256403}
}

@article{Damascelli2003,
  author  = {Damascelli, Andrea and Hussain, Zahid and Shen, Zhi-Xun},
  title   = {Angle-resolved photoemission studies of the cuprate superconductors},
  journal = {Rev. Mod. Phys.},
  volume  = {75},
  number  = {2},
  pages   = {473--541},
  year    = {2003},
  doi     = {10.1103/RevModPhys.75.473}
}

@book{Mahan2000,
  author    = {Mahan, Gerald D.},
  title     = {Many-Particle Physics},
  edition   = {3rd},
  series    = {Physics of Solids and Liquids},
  publisher = {Kluwer Academic / Plenum Publishers},
  address   = {New York},
  year      = {2000},
  isbn      = {0306463385}
}

@article{Ranninger1997,
  author  = {de Mello, E. V. L. and Ranninger, J.},
  title   = {Dynamical properties of small polarons},
  journal = {Phys. Rev. B},
  year    = {1997},
  volume  = {55},
  pages   = {14872--14885},
  doi     = {10.1103/PhysRevB.55.14872}
}

@article{Bonca2019,
  author  = {Bon{\v c}a, J. and Trugman, S. A. and Berciu, M.},
  title   = {Spectral function of the Holstein polaron at finite temperature},
  journal = {Phys. Rev. B},
  year    = {2019},
  volume  = {100},
  number  = {9},
  pages   = {094307},
  doi     = {10.1103/PhysRevB.100.094307}
}

@article{Meisner2020,
  author  = {Jansen, David and Bon{\v c}a, Janez and Heidrich-Meisner, Fabian},
  title   = {Finite-temperature density-matrix renormalization group method for electron-phonon systems: Thermodynamics and Holstein-polaron spectral functions},
  journal = {Phys. Rev. B},
  year    = {2020},
  volume  = {102},
  number  = {16},
  pages   = {165155},
  doi     = {10.1103/PhysRevB.102.165155}
}

\section*{End Matter}

\begin{figure}[b]
\includegraphics[width=0.7\columnwidth]{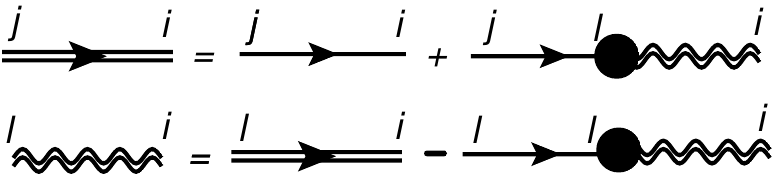}
\caption{Diagrammatic solution for Eq.~(\ref{p4}). The straight double-line is $G_{i-j}(\omega)$, the straight  line is $G^0_{i-j}(\omega)$, the filled circle is $\epsilon(\omega)$, and the wriggly double-line is the auxiliary Green's function defined in the bottom panel. }
\label{figS1}
\end{figure}

\begin{figure*}[t]
    \centering
    \includegraphics[width=0.67\columnwidth]{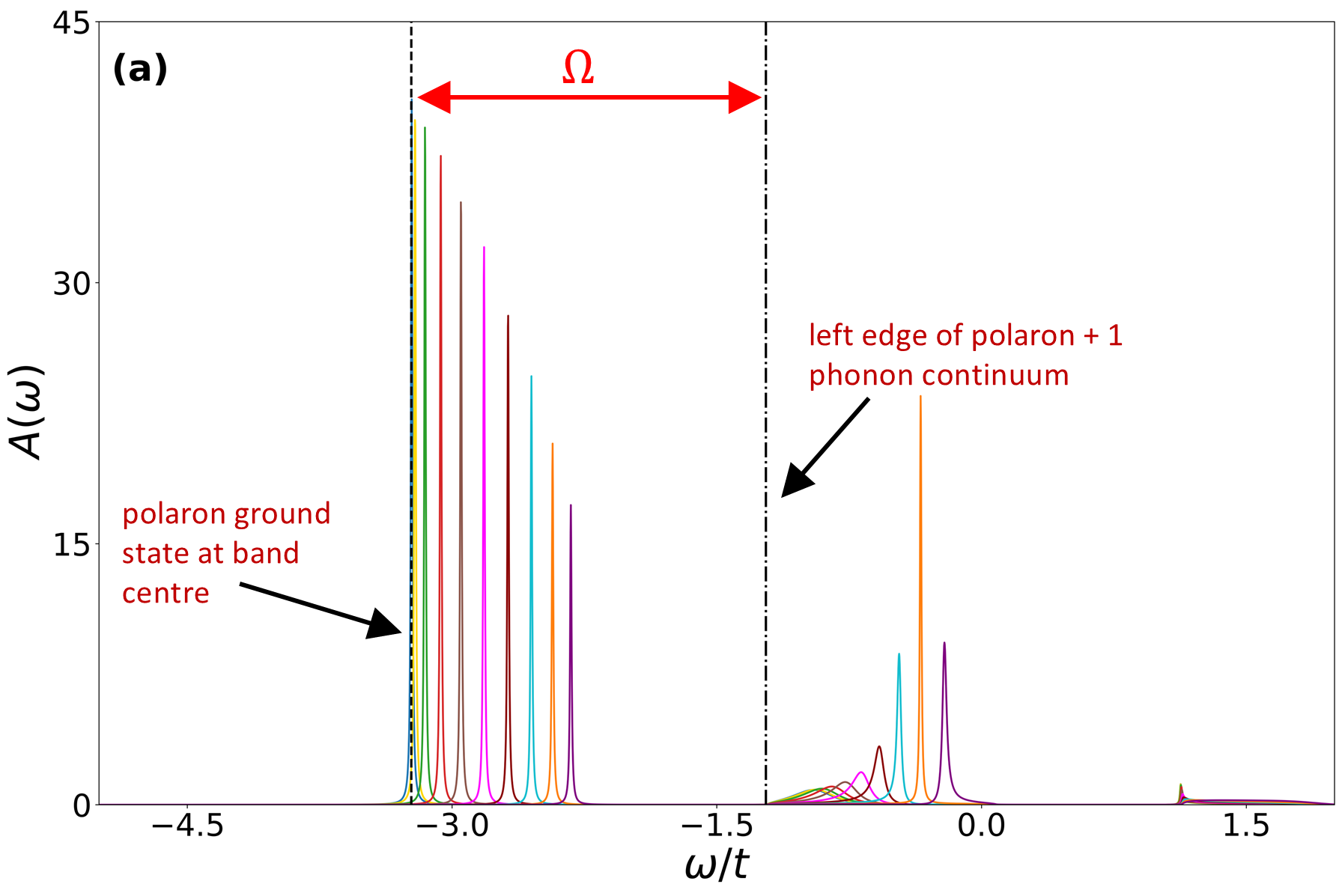}
    \includegraphics[width=0.67\columnwidth]{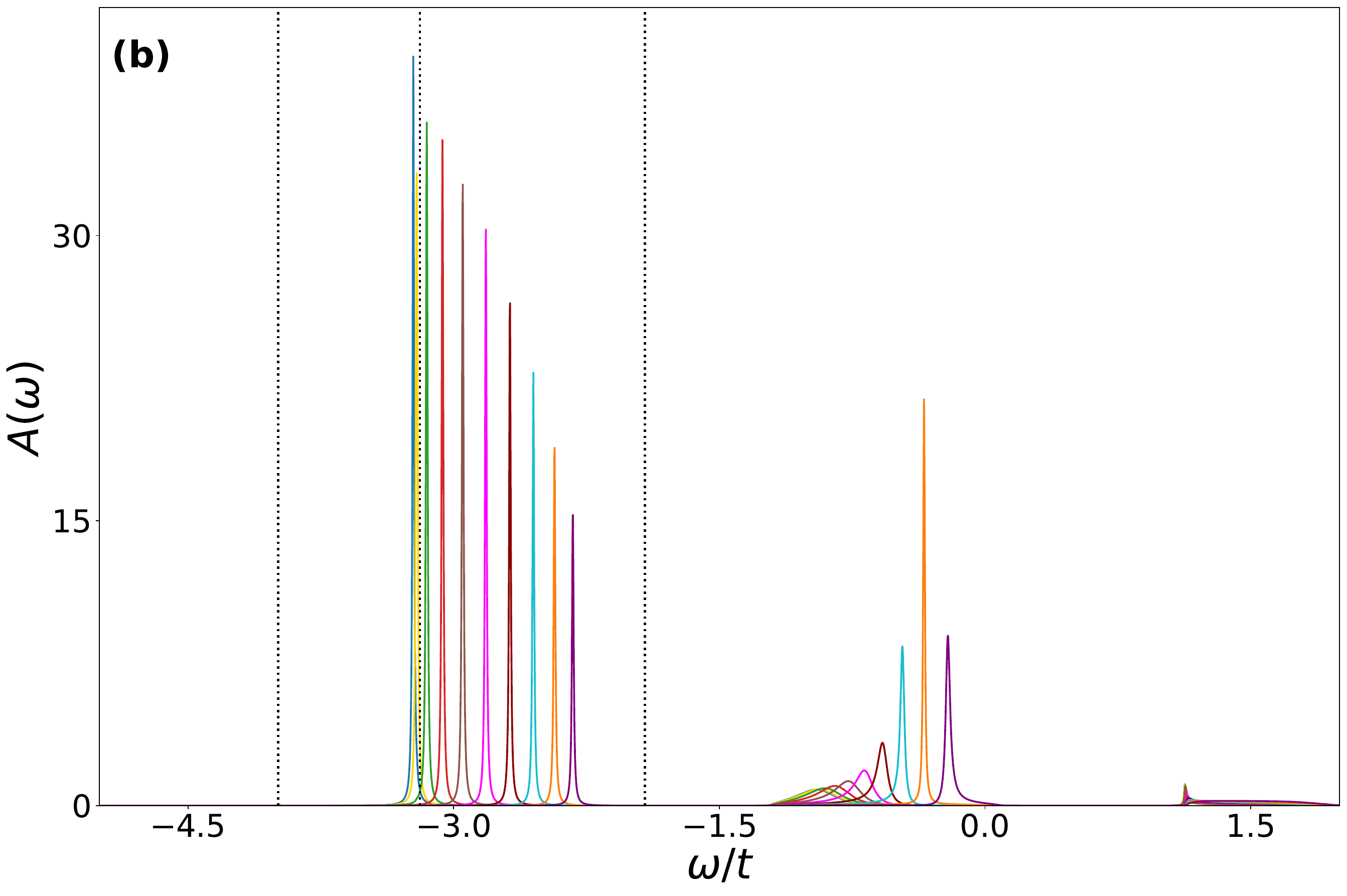}
    \includegraphics[width=0.67\columnwidth]{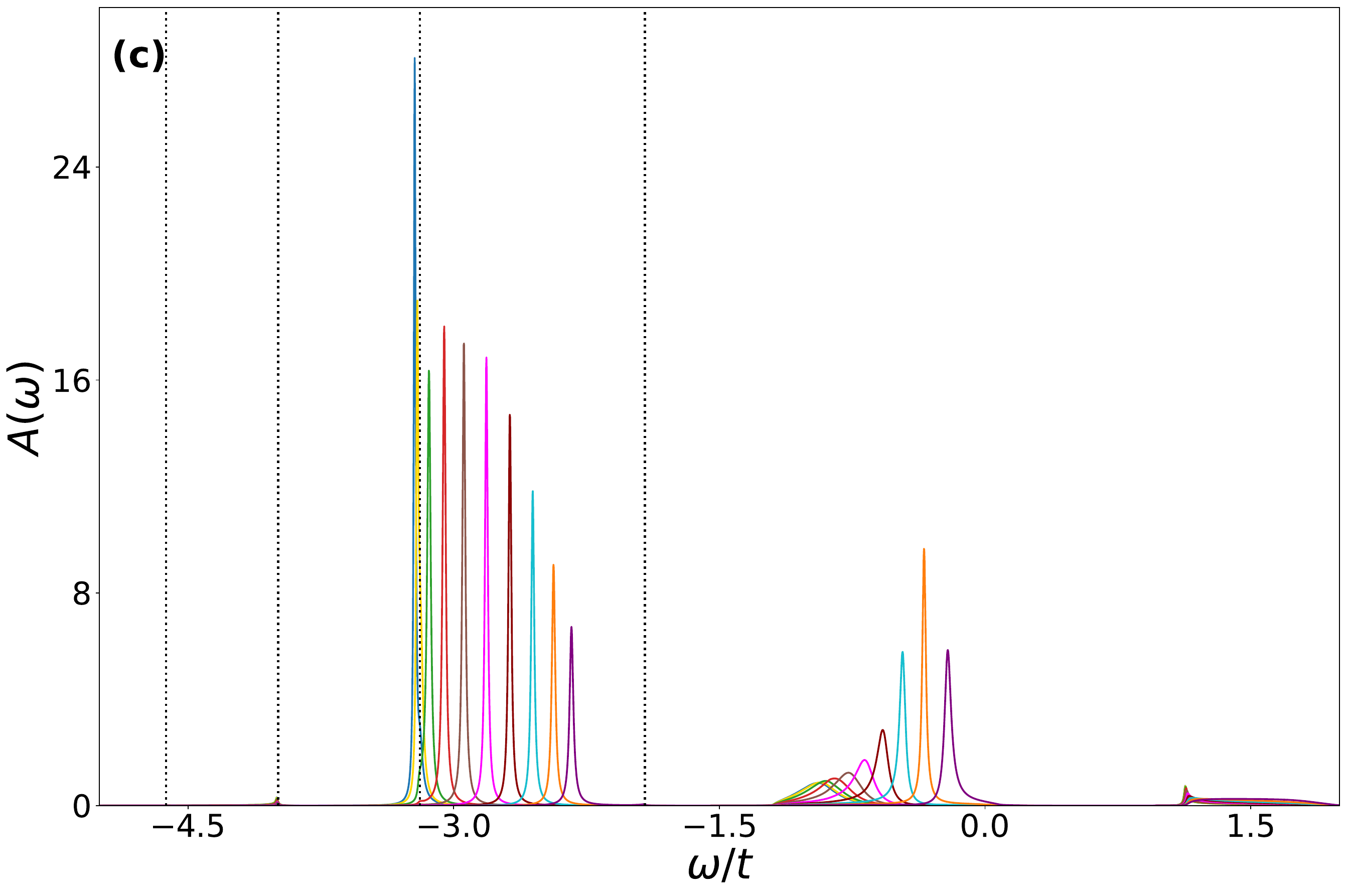}

    \vspace{0.5em}

    \includegraphics[width=0.67\columnwidth]{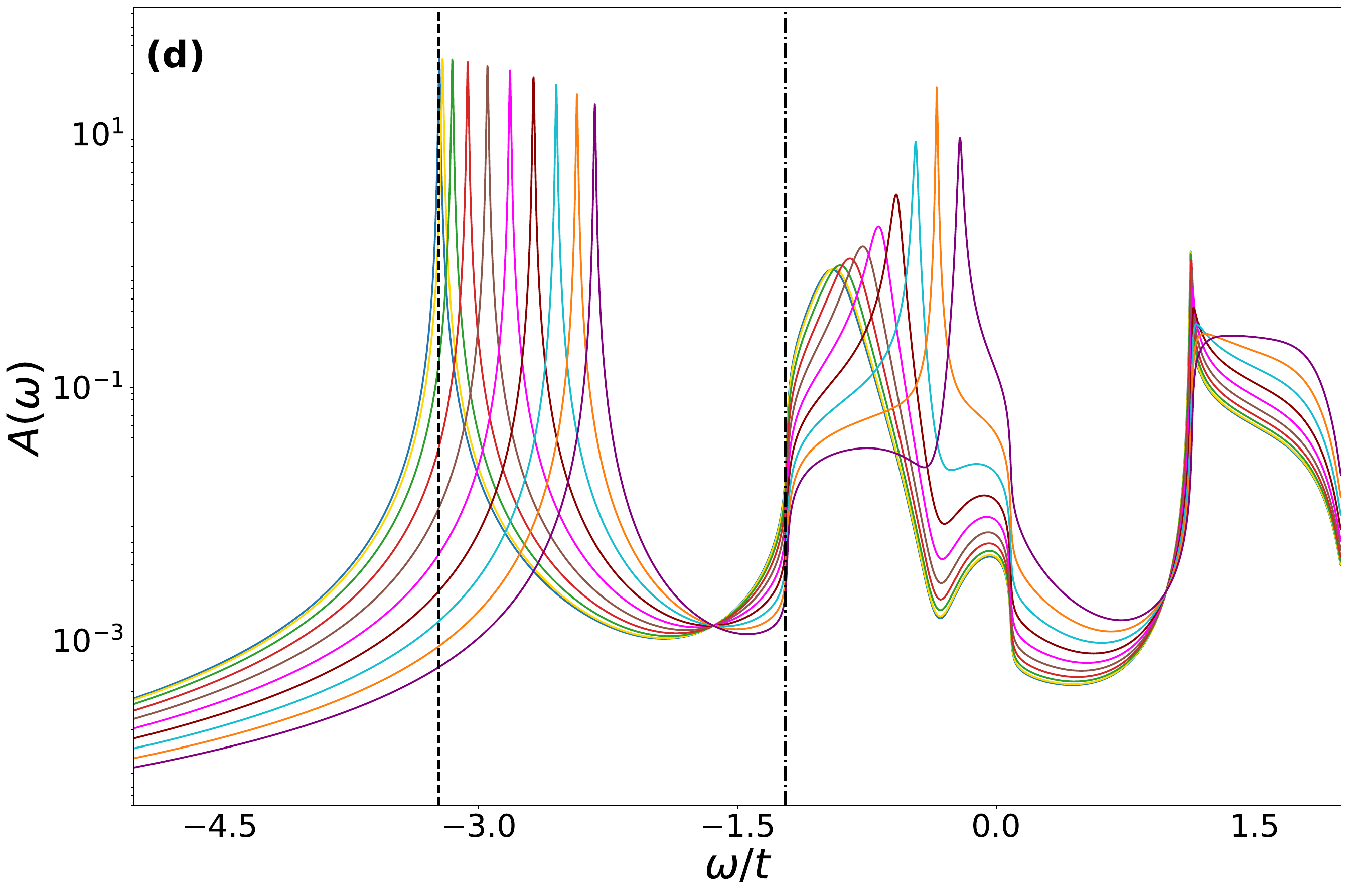}
    \includegraphics[width=0.67\columnwidth]{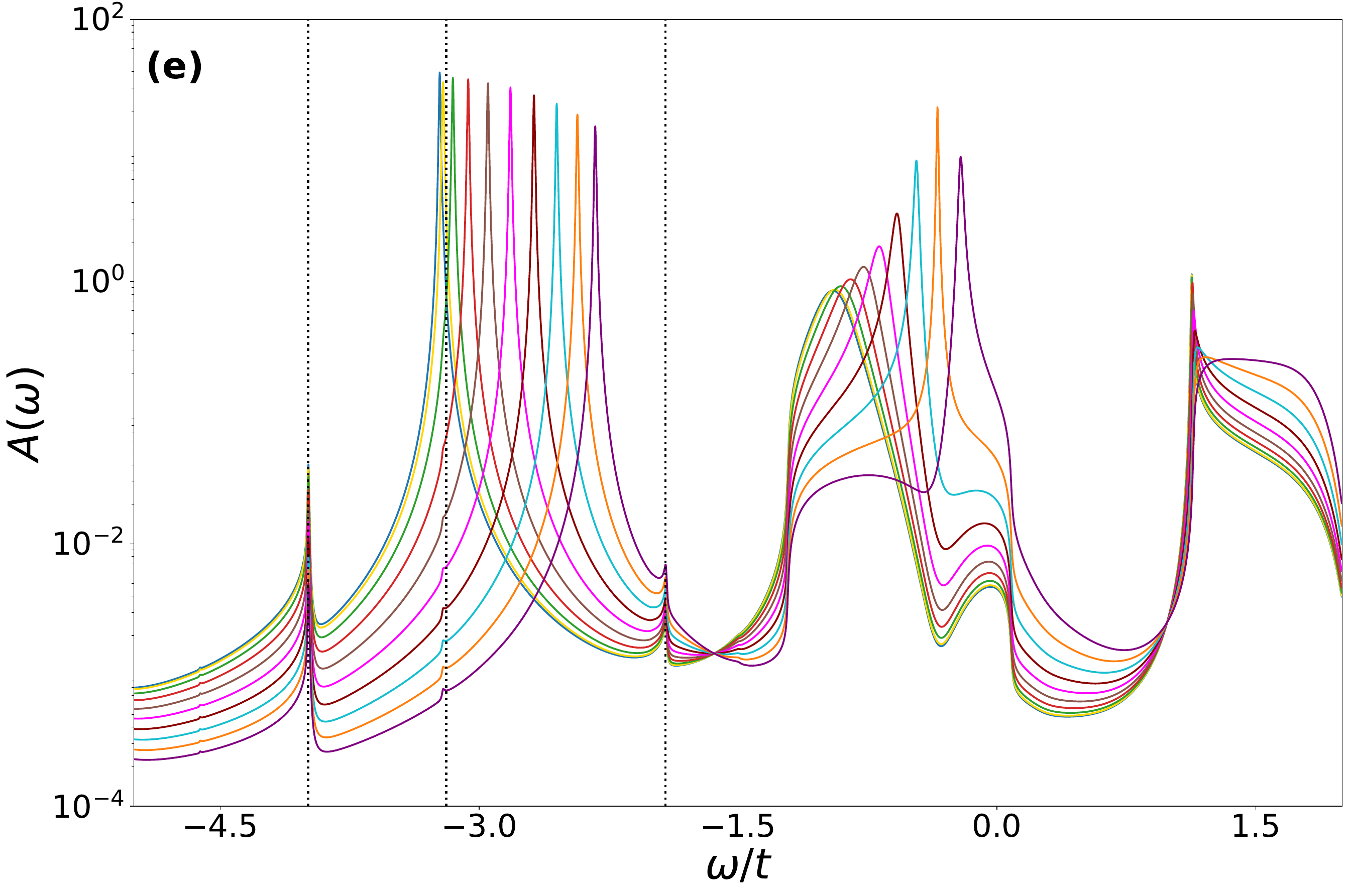}
    \includegraphics[width=0.67\columnwidth]{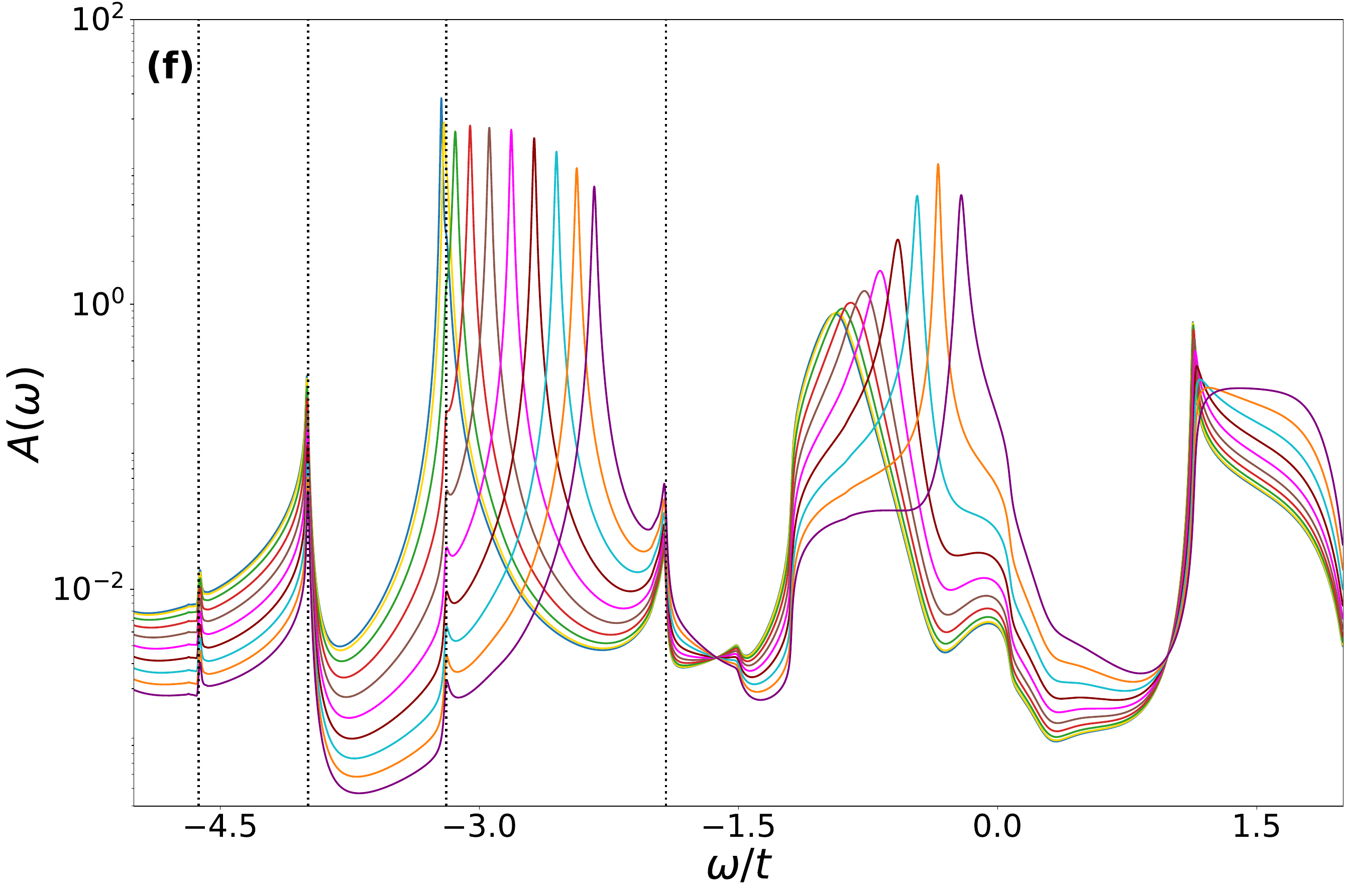}

    \caption{1D MA$^{(1)}$ $A(k, \omega)$ for $0 \le k \le \pi/2$. The upper row shows the linear-scale plots for $T=0.001$ (left), $0.3$ (middle), and $0.5$ (right),  while the lower row shows the corresponding logarithmic-scale plots.
    For $T=0$ [panels (a) and (d)], the lowest-energy feature is the dispersing \textit{qp} Lorentzian peak, followed by the polaron+one-phonon continuum. For clarity, $E_P(k=0)$ is marked by a dashed vertical line and the left edge of the continuum is marked by a dash-dot vertical line. At finite $T$ [panels (b), (c), (e), and (f)], the dispersing polaron peak is a resonance inside an overlapping thermal continuum, marked by dotted vertical lines in (e) and (f). As a result, a fit to a sum of Lorentzians  becomes increasingly poor with increasing $T$. Note also the appearance of a lower-energy thermal continuum below the \textit{qp} features.
    Other parameters are $\lambda=1.0$, $\Omega=2t$, $\eta_0=0.005$.}
    \label{figEM2}
\end{figure*}

\textit{Derivation of the MA$^{(0)}$ self-energy:} We outline here the key steps in deriving the one-site MA$^{(0)}$ self-energy at finite $T$. The diagrammatic resummation strategy is new, and yields a result that is exact in the atomic limit at all temperatures.

We define $G_{ij}(\{ n\}, \omega) \equiv \langle \{ n\}|c_i \hat{G}(\omega + \Omega \sum_{i}^{}n_i) c_j^\dagger|\{n\}\rangle$ and note that $G({\bf k},\omega) = \sum_{i} \exp(-i {\bf k}{\bf R}_i) G_i(\omega) $, where $G_{i-j}(\omega) = \langle G_{ij}(\{ n\}, \omega)\rangle_{\rm th}$ is the thermal average.

Dyson's identity  $\hat{G}(\omega) = \hat{G}_0(\omega) + \hat{G}(\omega) \hat{H}_\mathrm{ep} \hat{G}_0(\omega)$ where $\hat{G}_0(\omega)=[\omega + i\eta_0 - \hat{H}_0]^{-1}$ and $\hat{H}_0=\hat{H}_\mathrm{e}+\hat{H}_\mathrm{ph}$, leads to
\begin{align}
  & G_{ij}(\{ n\}, \omega) = G^0_{i-j}(\omega) \nonumber \\
  & + g \sum_{l}^{} \left[f^{il}_1(\{ n\}, \omega) + \tilde{f}^{il}_1(\{ n\},\omega)\right] G^0_{l-j}(\omega).
\label{eqn:dyson_gij}
\end{align}
Here $G^0_{i-j}(\omega)=\langle 0 |c_i \hat{G}_0(\omega) c_j^\dagger|0\rangle$ is the bare propagator, and  $f^{il}_n(\{ n\}, \omega) \equiv \langle \{ n\}|c_i \hat{G}(\omega + \Omega \sum_{i}^{}n_i) c_l^\dagger b_l^{\dagger n}|\{n\}\rangle$ and  $\tilde{f}^{il}_n(\{ n\}, \omega) \equiv \langle \{ n\}|c_i \hat{G}(\omega + \Omega \sum_{i}^{}n_i) c_l^\dagger b_l^{n}|\{n\}\rangle$. If the particle only affects the phonon distribution at one site (the MA$^{(0)}$ assumption), then $f^{il}_n(\{ n\},\omega) = a_n(n_l,\omega) f^{il}_{n-1}(\{ n\},\omega)$ and  $\tilde{f}^{il}_n(\{ n\},\omega) = b_n(n_l,\omega) \tilde{f}^{il}_{n-1}(\{ n\},\omega)$, where $a_n(n_l,\omega)$ and $b_n(n_l,\omega)$ are continued fractions listed in Sec.~I of the Supplemental Material\cite{SM}. We then find
\begin{equation}
G_{ij}(\{ n\}, \omega) = G^0_{i-j}(\omega) + \sum_{l}^{} G_{il}(\{ n\},\omega)A(n_l,\omega) G^0_{l-j}(\omega),
\label{eqn:gij_A_iteration}
\end{equation}
where $A(n_l,\omega) = g[a_1(n_l,\omega)+b_1(n_l,\omega)]$ depends only on the number $n_l$ of thermal phonons at site $l$. Because of this, $\langle G_{il}(\{ n\},\omega)A(n_l,\omega)\rangle_{\rm th}\ne \langle G_{il}(\{ n\},\omega)\rangle_{\rm th}\langle A(n_l,\omega)\rangle_{\rm th}$ and we cannot carry out the thermal average of Eq. (\ref{eqn:gij_A_iteration}) directly.

Instead, we use iterations to formally expand $G_{ij}(\{ n\}, \omega) = G^0_{i-j}(\omega) + \sum_{m=1}^{\infty} T^{ij}_m(\{ n\},\omega)$, where
\begin{align}
&  T^{ij}_m(\{ n\},\omega) = \sum_{l_1,...,l_m}^{} G^0_{i-l_1}(\omega)A(n_{l_1},\omega)G^0_{l_1-l_2}(\omega)\nonumber
  \\ &\times A(n_{l_2},\omega)G^0_{l_2-l_3}(\omega)\cdots A(n_{l_m},\omega)G^0_{l_m-j}(\omega).
  \label{ex}
  \end{align}
At energies $\omega\sim E_P({\bf k})$ which lie well below the free particle continuum, $G^0_{i-j}(\omega)$ decreases exponentially with the distance $|{\bf R}_i-{\bf R}_j|$~\cite{Economou2006}. The largest contributions to  $T^{ij}_m(\{ n\}, \omega)$ are therefore when $l_1=\cdots=l_m$, followed by those with $l_1=\cdots=l_p \ne l_{p+1}=\cdots=l_m$, then  $l_1=\cdots=l_p \ne l_{p+1}=\cdots=l_{p'}\ne l_{p'+1}=\cdots=l_m$, etc. This re-ordering allows us to accurately carry out the sum over $m$ and then the thermal average (see Sec.~II in the Supplemental Material~\cite{SM} for more details), to find
\begin{align}
\label{p4}
  G_{i-j}(\omega) = G^0_{i-j}(\omega) + \sum_{l_1}^{} G^0_{i-l_1}(\omega)\epsilon(\omega) G^0_{l_1-j}(\omega)\nonumber\\
  +\sum_{l_1, l_2\ne l_1}^{} G^0_{i-l_1}(\omega) \epsilon(\omega) G^0_{l_1-l_2}(\omega)\epsilon(\omega) G^0_{l_2-j}(\omega) +\dots,
  \end{align}
where
\begin{equation}
\label{p3}
\epsilon(\omega) = \left\langle \frac{A(n_l,\omega)}{1-G^0_0(\omega)A(n_l,\omega)}\right\rangle_{\rm th}.
\end{equation}
Eq.~(\ref{p4}) is resummed diagrammatically as shown in the top panel of Fig.~\ref{figS1}, by introducing an auxiliary propagator (wriggly double-line) defined in the lower panel (an alternative solution is given in Sec.~II in the Supplemental Material \cite{SM}). These equations are solved  to find the self-energy of Eq.~(\ref{eqn:SE_MA0}), where $g_0(\omega)\equiv G_0^0(\omega)$ is a simpler notation. It is straightforward to verify that this result is {\em exact in the atomic limit $t=0$ for any $T$}, therefore our results are very accurate for strong coupling $\lambda \gg 1$. For $t\ne 0$ and  $T\rightarrow 0$, we find $\Sigma^{(0)}_T(\omega)= A(0,\omega)$, the expected  MA$^{(0)}$ solution~\cite{Mona2006}.

\textit{Lorentzian approximation for the polaron feature:} Writing the self-energy in terms of its real and imaginary parts  $\Sigma_T = \Sigma^{'}_T + i\Sigma^{''}_T$, and expanding it to first order in $\omega$ near the polaron energy $E_{P}(\mathbf{k},T) = \epsilon_{\mathbf{k}}+\Sigma^{'}_T\!\left(\mathbf{k},E_P(\mathbf{k},T)+i0^+\right)$, the propagator approximates to:
\begin{equation}
G({\mathbf{k},\omega})\approx \frac{Z_{\mathbf{k},T}}{\omega- E_P(\mathbf{k},T) + i \eta_T(\mathbf{k},\omega)},
\label{approx}
\end{equation}
with $Z_{\mathbf{k},T}= [1-\left.
\partial_{\omega}\Sigma^{'}_T(\mathbf{k},z)
\right|_{z=E_P}]^{-1}$ and $\eta_T(\mathbf{k},\omega) =\eta_0 - Z_\mathbf{k,T}[\Sigma^{"}_T(\mathbf{k}, E_P)+ (\omega-E_P)\left.
\partial_{\omega}\Sigma^{"}_T(\mathbf{k},z)
\right|_{z=E_P}] $, where in the latter we suppressed $E_P$'s arguments for simplicity.

From Eq.~(\ref{eqn:SE_MA0}), $\Sigma^{(0)}_{T=0}(\omega)\propto g_0(\omega-\Omega)$. This bare propagator has vanishing imaginary part if $\omega \le \Omega-2Dt$ and $\eta_0\rightarrow0$, therefore near the polaron peak we find $\eta_{T=0} =\eta_0$ as expected from the Lehmann representation, and the spectral weight obtained from Eq.~(\ref{approx}) is a standard Lorentzian
of width $\eta_0$. At finite-$T$, however, Eq.~(\ref{approx})  no longer leads to a  standard Lorentzian because the half-width $\eta_T(\mathbf{k},\omega)$ varies across the frequency range of the peak; this is a direct consequence of the frequency-dependence of $\Sigma^{''}_T({\bf k},\omega)$. Physically, the polaron peak is now a resonance inside a continuum controlled by the polaron bandwidth, as shown in Fig.~\ref{figEM2}. Only at sufficiently low $T/\Omega$ where $\Sigma^{''}_T$ is roughly constant near $E_P$ because the weight of the continuum is very small, is the \textit{qp} feature well-approximated by a Lorentzian whose half-width $\eta_T$ can be extracted with confidence.

\textit{Details for the multi-Lorentzian fitting of finite-$T$ spectral features:} For each parameter set, the finite-$T$ values of $\eta^\ast$ and $m^\ast=1/Z_T$ were extracted from the \textit{qp } peak in $A({\bf k=0}, \omega)$  through a multi-Lorentzian fitting algorithm,  as follows:

(i) Each data set contains a baseline  $T\approx0$ case. The algorithm identifies its lowest-energy feature as the \textit{qp} peak. A single Lorentzian is fitted to this peak and the $T=0$ peak energy $E_P({\bf k=0})$,\textit{  qp }weight $Z_{\textbf{k}=0}$, and width $\eta_0$ are recorded.

 (ii) For every $T>0$, an energy window of half-width $\sim 0.2t-0.5t$ is centered at the previous-$T$ polaron energy, so as to include nearby thermal features that may appear on either side of the \textit{qp} peak as $T$ increases, while excluding the higher-energy excited states of the $T=0$ spectrum.

(iii) The spectrum in this window is fitted with a sum of four Lorentzians: the principal Lorentzian for the \textit{qp} peak and the others for thermal continua features. At each value of $T$, the initial guesses for the fitting parameters are the previous-$T$ values. This is physically motivated because for small $\delta T$ steps, the evolution of the spectrum must be continuous. Moreover, the fitting parameters are constrained so that the thermal features are broader than the \textit{qp} peak, and the peak positions of all Lorentzians are restricted to the aforementioned window.  We used fitting routines from LMFIT, based on non-linear least squares minimization~\cite{LMFIT}.

(iv) The fitting process is attempted for all available $T$ values in the dataset (typically up to $T\sim\Omega$). The cut-off temperature up to which we show results is based on our confidence level, which is determined by three metrics: the goodness of fit $R^2>0.99$; the absence of sudden jumps or kinks in the fitted $\eta^\ast$ or $m^\ast$ data; and visual inspection of the fit quality, especially when the spectrum develops too many thermal features for the polaron pole to be reliably isolated. The combination of the above resulted in the discarding of data-sets corresponding to  difficult-to-fit parameters.

This algorithm  assumes that a temperature step of roughly $\delta T = 0.025\Omega$ resolves the continuous evolution of the\textit{ qp} peak and its nearby thermal features within the relevant window; the thermal features are sufficiently `pole-like' that they are approximated as Lorentzians.

The ability of MA to cheaply generate a large amount of high-resolution data (the frequency grid was resolved at $\Delta\omega/t \sim 10^{-4}$), with fine $\delta T$ steps and an arbitrarily low $\eta_0$, allowed us to access the whole parameter regime and reliably extract the Lorentzian fit parameters up to $T/\Omega = 0.75$, in some cases. This establishes the ability of MA to access \textit{qp} properties at finite-$T$, hitherto out of reach without extensive computational resources. All simulations presented here were performed on a 16-core personal computer.

\end{document}


\preprint{APS/123-QED}

\title{Supplemental material for ``Universal thermal breakdown of polaron coherence in one, two, and three dimensions''}

\author{Jeet Shannigrahi}
\affiliation{Department of Physics and Astronomy, University of British Columbia, Vancouver, British Columbia, Canada, V6T 1Z1}
\affiliation{Quantum Matter Institute, University of British Columbia, Vancouver, British Columbia, Canada, V6T 1Z4}

\author{Janez Bon\v{c}a}
\affiliation{Faculty of Mathematics and Physics, University of Ljubljana, 1000 Ljubljana, Slovenia}
\affiliation{Josef Stefan Institute, 1000 Ljubljana, Slovenia}

\author{Mona Berciu}
\affiliation{Department of Physics and Astronomy, University of British Columbia, Vancouver, British Columbia, Canada, V6T 1Z1}
\affiliation{Quantum Matter Institute, University of British Columbia, Vancouver, British Columbia, Canada, V6T 1Z4}

\begin{abstract}
Here we supplement the main text with additional analytical details of the MA$^{(0)}$ and MA$^{(1)}$ approximations, a physical interpretation of the finite-$T$ spectral features, and further numerical comparisons. Sec.~\ref{SM:CF} lists the continued fractions for the MA$^{(0)}$ phonon cloud. Sec.~\ref{SM:MA0_derivation} gives an alternate algebraic derivation of the MA$^{(0)}$ self-energy, complementing the diagrammatic derivation in the End Matter. Sec.~\ref{SM:MA1_derivation} summarizes the MA$^{(1)}$ self-energy. Sec.~\ref{SM:Physical} provides a physical interpretation of the finite-$T$ spectral features via the Born approximation. Sec.~\ref{SM:Additional_data} is split across several subsections presenting additional finite-$T$ spectral data in different dimensions across different regimes, and the finite-$T$ effective polaron mass $m^\ast$ and thermal broadening $\eta^\ast$ for $\Omega = 1.0$ and $\Omega = 0.75$. Sec.~\ref{SM:comparison} presents additional benchmarking against FTLM-VED, DMFT and DMRG.
\end{abstract}

\maketitle

\section{\texorpdfstring{Continued fractions for the MA$^{(0)}$ phonon cloud}{Continued fractions for the MA(0) phonon cloud}}
\label{SM:CF}

Within MA$^{(0)}$, the continued fractions for the phonon emission branch $a_m(n_l,\omega)$ and annihilation branch $b_m(n_l,\omega)$ are:
\begin{equation}
a_m(n_l,\omega) = \frac{(m+n_l)\,g\,g_0(\omega-m\Omega)}{1-g\,g_0(\omega-m\Omega)\,a_{m+1}(n_l,\omega)}
\label{eqn:am_cf}
\end{equation}
\begin{equation}
b_m(n_l,\omega) = \frac{(n_l-m+1)\,g\,g_0(\omega+m\Omega)}{1-g\,g_0(\omega+m\Omega)\,b_{m+1}(n_l,\omega)}
\label{eqn:bm_cf}
\end{equation}
This results in $A(n_l,\omega) = g\bigl[a_1(n_l,\omega)+b_1(n_l,\omega)\bigr]$, with
\begin{equation}
\resizebox{\columnwidth}{!}{$\displaystyle a_1(n_l,\omega)= \cfrac{(1+n_l)\,g\,g_0(\omega-\Omega)}{1- \cfrac{g^2g_0(\omega-\Omega)\,(2+n_l)g_0(\omega-2\Omega)}{1- \cfrac{g^2g_0(\omega-2\Omega)\,(3+n_l)g_0(\omega-3\Omega)}{1-\cdots}}} $}
\label{a1}
\end{equation}
\begin{equation}
\resizebox{\columnwidth}{!}{$\displaystyle b_1(n_l,\omega)= \cfrac{\,n_l\,g\,g_0(\omega+\Omega)}{1- \cfrac{g^2g_0(\omega+\Omega)\,(n_l-1)g_0(\omega+2\Omega)}{1- \cfrac{g^2g_0(\omega+2\Omega)\,(n_l-2)g_0(\omega+3\Omega)}{1-\cdots}}} $}
\label{b1}
\end{equation}
At $T=0$, only the $n_l=0$ term contributes. In this case,  $a_1(n_l=0,\omega)$ matches the $T=0$  continued fraction Eq.~(12) in Ref.~\cite{Mona2007}, while $b_1(n_l=0,\omega)\equiv 0$, recovering the expected $T=0$ MA$^{(0)}$ self-energy.

\section{\texorpdfstring{Alternate derivation of the MA$^{(0)}$ self-energy at finite $T$}{Alternate derivation of the MA(0) self-energy at finite T}}
\label{SM:MA0_derivation}

Here we re-derive Eq.~(2) of the main text algebraically, as a complement to the diagrammatic approach in the End Matter. The starting point is the thermal average of the expansion $G_{i-j}(\omega) = G^0_{i-j}(\omega) + \sum_{m=1}^{\infty} T^{ij}_m(\omega)$, where
\begin{align}
    & T_p^{ij}(\omega) = \sum_{l_1,\ldots,l_p} G_0(i-l_1,\omega)\,G_0(l_1-l_2,\omega)\cdots \nonumber \\ &\times G_0(l_p-j,\omega)\, \bigl\langle A(n_{l_1},\omega)\cdots A(n_{l_p},\omega)\bigr\rangle_\mathrm{th}
\label{eqn:Tp_thermal_average}
\end{align}
is the thermal average of Eq.~(5) in the main text. At energies $\omega \sim E_P(\mathbf{k})$ below the free-particle continuum, $G^0_{i-j}(\omega)$ decays exponentially with $|R_i-R_j|$ (more steeply in higher dimensions, explaining why MA becomes more accurate with increasing $D$). The dominant contribution to $T^{ij}_p$ thus comes from $l_1=\cdots=l_p$ (single site), giving
\begin{equation}
T_p \approx \bigl\langle A^p(n_l,\omega)\bigr\rangle_\mathrm{th} g_0^{\,p-1}(\omega)\,G_0^2(k,\omega),
\label{eqn:Tp_single_site_approx}
\end{equation}
where we use the simpler notation $g_0(\omega)\equiv G^0_0(\omega)$. This result is exact for $p=1$ and becomes exact for all $p$ in the atomic limit $t=0$ where $G^0_{i-j}(\omega)= \delta_{i,j}/(\omega+i\eta_0)$. Summing over $p$  (a geometric series) these contributions add up to
\begin{equation}
\epsilon(\omega) = \sum_{p=1}^{\infty}\bigl\langle A^{p}\bigr\rangle g_0^{\,p-1} = \Bigl\langle \frac{A(n_l,\omega)}{1-g_0(\omega)A(n_l,\omega)} \Bigr\rangle_\mathrm{th},
\label{eqn:e(w)_defn}
\end{equation}
which is Eq.~(7) of the End Matter. Truncating at this (single-site) level gives a self-energy  that is exact for $t=0$  but has the wrong $T\to 0$ limit for $t\neq 0$.

To recover the correct $T\to 0$ limit, we include contributions from paths visiting two or more distinct sites. Using the shorthand $A_{p,m}$ for the sum of thermal averages obtained by partitioning $p$ factors of $A$ into $m$ groups of equal-site indices (e.g.\ $A_{3,1}=\langle A^3\rangle$, $A_{3,2}=2\langle A\rangle\langle A^2\rangle$, $A_{3,3}=\langle A\rangle^3$), and approximating closed-loop contributions by their local equivalents (valid because hopping away from a site is exponentially suppressed at low $\omega$), the full $T_p$ takes the compact form
\begin{equation}
T_{p}(\omega)= \sum_{l=1}^{p} A_{p,l}\,g_{0}^{\,p-l}(\omega)\, G_{0}^{2}(k,\omega)\, \bigl[G_{0}(k,\omega)-g_{0}(\omega)\bigr]^{l-1}.
\label{eqn:Tp_final}
\end{equation}
Substituting Eq.~(\ref{eqn:Tp_final}) into $G_{i-j}(\omega) = G^0_{i-j}(\omega)+\sum_p T^{ij}_p$, doing the sum over $p$ and Fourier transforming  gives
\begin{align}
&G(k,\omega) =G_{0}(k,\omega) +\frac{G_{0}^{2}(k,\omega)\epsilon(\omega)} {1-\epsilon(\omega) \bigl[G_{0}(k,\omega)-g_{0}(\omega)\bigr]}
\label{eqn:G_resummed_epsilon}
\end{align}
from which the MA$^{(0)}$ self-energy follows after some straightforward algebra:
\begin{equation}
\Sigma_T^{(0)}(\omega) = \frac{\epsilon(\omega)}{1+g_0(\omega)\,\epsilon(\omega)},
\label{eqn:MA0_SE_supp}
\end{equation}
recovering Eq.~(2) of the main text. This result is exact at $t=0$ for any $T$, and reduces to the known MA$^{(0)}$ self-energy \cite{Glenn2006} as $T\to 0$.

\section{\texorpdfstring{MA$^{(1)}$ self-energy at finite $T$}{MA(1) self-energy at finite T}}
\label{SM:MA1_derivation}

Within MA$^{(1)}$, an additional phonon may be absorbed or emitted outside the single-site cloud \cite{Mona2007}, enlarging the variational space and correcting the polaron bandwidth at all $\lambda$. The MA$^{(1)}$ self-energy has the same form as Eq.~(2) of the main text but with $\epsilon(\omega)\to\widetilde\epsilon(\omega)$, obtained by renormalizing the continued fractions $a_1$ and $b_1$ as follows.

The annihilation branch $b_1(n_l,\omega)$ is replaced by $\widetilde b_1(n_l,\omega)$, derived from
\begin{equation}
b(n_l,\omega) = g \!\Biggl[ \frac{g (1+n_l)\,F_{1}(\omega-\Omega)}{1 \;-\; g\,a_{2}(n_l,\omega)\,F_{1}(\omega-\Omega)} \;+\; b_{1}(n_l,\omega) \Biggr]
\label{eqn:b(n,w) in MA1}
\end{equation}
where
\begin{equation}
F_1(\omega) = \frac{g_0\bigl(\omega-\Sigma^{(0)}_T(\omega)\bigr)}{1 + g_0\bigl(\omega-\Sigma^{(0)}_T(\omega)\bigr)\Sigma^{(0)}_T(\omega)}
\label{eqn:F1_MA1}
\end{equation}
and $\Sigma^{(0)}_T(\omega)$ is from Eq.~(2) of the main text. Taking the thermal average, we find:
\begin{equation}
\tilde b(\omega) = \biggl\langle \frac{b(n_l,\omega)}{1 \;-\; g_0(\omega)\,b(n_l,\omega)} \biggr\rangle_\mathrm{th}
\label{eqn:btilde_avg}
\end{equation}
Defining $B(\omega) = \tilde b(\omega)/[1+g_0(\omega)\,\tilde b(\omega)]$
and
$G_{B}(\omega) = {g_0\bigl(\omega - B(\omega)\bigr)}/[1 \;+\; B(\omega)\,g_0\bigl(\omega - B(\omega)\bigr)]$
we find
\begin{equation}
\widetilde b_1(n_l,\omega) = \frac{g n_l  G_B(\omega+\Omega)}{1 - g  b_2(n_l,\omega) G_B(\omega+\Omega)}.
\label{eqn:btilde1_MA1}
\end{equation}
This mirrors Eq.~(\ref{b1}) but with the  $G_B$ replacing $g_0$ on the upper two floors of the continuum fraction.

Similarly, the emission branch $a_1(n_l,\omega)$ is replaced by $\widetilde a_1(n_l,\omega)$, derived from
\begin{equation}
\widetilde a_{1}(n_{l},\omega) = \frac{g\,(1 + n_{l})\,G_E(\omega-\Omega)}{1 \;-\; g a_{2}(n_{l},\omega)\,G_E(\omega-\Omega)},
\label{eqn:atilde1_MA1}
\end{equation}
similar to Eq.~(\ref{a1}) but with a renormalized
$G_E(\omega) = g_0\bigl(\omega - A(\omega)\bigr)/[1 + A(\omega)\,g_0\bigl(\omega - A(\omega)\bigr)]$, where
$A(\omega) = \tilde a(\omega)/[1+g_0(\omega)\,\tilde a(\omega)]$,
\begin{equation}
\tilde a(\omega) = \biggl\langle \frac{a(n_l,\omega)}{1 - g_0(\omega)\,a(n_l,\omega)} \biggr\rangle_\mathrm{th},
\label{eqn:atilde_avg}
\end{equation}
and
\begin{equation}
a(n_l,\omega) = g \!\Biggl[\, a_{1}(n_l,\omega) + \frac{g  n_l\,F_{1}(\omega+\Omega)}{1 \;-\; g\,b_{2}(n_l,\omega)\,F_{1}(\omega+\Omega)} \Biggr],
\label{eqn:a_branch_MA1}
\end{equation}

Combining the two, $\widetilde A(n_l,\omega) = g\bigl[\widetilde a_1(n_l,\omega) + \widetilde b_1(n_l,\omega)\bigr]$, and the MA$^{(1)}$ self-energy is
\begin{equation}
\Sigma_T^{(1)}(\omega) = \frac{\widetilde\epsilon(\omega)}{1 + g_0(\omega)\,\widetilde\epsilon(\omega)}
\label{eqn:MA1_SE_supp}
\end{equation}
where
\begin{equation}
\widetilde\epsilon(\omega) = \Bigl\langle \frac{\widetilde A(n_l,\omega)}{1 - g_0(\omega)\,\widetilde A(n_l,\omega)} \Bigr\rangle_\mathrm{th}.
\label{eqn:epsilon_tilde_MA1}
\end{equation}
One can verify that this reduces to the $T=0$ MA$^{(1)}$ self-energy (Eq.~(16) in Ref.~\cite{Mona2007}) as $T\to 0$.

\section{\texorpdfstring{Physical interpretation of finite-$T$ spectral features}{Physical interpretation of finite-$T$ spectral features}}
\label{SM:Physical}

\subsection{Lehmann representation and thermal poles}

The Lehmann representation of the finite-$T$ propagator (Eq.~(1) of the main text) is obtained by inserting a complete set of eigenstates $|\alpha \rangle$ of $\hat{H}$, with eigen-energies $E_\alpha$, giving

\begin{equation}
G({\bf k},\omega)=\sum_{\{n\}}\frac{e^{-\beta\Omega\sum_i n_i}}{\mathcal{Z}}\sum_{\alpha}\frac{\langle\{n\}|c_{\bf k}|\alpha\rangle\langle\alpha|c_{\bf k}^\dagger|\{n\}\rangle}{\omega+i\eta_0-(E_{\alpha}-\Omega\sum_i n_i)}.
\label{eq:lehmann_main}
\end{equation}
For $T>0$, contributions from terms with $n_i>0$ lead to new poles appearing at energies \emph{below} the polaron ground state, arising from the electron absorbing thermal phonons already present in the lattice; these thermal poles form incoherent continua with finite lifetimes. This is in sharp contrast to $T=0$, where there are no thermal phonons ($n_i=0$, $\forall i$) and the poles of the propagator match the eigenenergies of the Hamiltonian. As $T$ increases toward $\sim\Omega$, the thermal continua gain sufficient spectral weight that the quasiparticle feature becomes hard to separate from the background, with direct consequences for the temperature dependence of the quasiparticle properties discussed in the \textit{Results} of the main text.

\subsection{Thermal continua from the Born approximation}

To understand the origin of the thermal continua analytically, it is instructive to examine the imaginary part of the self-energy within the Born approximation (BA; also called the Migdal approximation in Ref.~\cite{Mitric2023}), where only the first-order (single-phonon) diagram is retained:
\begin{equation}
\resizebox{\columnwidth}{!}{$\displaystyle \Sigma^{(\mathrm{BA})}_T(\omega) = \frac{g^{2}}{N}\sum_{q} \left[ \frac{n_{B}(\Omega)+1}{\omega-\Omega-\epsilon_{k-q}+i0^{+}} + \frac{n_{B}(\Omega)}{\omega+\Omega-\epsilon_{k-q}+i0^{+}} \right], $}
\label{eqn:SE_BA}
\end{equation}
where $n_B(x)=1/(e^{\beta x}-1)$ is the Bose-Einstein factor. Expressed in terms of the free-electron density of states (DOS) $\rho(\omega)=\frac{1}{N}\sum_k\delta(\omega-\epsilon_k)$, the imaginary part is:
\begin{equation}
\Sigma^{{''}(\mathrm{BA})}_T(\omega) = -\pi g^{2}\bigl[(n_B+1)\,\rho(\omega-\Omega) + n_B\,\rho(\omega+\Omega)\bigr].
\label{eqn:DOS_BA}
\end{equation}
For a nearest-neighbor dispersion $\epsilon_{k}=-2t\cos k$ in 1D, the DOS is \(\rho(\epsilon)=\frac{1}{\pi}\,\frac{1}{\sqrt{4t^{2}-\epsilon^{2}}}\) for $|\epsilon|<2t$, and zero outside, so Eq.~\ref{eqn:DOS_BA} becomes:
\begin{equation}
\resizebox{\columnwidth}{!}{$\displaystyle
\begin{aligned}
\Sigma^{''\,(\mathrm{BA})}_T(\omega)
&=
-g^{2}\left[
\frac{n_{B}+1}{\sqrt{4t^{2}-(\omega-\Omega)^{2}}}\,
\Theta\!\big(2t-|\omega-\Omega|\big)
\right.\\
&\qquad\left.
+\frac{n_{B}}{\sqrt{4t^{2}-(\omega+\Omega)^{2}}}\,
\Theta\!\big(2t-|\omega+\Omega|\big)
\right].
\end{aligned}
$}
\label{eqn:DOS_BA_1D}
\end{equation}

\begin{figure}[t]
    \centering
\includegraphics[width=0.7\columnwidth]{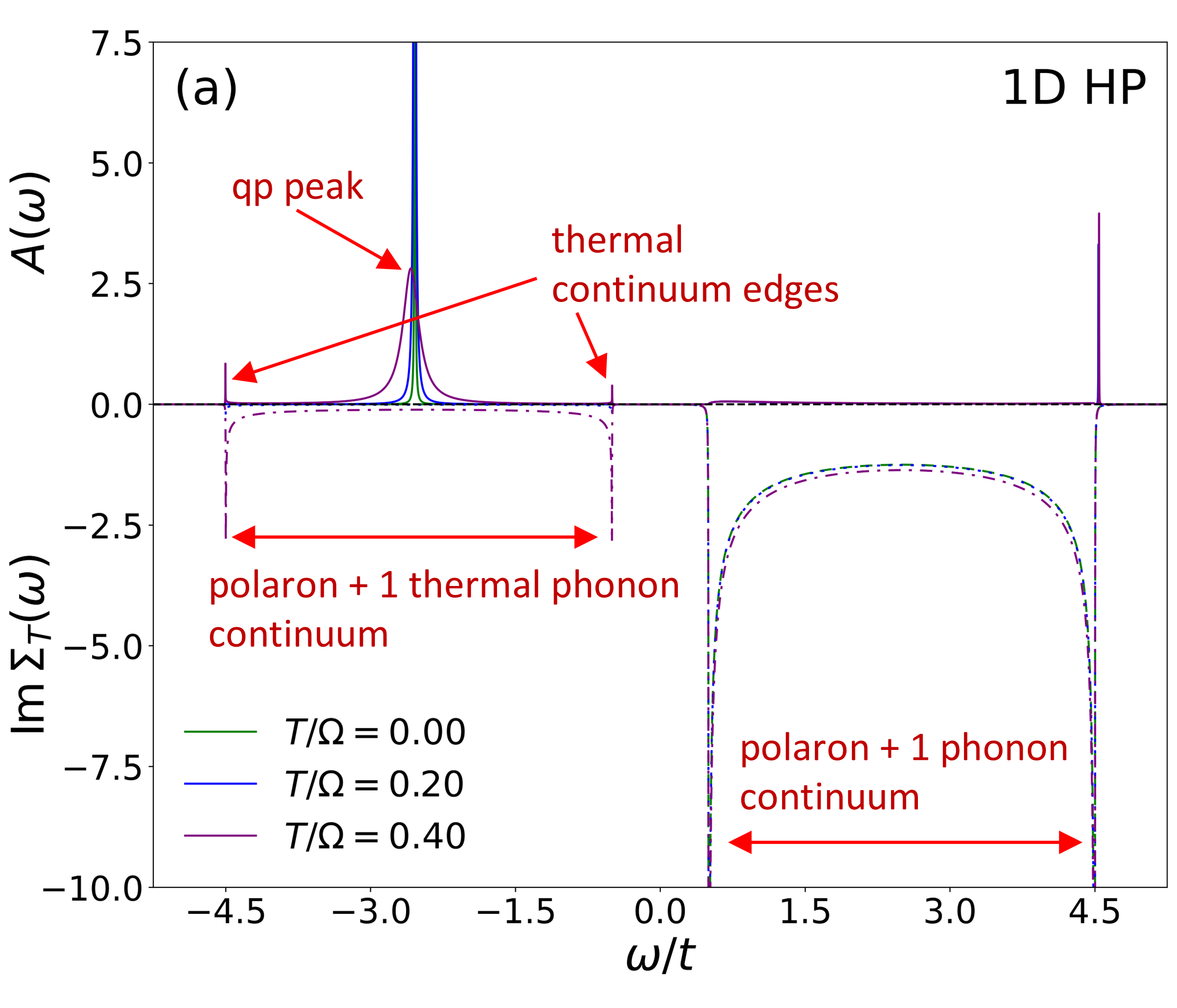}
   \includegraphics[width=0.7\columnwidth]{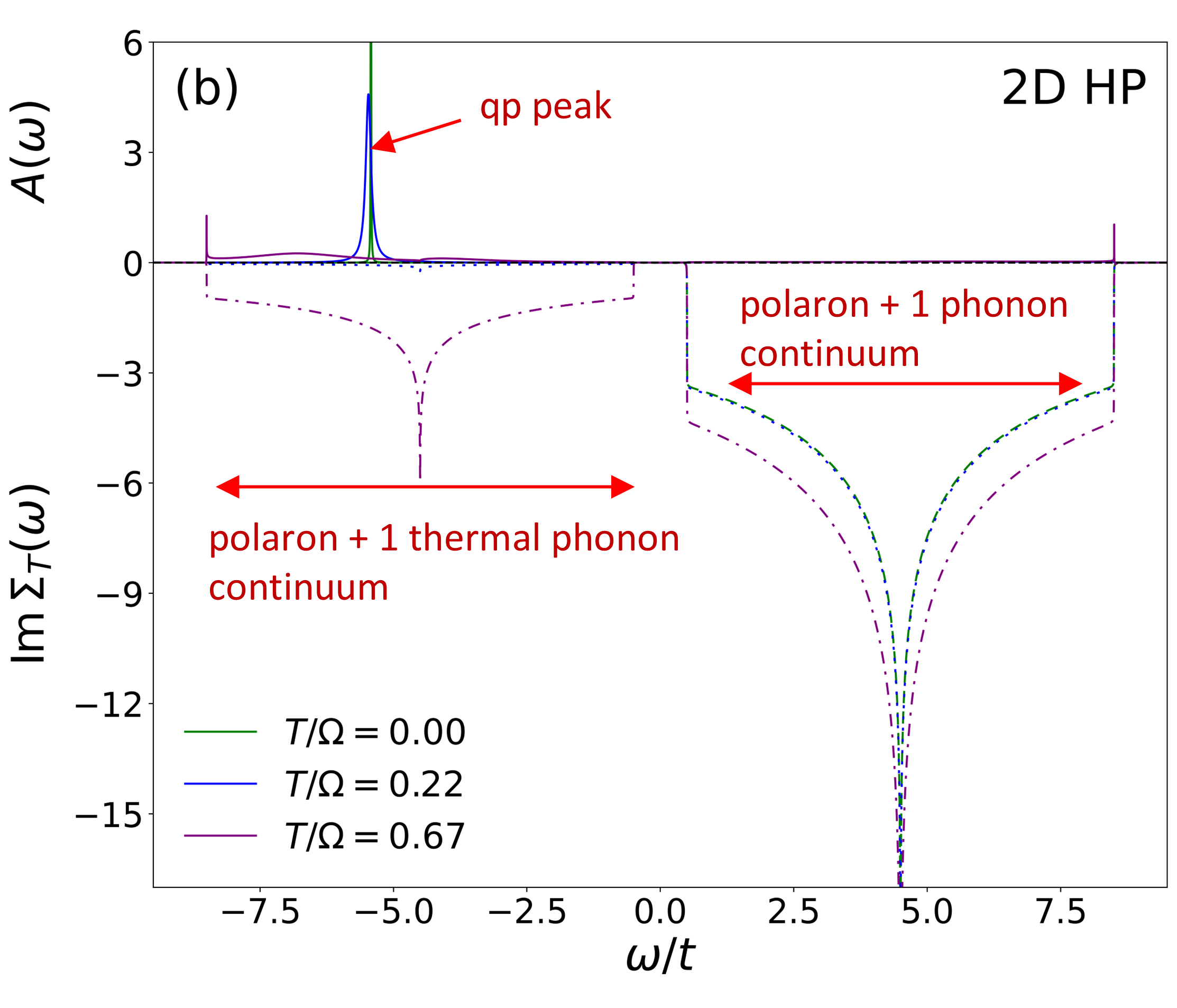}
\includegraphics[width=0.7\columnwidth]{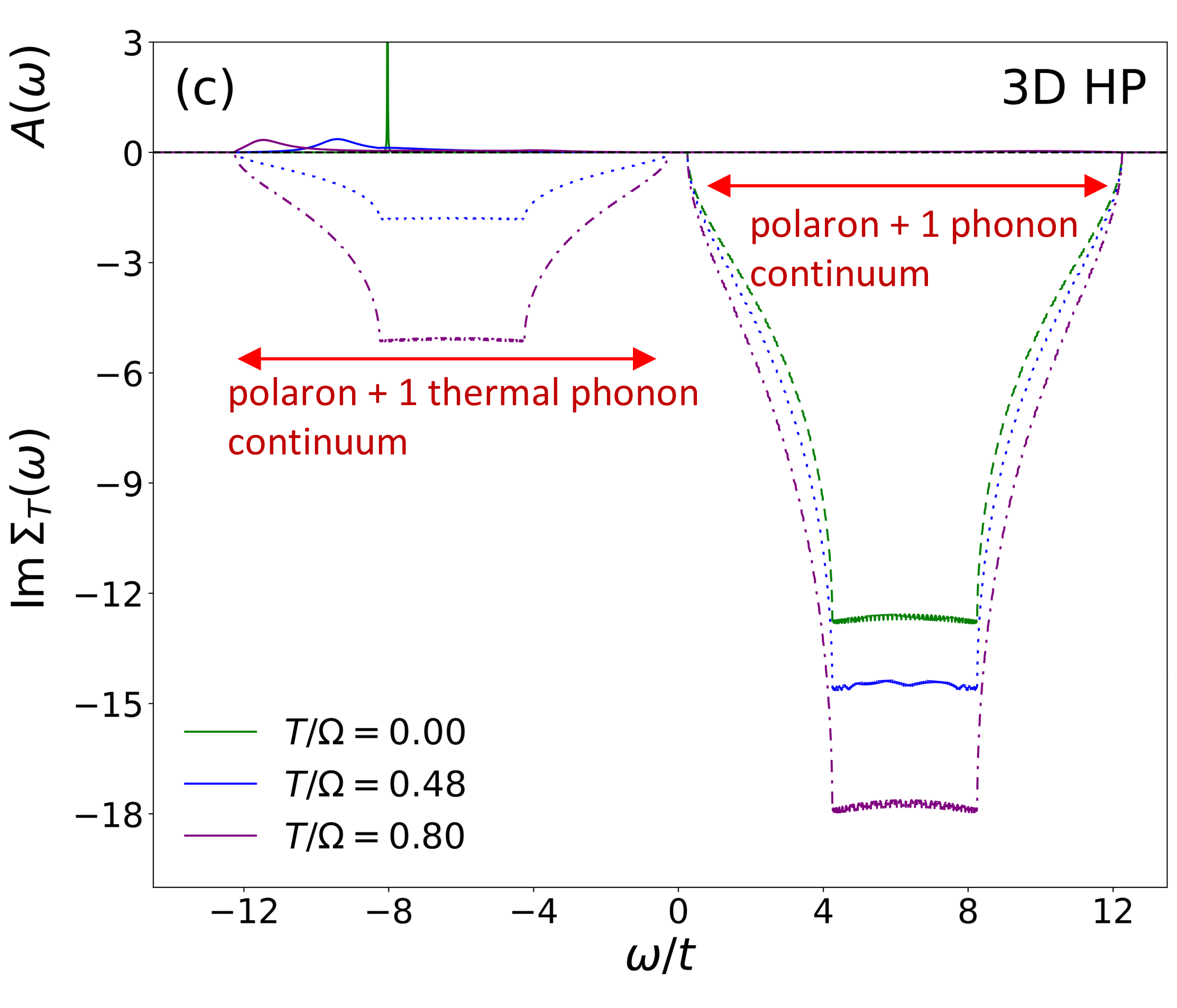}
    \caption{(color online) Spectral function $A(\omega,k=0)$ (solid lines) and the imaginary part of the self-energy $\Sigma^{"\,(\mathrm{BA})}_T(\omega)$ (dotted lines) from the finite-$T$ Born approximation, plotted on a common vertical axis at different temperatures. Panels (a)--(c) show 1D, 2D, and 3D, respectively. Parameters: (a) $\lambda=0.5$, $\Omega=2.5, t=1$; (b) $\lambda=0.75$, $\Omega=4.5, t=1$; (c) $\lambda=0.75$, $\Omega=6.25, t=1$. The large $\Omega$ is chosen to separate the polaron$+$1ph continuum (right) from the thermal continuum (left). $\eta_0=0.001$.}
    \label{fig:BA_SE}
\end{figure}

The first term of Eq.~\ref{eqn:DOS_BA_1D} is non-zero for $\omega\in[\Omega-2Dt,\Omega+2Dt]$ even at $T=0$, and represents the well-known  polaron$+$1-phonon continuum in the BA approximation. The second term is a purely thermal contribution, non-zero for $\omega\in[-\Omega-2Dt,-\Omega+2Dt]$, and explains the finite-$T$ spectral features appearing below $E_P(k=0)$, visible in Fig.~\ref{fig:BA_SE}. Since only the lowest-order diagram is kept, higher-order thermal continua shifted by $\pm 2\Omega,\pm 3\Omega,\ldots$ are absent within this approximation, and the imaginary part of the self-energy simply follows the bare electronic DOS. These consequences of retaining only the lowest diagram will change when using MA.

\begin{figure}[t]
    \centering
    \includegraphics[width=0.48\columnwidth]{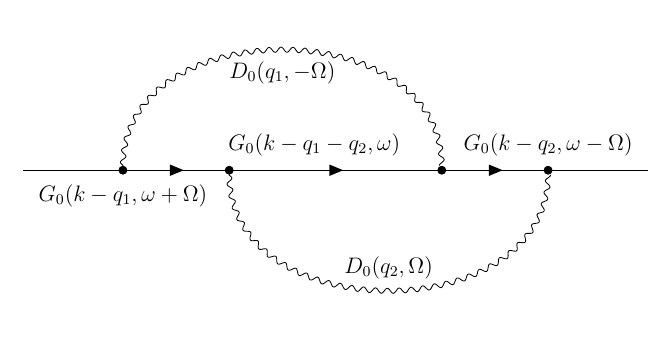}
    \hfill
    \includegraphics[width=0.48\columnwidth]{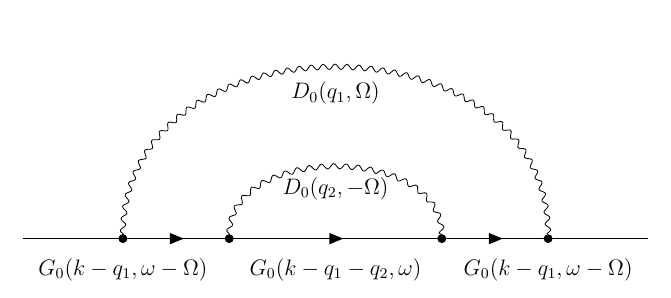}
    \caption{Feynman diagrams showcasing the second order ($g^4$) diagrams of the HP in the electron self-energy, with phonon propagators representing the thermal phonon with $-\Omega$ and the remote phonon with $\Omega$. }
    \label{fig:g4_diag}
\end{figure}

\begin{figure}[!t]
    \centering
\includegraphics[width=0.95\columnwidth]{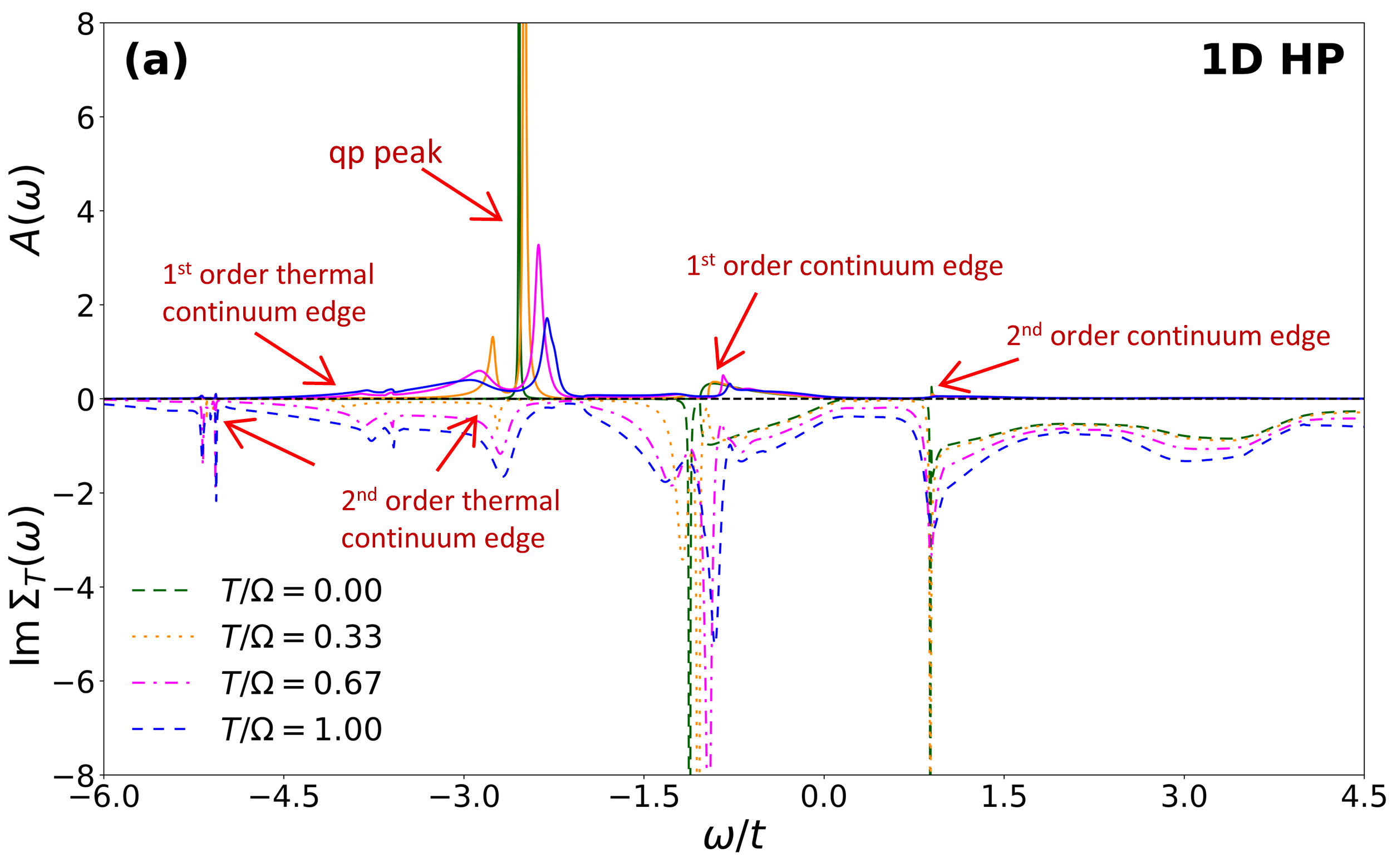}
\includegraphics[width=0.95\columnwidth]{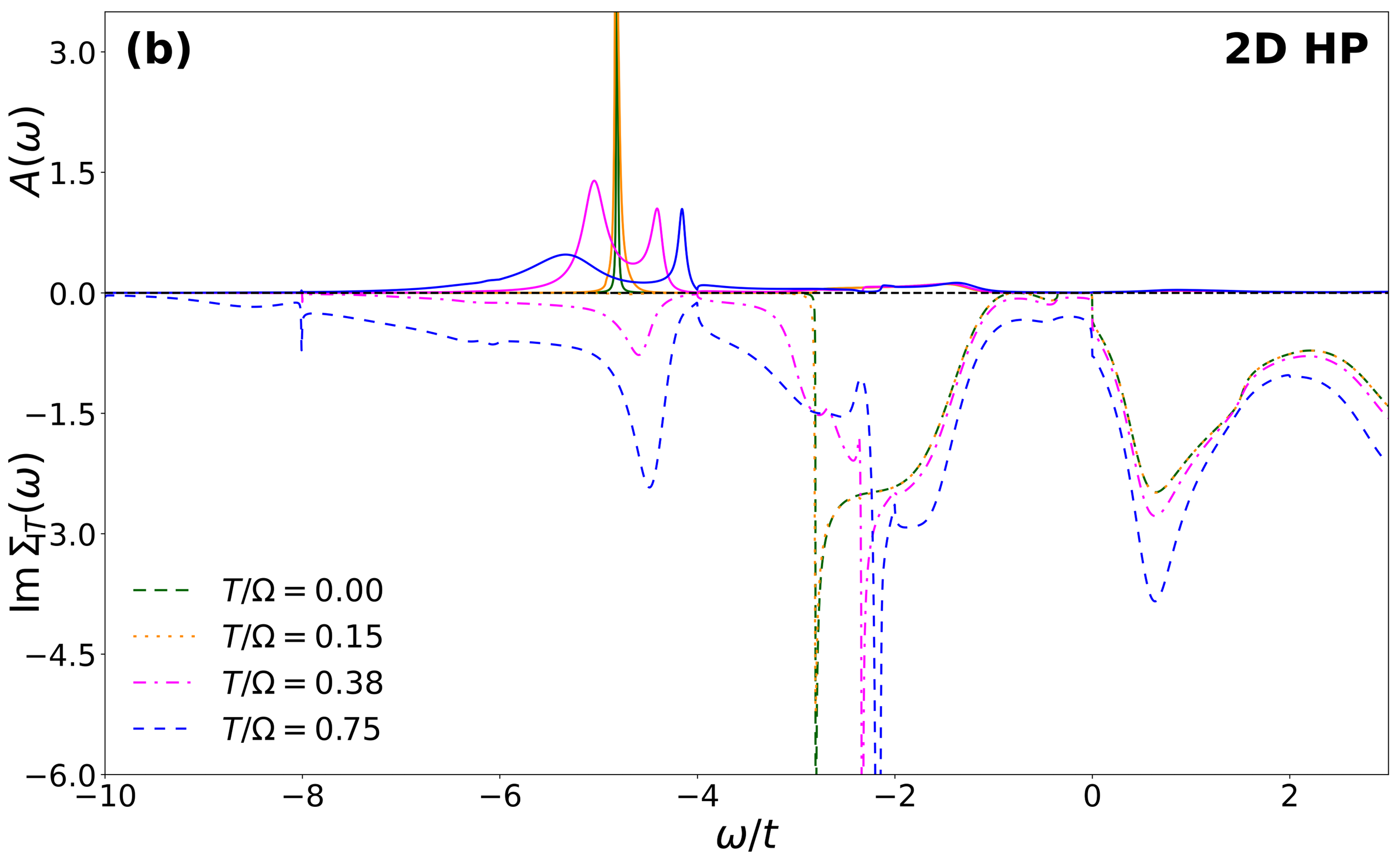}
\includegraphics[width=0.95\columnwidth]{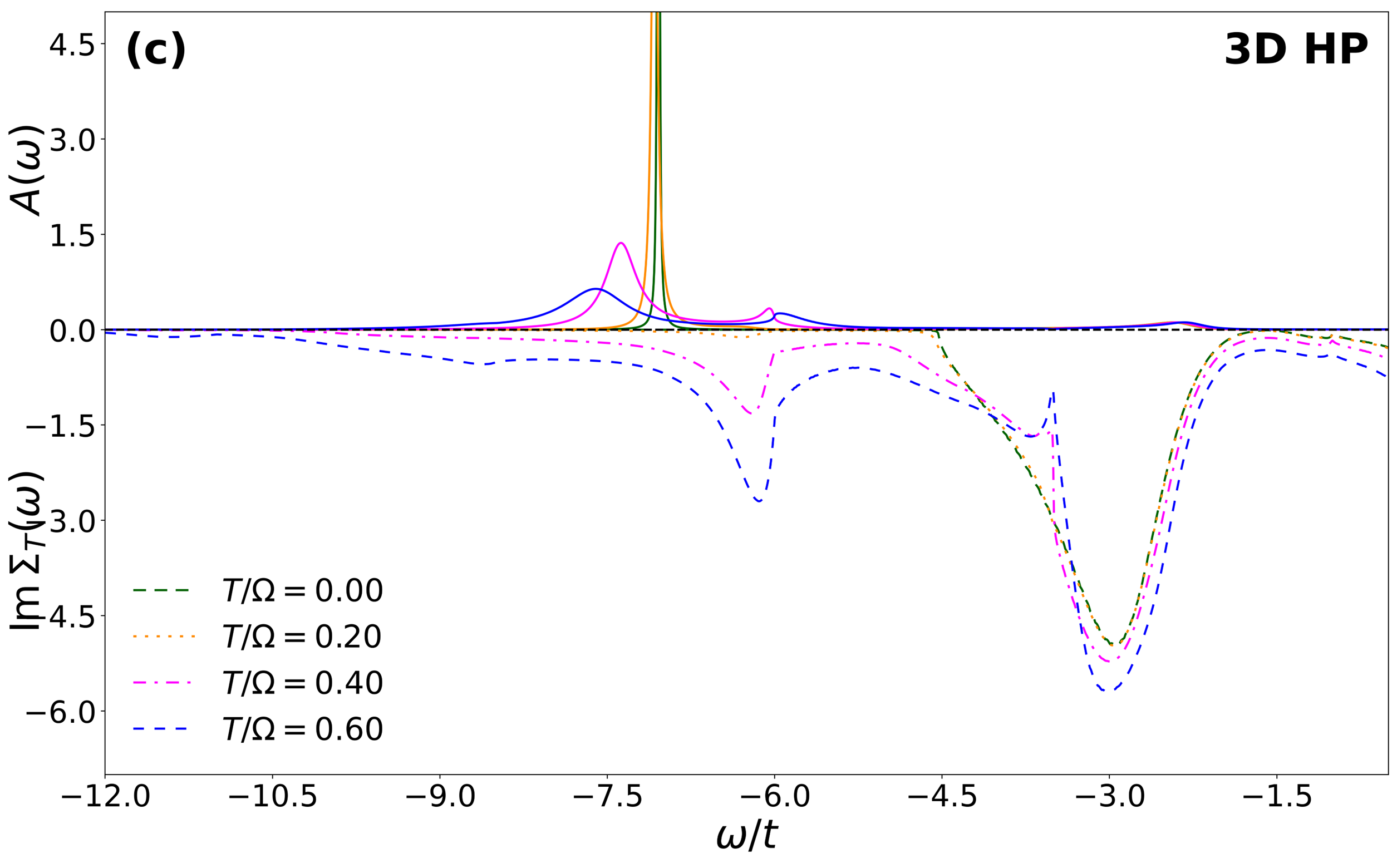}
    \caption{(color online) $A(\omega,k=0)$ (solid) and imaginary part $\Sigma^{"}_T(\omega)$ (dotted) from MA$^{(1)}$ at different temperatures, for 1D (a), 2D (b), and 3D (c). Parameters: (a) $\lambda=0.5$, $\Omega=1.5, t=1$; (b) $\lambda=0.5$, $\Omega=2.0, t=1$; (c) $\lambda=0.5$, $\Omega=2.5, t=1$; $\eta_0=0.001$ (1D, 2D) and $0.005$ (3D); Parameters chosen to separate the polaron$+$1ph and different orders of thermal continua.}
    \label{fig:MA1_SE}
\end{figure}

Beyond BA, higher-order diagrams produce three additional effects: (a) higher-order thermal continua spaced by multiples of $\Omega$ from the polaron ground state (or a more complex structure at strong coupling, from  2nd bound states and/or polaron+one-phonon continuum); (b) the correct placement of the first-order continuum at $E_P(k=0)\pm\Omega$ rather than $-2Dt\pm\Omega$; and (c) the width of the continua is set by the polaron bandwidth rather than the free-electron bandwidth, so the continuum edges are determined by $E_P(k=0)$ and $E_P(k=\pi)$ rather than $\pm 2Dt$. At strong coupling, further continua arise from higher energy bound states  (when they exist in the spectrum), with widths set by these bound states' bandwidths, which can exceed the polaron bandwidth. In particular, certain $g^4$ diagrams (Fig.~\ref{fig:g4_diag}), where a thermal phonon is absorbed together with emission of a free phonon, place a thermal continuum directly over the polaron ground-state energy, as wide as the polaron bandwidth, as is clearly visible in Fig.~\ref{fig:MA1_SE} and well reproduced by MA$^{(1)}$ but not MA$^{(0)}$.

\subsection{\texorpdfstring{MA$^{(0)}$ vs.\ MA$^{(1)}$: self-energy and spectral function}{MA(0) vs. MA(1): self-energy and spectral function}}

\begin{figure}[!t]
    \centering
\includegraphics[width=0.95\columnwidth]{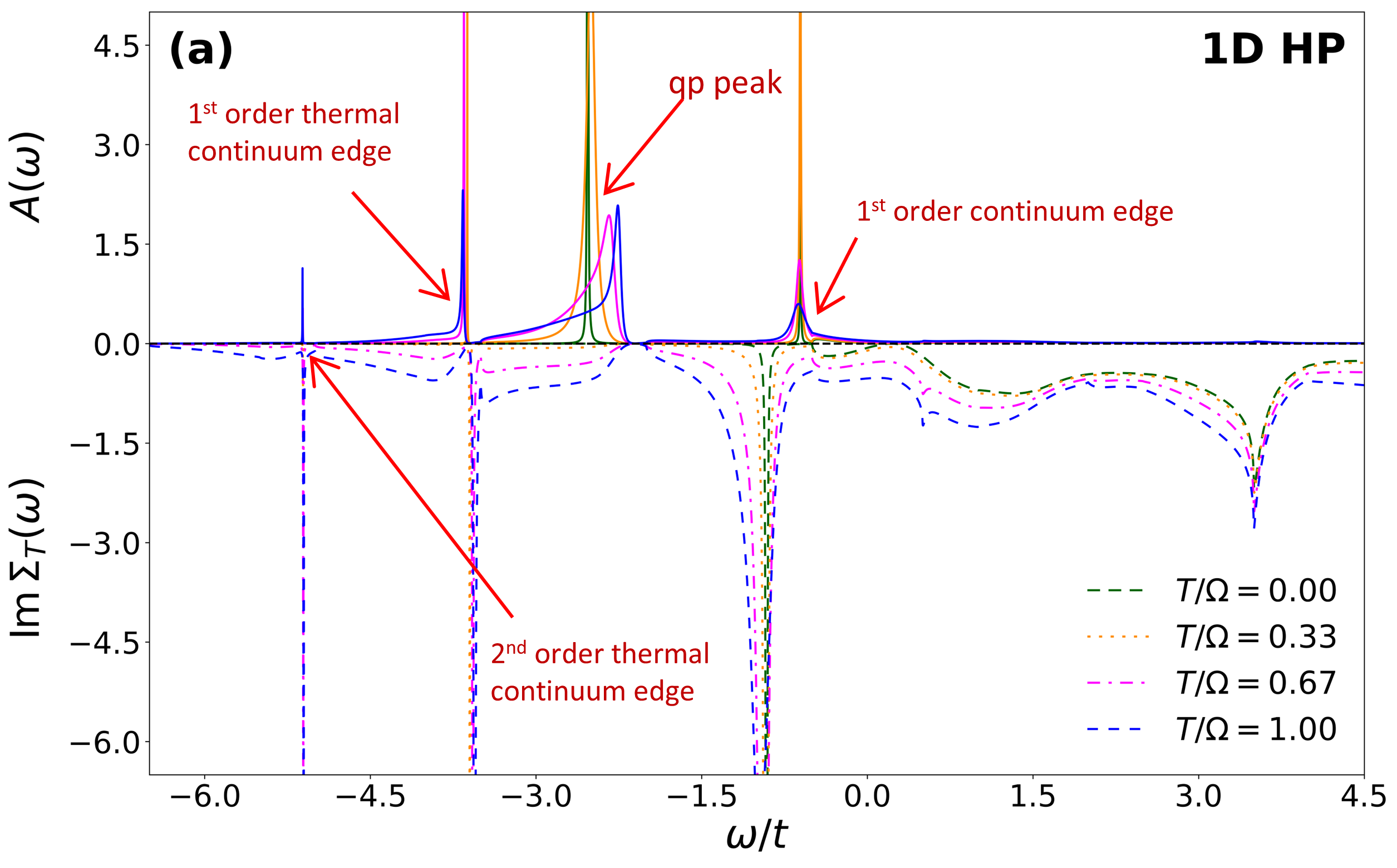}
\includegraphics[width=0.95\columnwidth]{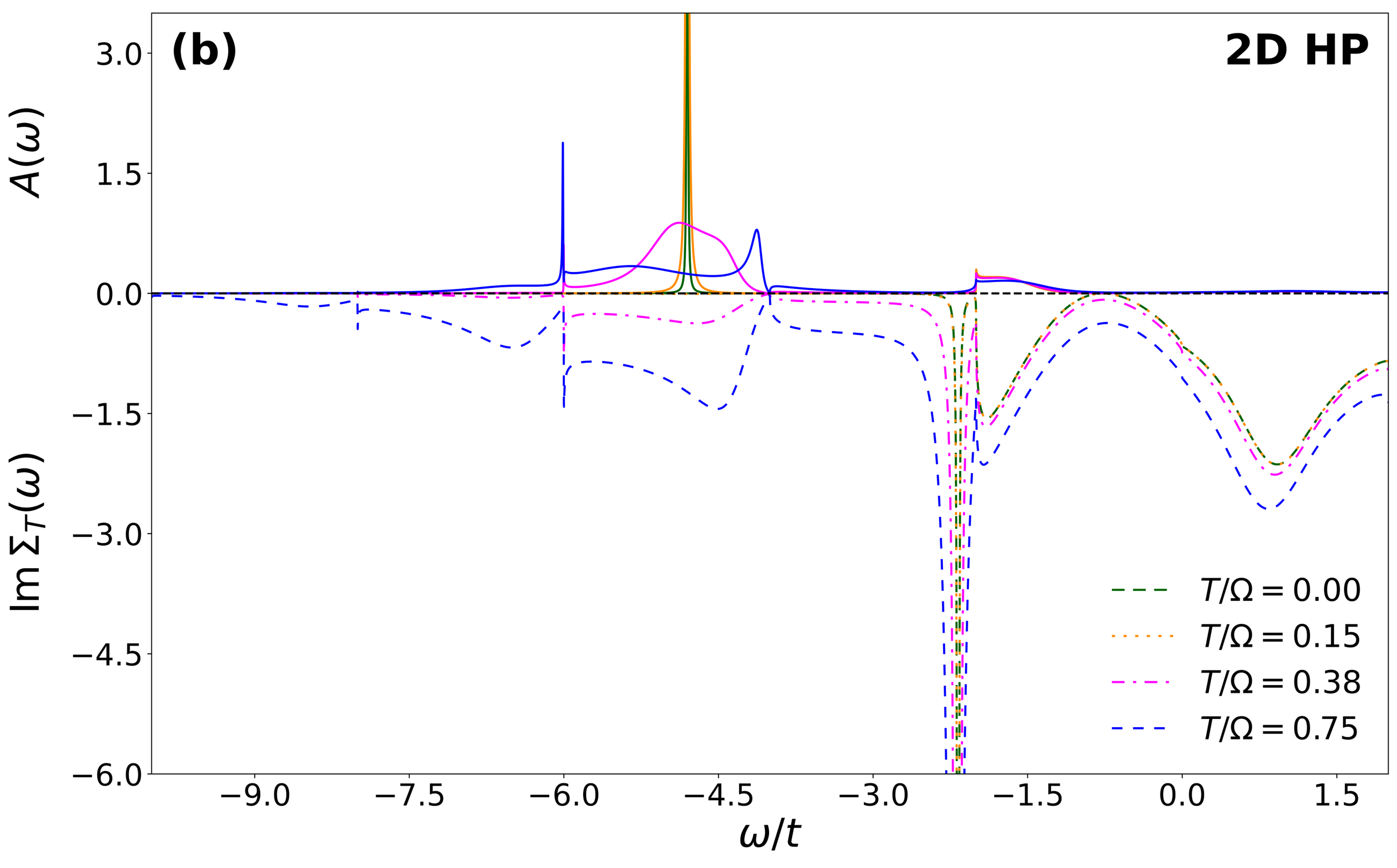}
\includegraphics[width=0.95\columnwidth]{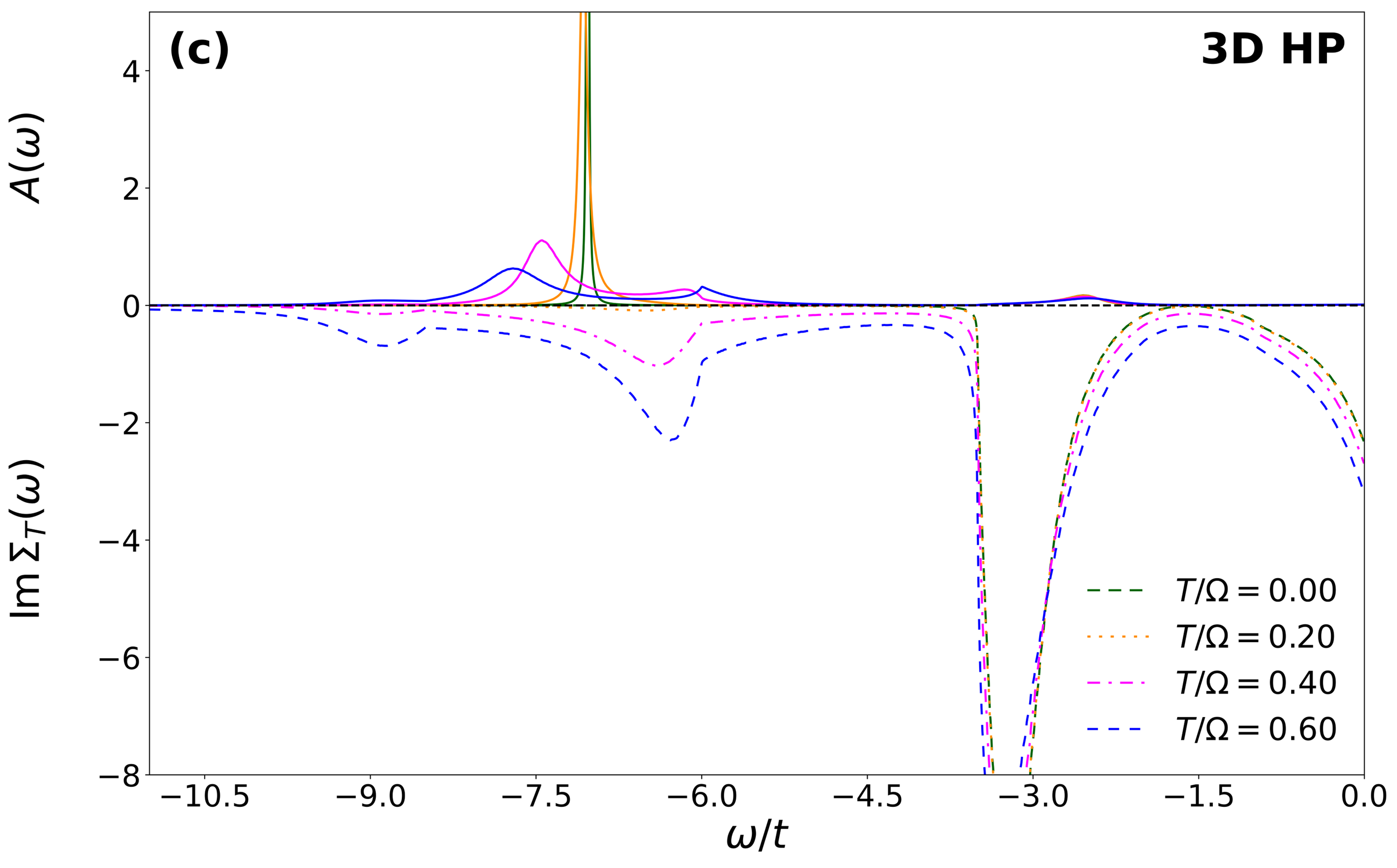}
    \caption{(color online) Same as Fig.~\ref{fig:MA1_SE} but for MA$^{(0)}$. Parameters identical to Fig.~\ref{fig:MA1_SE}.}
    \label{fig:MA0_SE}
\end{figure}

Figures~\ref{fig:MA1_SE} and \ref{fig:MA0_SE} show $A(\omega,k=0)$ and $\Sigma^{''}_T(\omega)$ for MA$^{(1)}$ and MA$^{(0)}$ respectively, at fixed anti-adiabatic parameters across 1D, 2D, and 3D. In both cases the imaginary part of the self-energy correctly follows the polaron density of states rather than the free-electron DOS, in contrast to the BA result of Fig.~\ref{fig:BA_SE}. Two key differences distinguish MA$^{(1)}$ from MA$^{(0)}$: first, the placement of the polaron$+$1 phonon continuum, which MA$^{(0)}$ gets wrong \cite{Glenn2006,Mona2007} but MA$^{(1)}$ corrects \cite{Mona2007}, an improvement that persists at finite $T$; second, and more important, the thermal continuum originating from the $g^4$ diagram (Fig.~\ref{fig:g4_diag}) that forms around the polaron GS; this is clearly seen in Fig.~\ref{fig:MA1_SE}(a) but missing in Fig.~\ref{fig:MA0_SE}.

\begin{figure}[!t]
    \centering
\includegraphics[width=0.95\columnwidth]{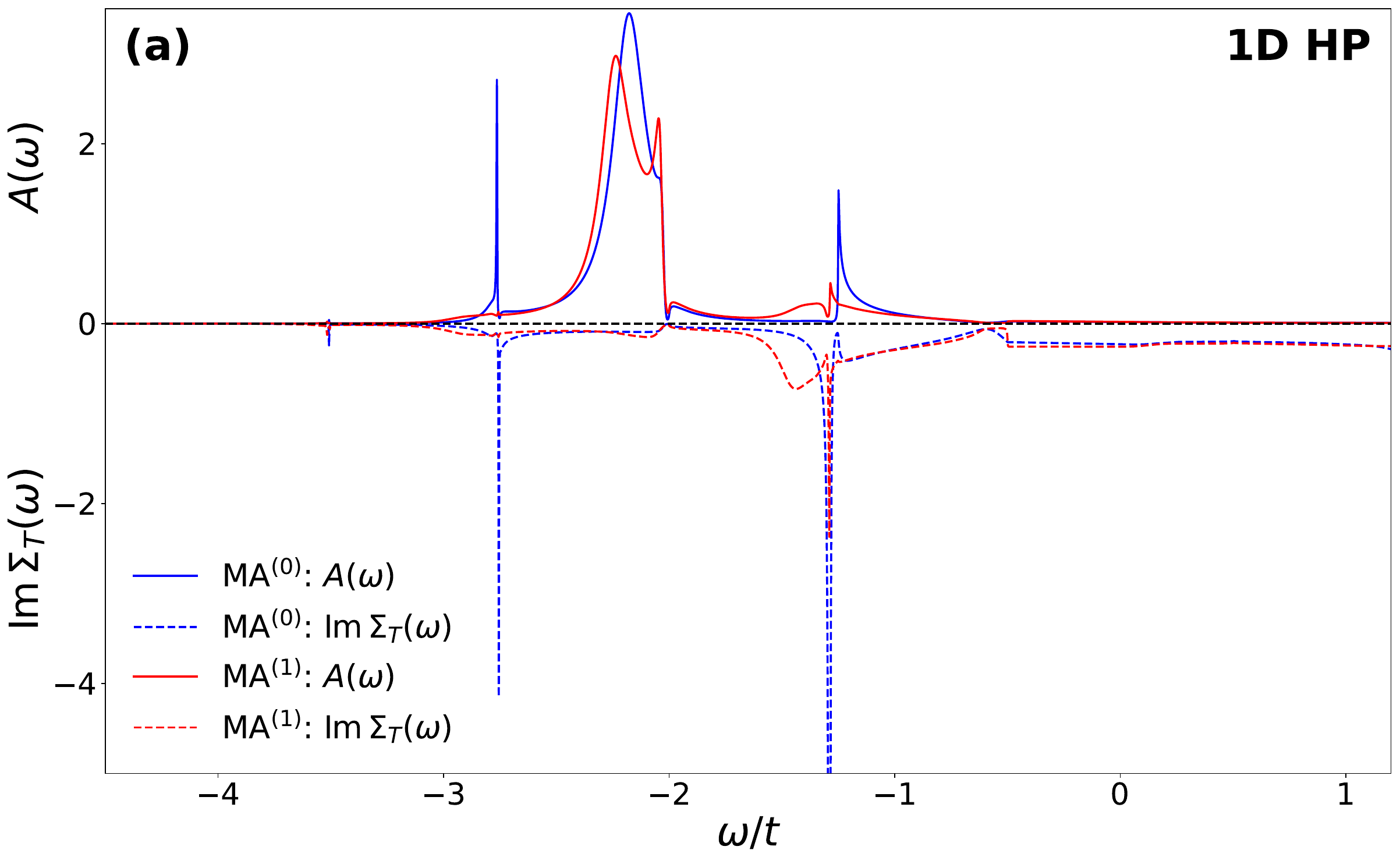}
\includegraphics[width=0.95\columnwidth]{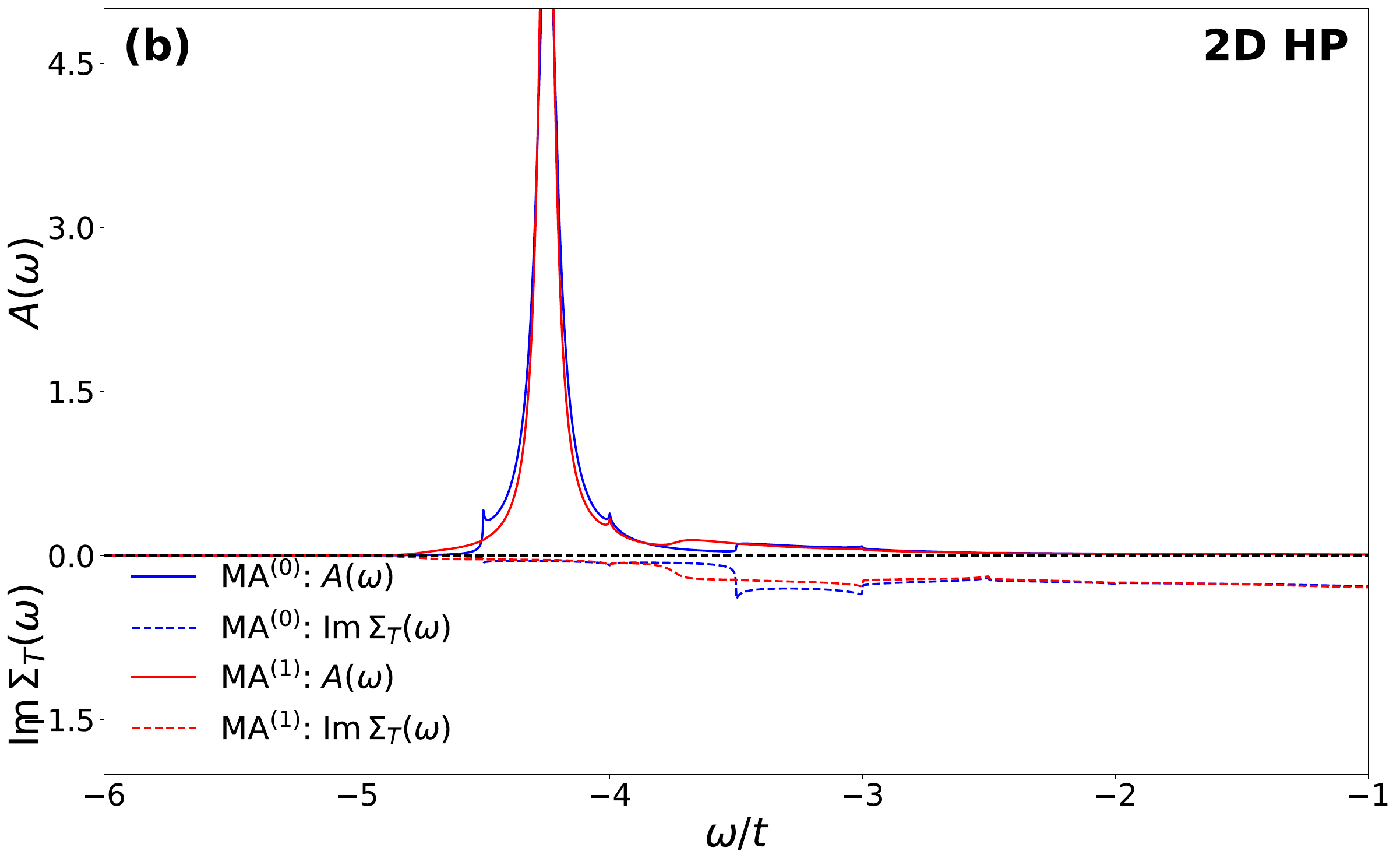}
\includegraphics[width=0.95\columnwidth]{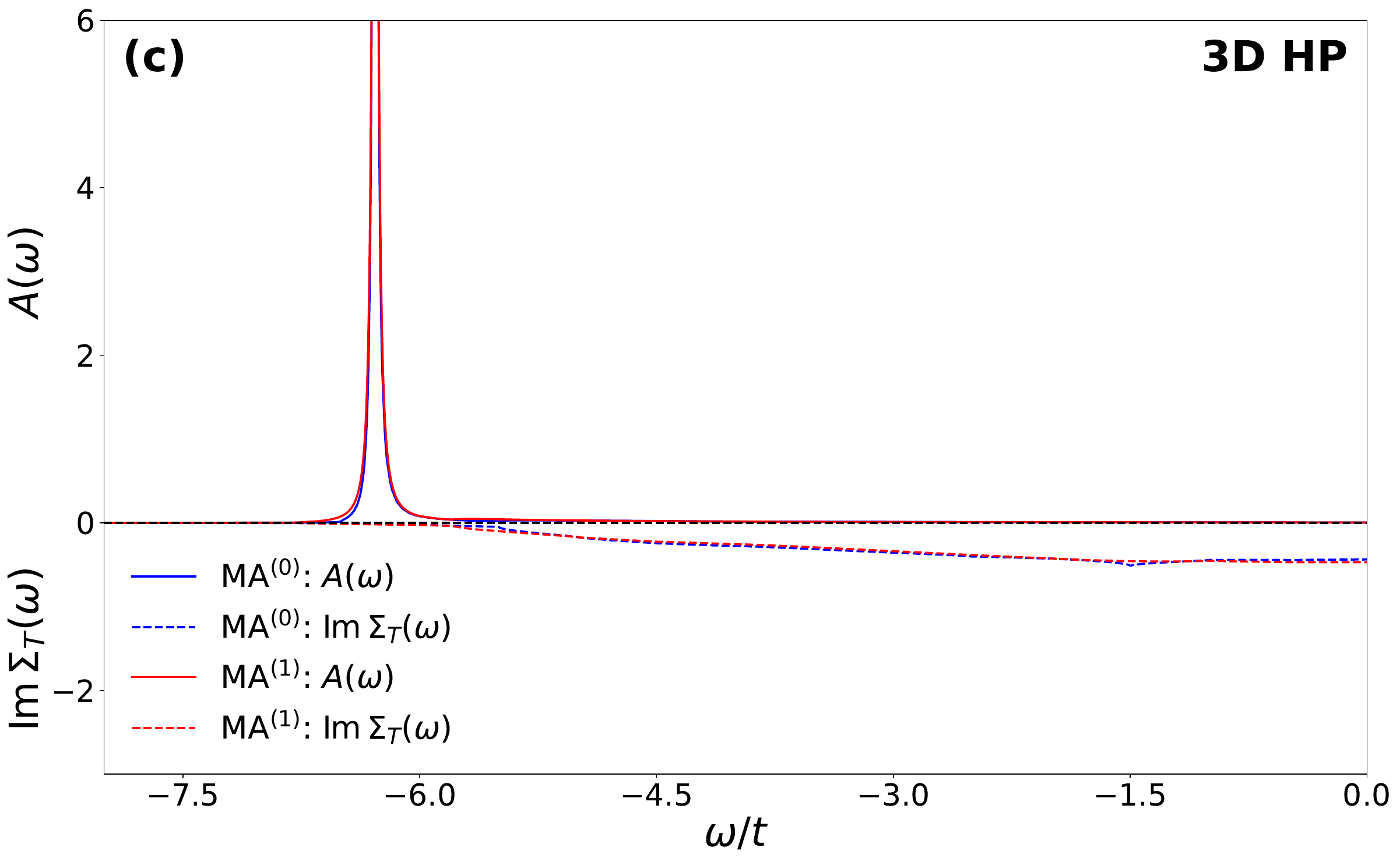}
    \caption{(color online) Direct comparison of MA$^{(1)}$ and MA$^{(0)}$ for $A(\omega,k=0)$ (solid) and $\Sigma^{"}_T(\omega)$ (dotted) in 1D (a), 2D (b), and 3D (c), in the adiabatic regime. Parameters: (a) $\lambda=0.2$, $\Omega=0.75\,t$; (b) $\lambda=0.25$, $\Omega=0.5\,t$; (c) $\lambda=0.3$, $\Omega=0.5\,t$; $\eta_0=0.001$.}
    \label{fig:MA0_vs_MA1_SE}
\end{figure}

For completeness, we have directly plotted MA$^{(1)}$ and MA$^{(0)}$ on top of each other for different parameter sets in Fig.~\ref{fig:MA0_vs_MA1_SE}, confirming that the two orders converge in higher dimensions \cite{Mona2007}, as expected from the steeper decay of $G^0_{i-j}(\omega)$ with distance.

\section{Additional results}
\label{SM:Additional_data}

\subsection{Role of coupling strength and phonon frequency in finite-$T$ behavior across dimensions}

\begin{figure*}[!th]
    \centering  \includegraphics[width=\textwidth]{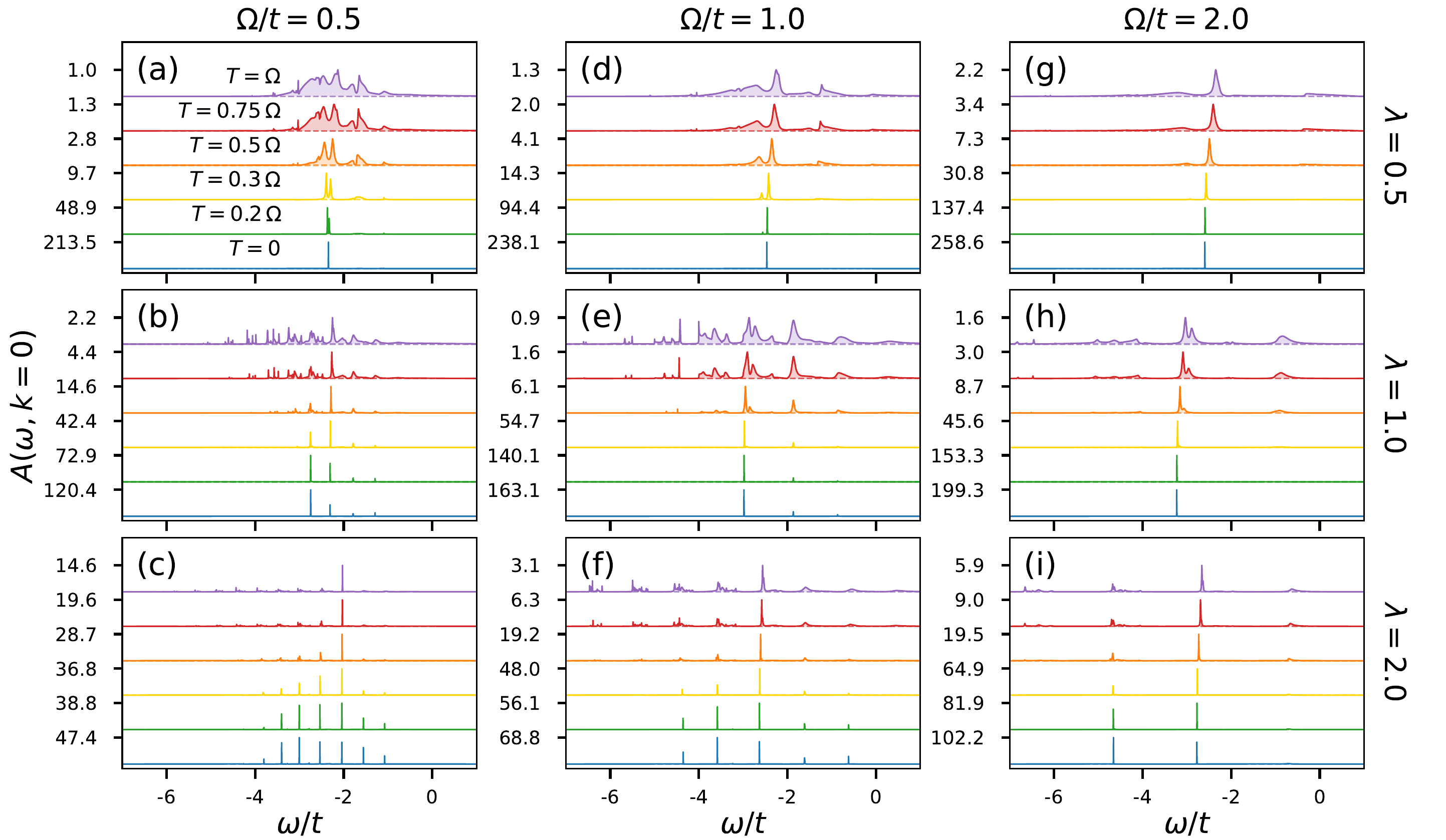} 
 \caption{(color online) 1D MA$^{(1)}$ spectral functions  $A(\omega,{\bf k}=0)$ of the HP. Columns correspond to phonon energies $\Omega/t=0.5$ (left), $\Omega/t=1.0$ (middle),  $\Omega/t=2.0$ (right). Rows corresponds effective couplings  $\lambda = 0.5$ (top), $\lambda = 1.0$ (middle), $\lambda = 2.0$ (bottom). For each ($\Omega,\lambda$) set, spectral plots are shown for the same set of temperatures $T/\Omega =  0.0, 0.2, 0.3, 0.5, 0.75, 1.0$ from bottom to top; the  curves are shifted and rescaled vertically for clarity. Other parameters are $t=1, \eta_0=0.001$.}
\label{fig:1D_Akw}
\end{figure*}

\begin{figure*}[!th]
    \centering  \includegraphics[width=\textwidth]{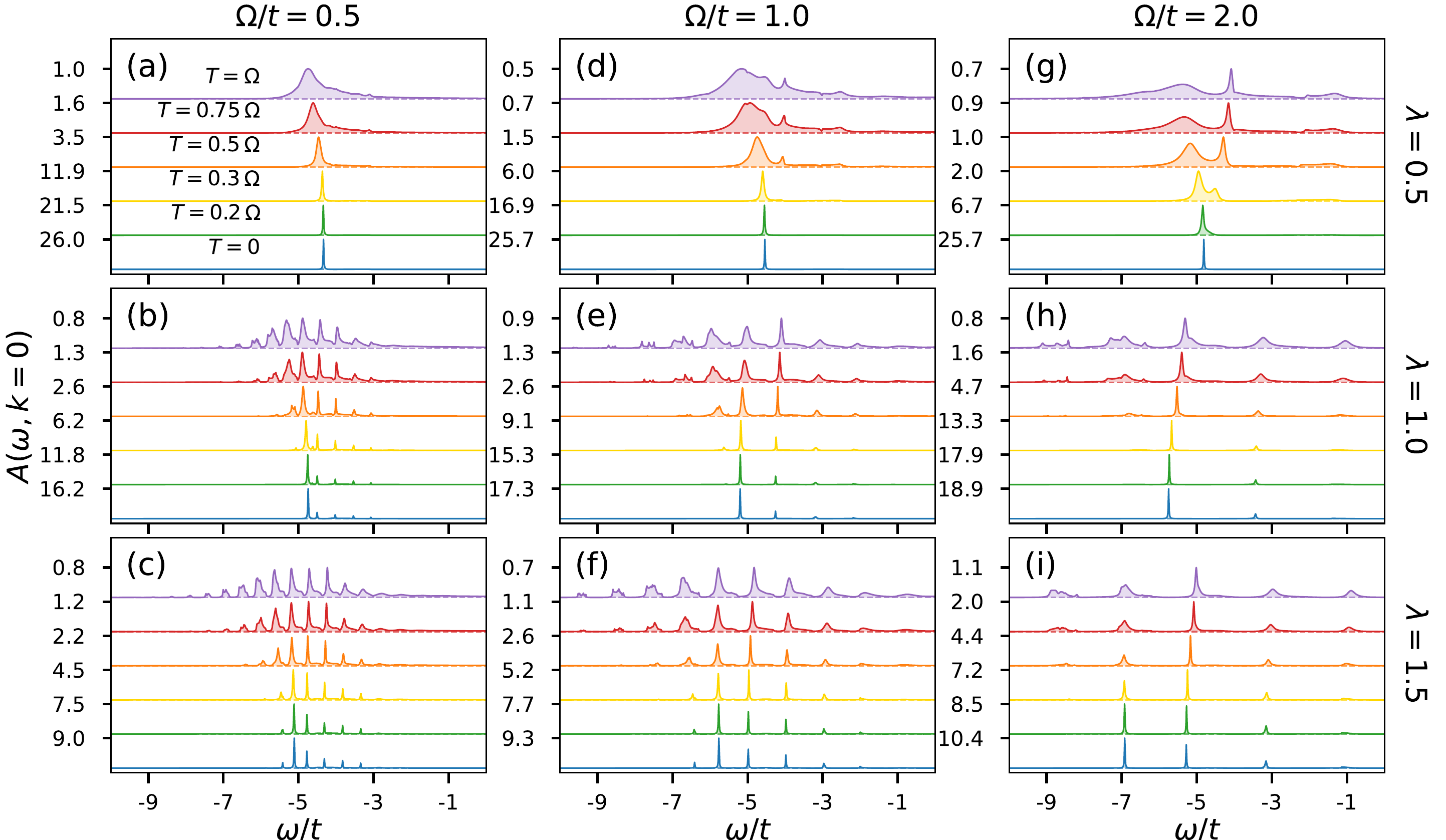} 
 \caption{(color online) 2D MA$^{(1)}$ spectral functions  $A(\omega,{\bf k}=0)$ of the HP. Columns correspond to phonon energies $\Omega/t=0.5$ (left), $\Omega/t=1.0$ (middle),  $\Omega/t=2.0$ (right). Rows corresponds effective couplings  $\lambda = 0.5$ (top), $\lambda = 1.0$ (middle), $\lambda = 1.5$ (bottom). For each ($\Omega,\lambda$) set, spectral plots are shown for the same set of temperatures $T/\Omega =  0.0, 0.2, 0.3, 0.5, 0.75, 1.0$ from bottom to top; the  curves are shifted and rescaled vertically for clarity. Other parameters are $t=1, \eta_0=0.01$.}
\label{fig:2D_Akw}
\end{figure*}

Complementing Fig. 3 of the main text, Figs.~\ref{fig:1D_Akw} and \ref{fig:2D_Akw} show the MA$^{(1)}$ spectral functions $A(\omega,{\bf k}=0)$ for the 1D and 2D HP at different coupling strengths and phonon frequencies, confirming the same qualitative behavior across dimensions. In  Fig.~\ref{fig:1D_Akw}, in contrast to higher dimensions, there is considerable loss of \textit{qp} coherence in the adiabatic regime even for the weakly coupled case as seen in  Figs.~\ref{fig:1D_Akw}(a).

\subsection{$A(\omega,k)$ contour plots at $T>0$}

\begin{figure}[!t]
    \centering
    \includegraphics[width=0.75\columnwidth]{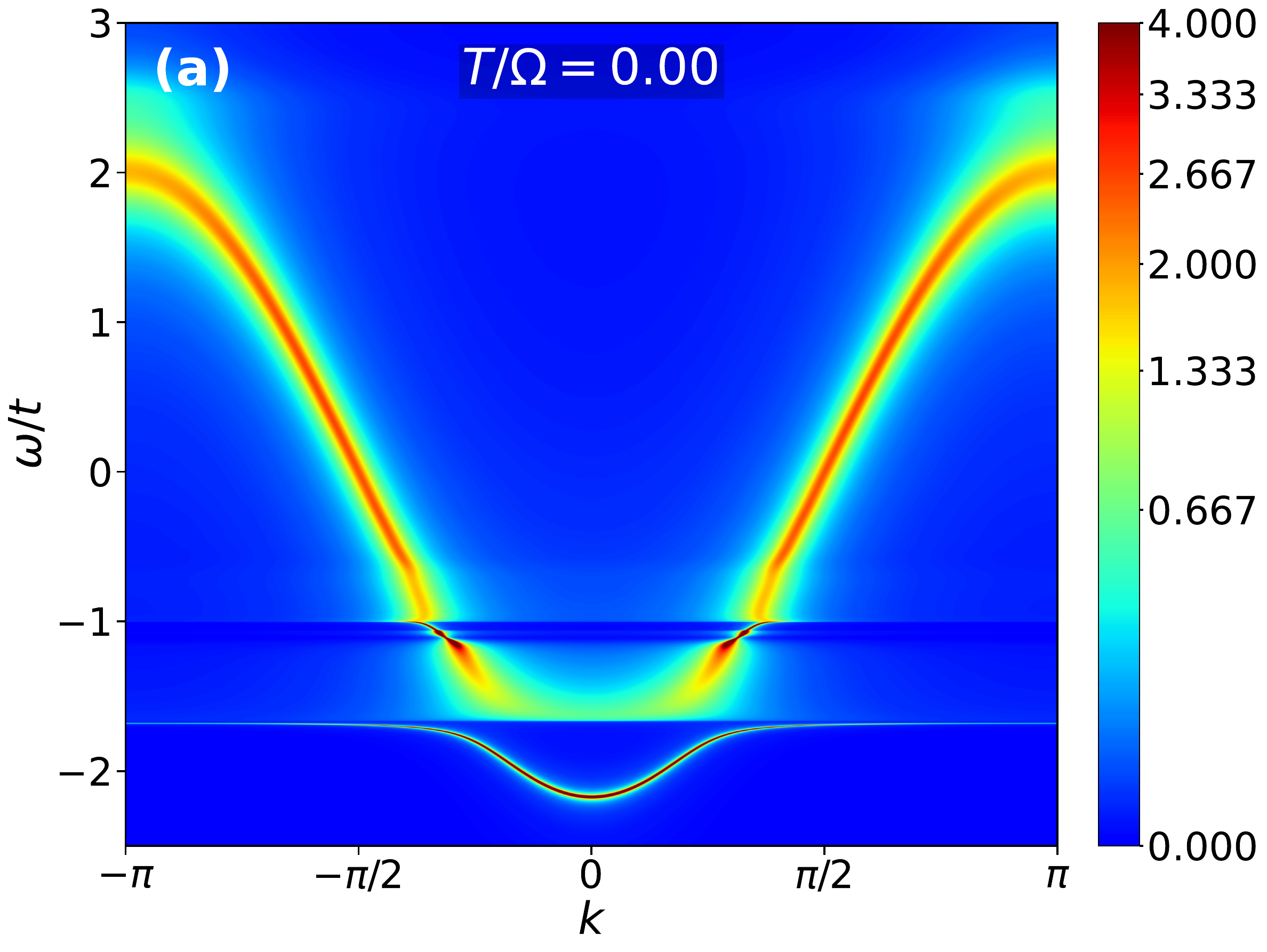}

    \includegraphics[width=0.75\columnwidth]{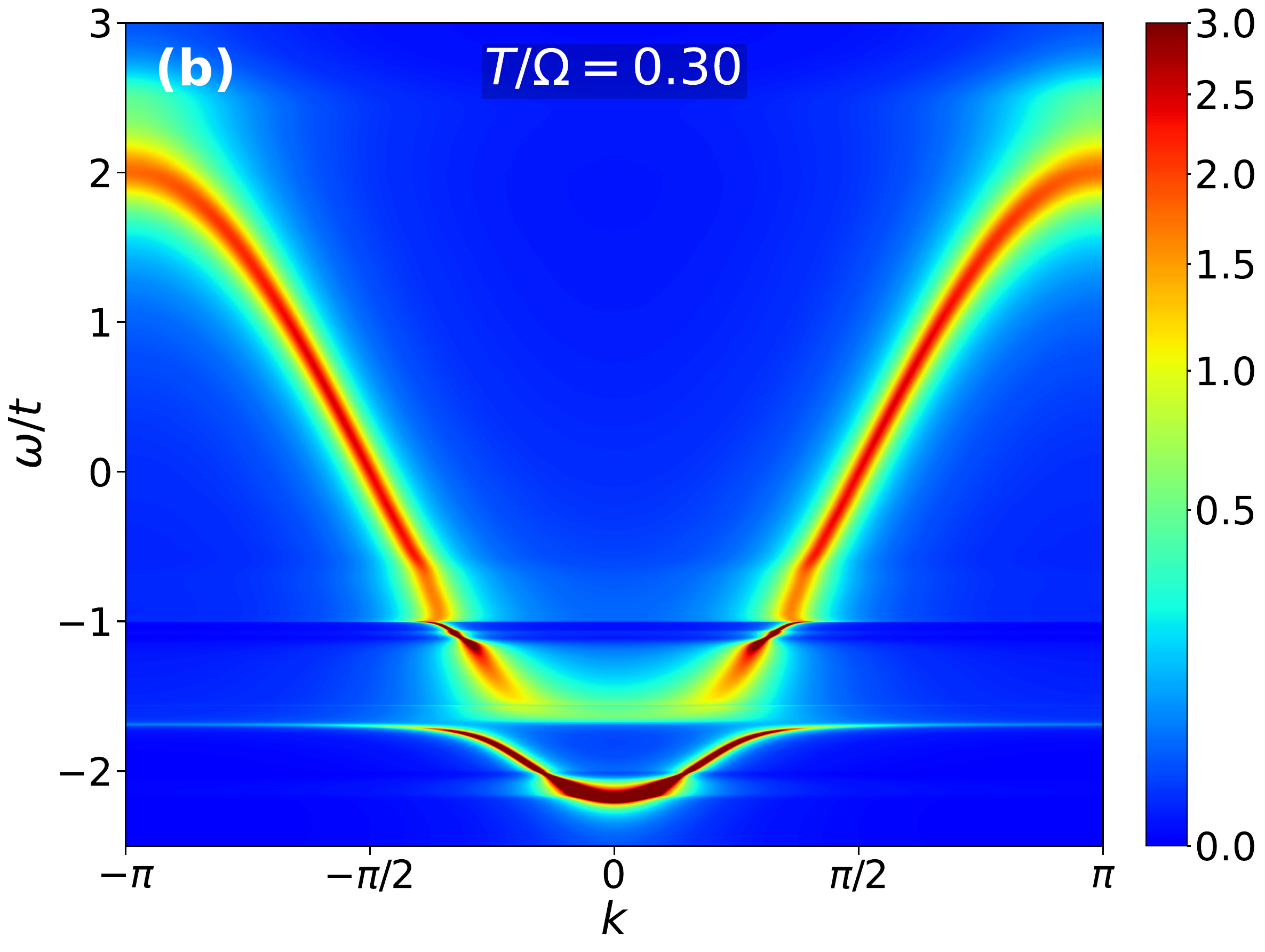}

    \includegraphics[width=0.75\columnwidth]{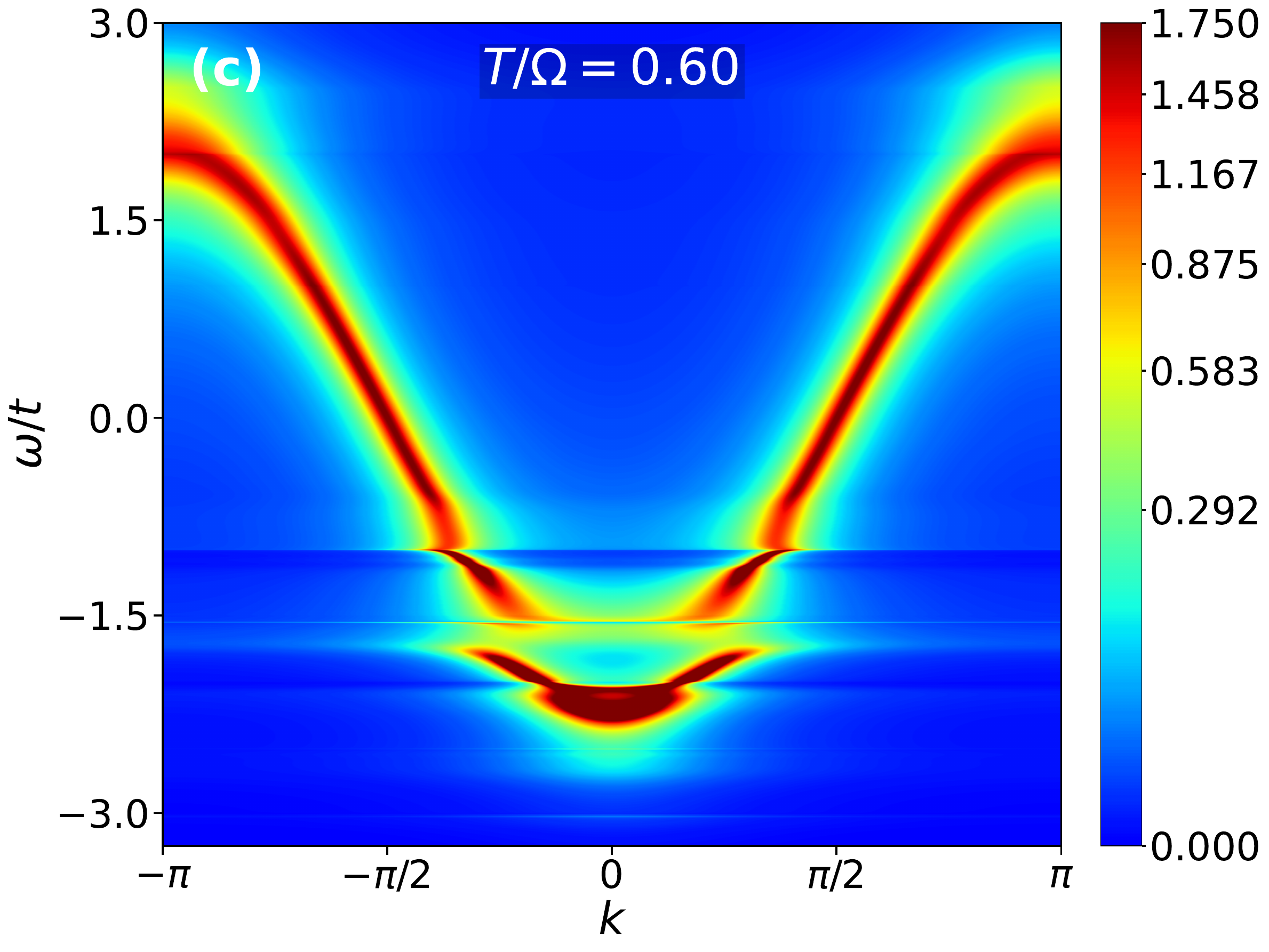}

        \includegraphics[width=0.75\columnwidth]{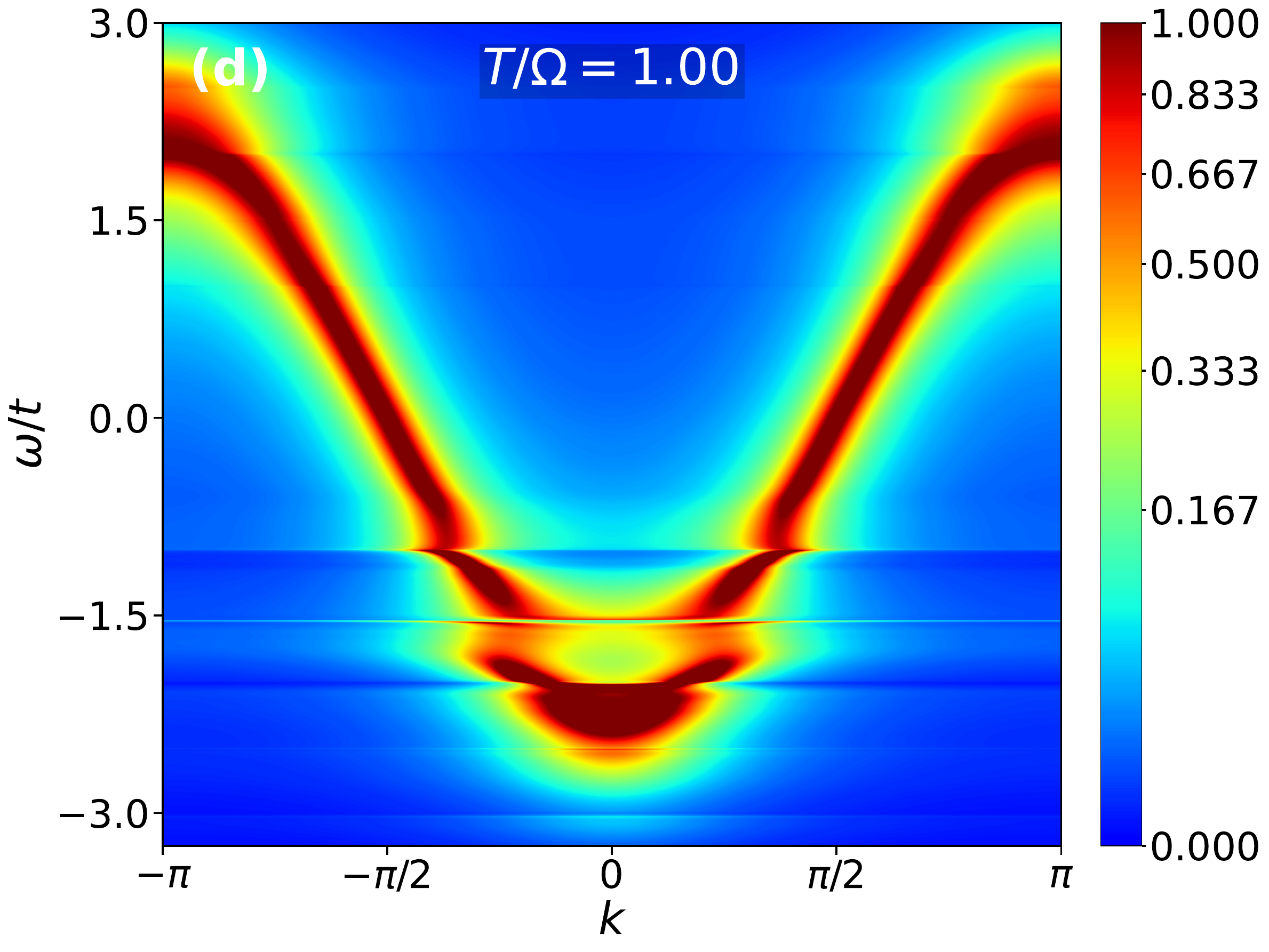}
 \caption{(color online) Evolution of the MA$^{(1)}$ $A(\omega,k)$ heat-map with  temperature for the 1D HP with parameters: $\Omega/t=0.5$, coupling strength $\lambda = 0.25$, $t=1$ and $\eta_0=0.001$.}

   \label{fig:heat_map_om_0.5}
\end{figure}

\begin{figure}[!t]
    \centering
      \includegraphics[width=0.75\columnwidth]{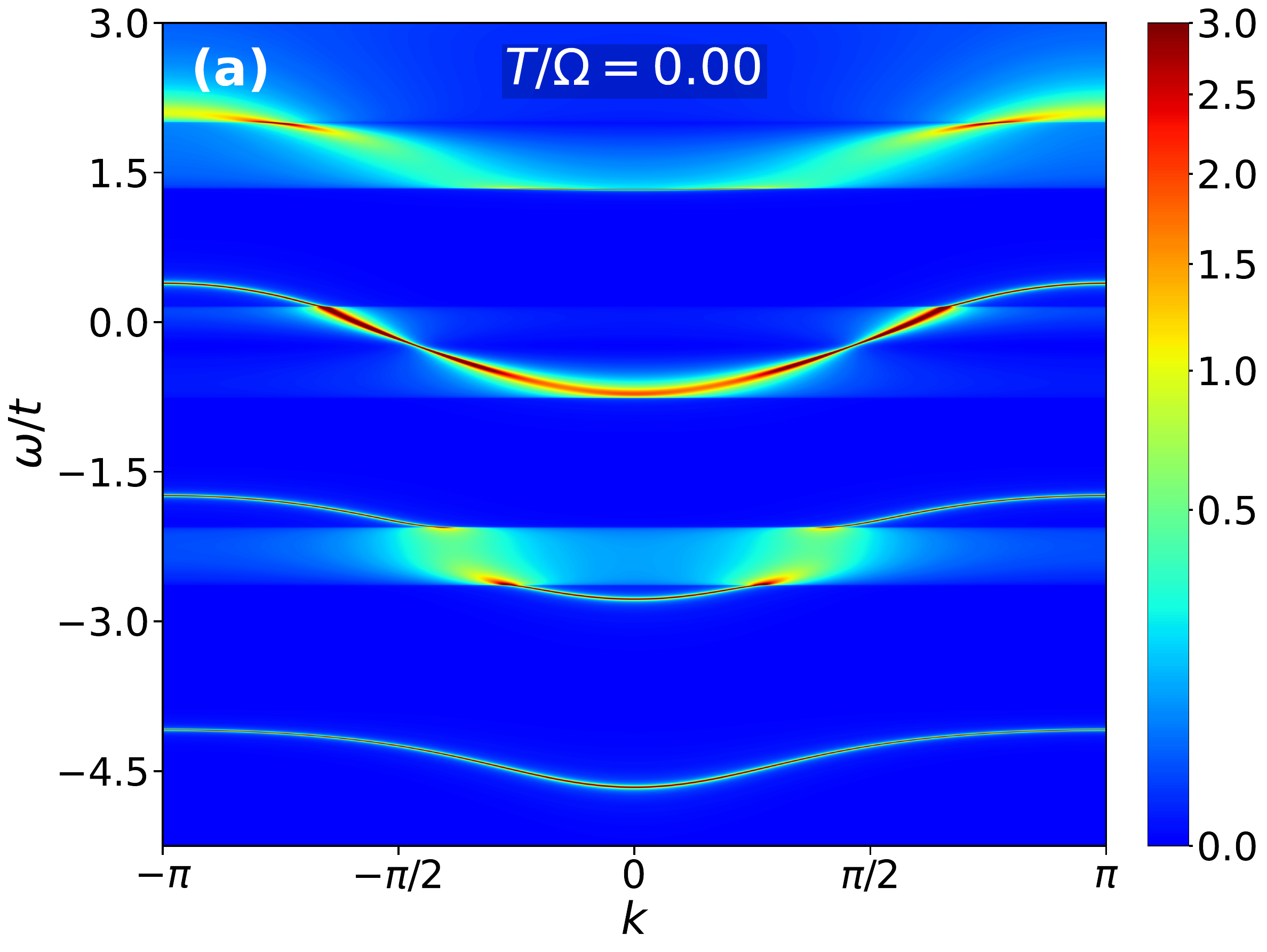}

    \includegraphics[width=0.75\columnwidth]{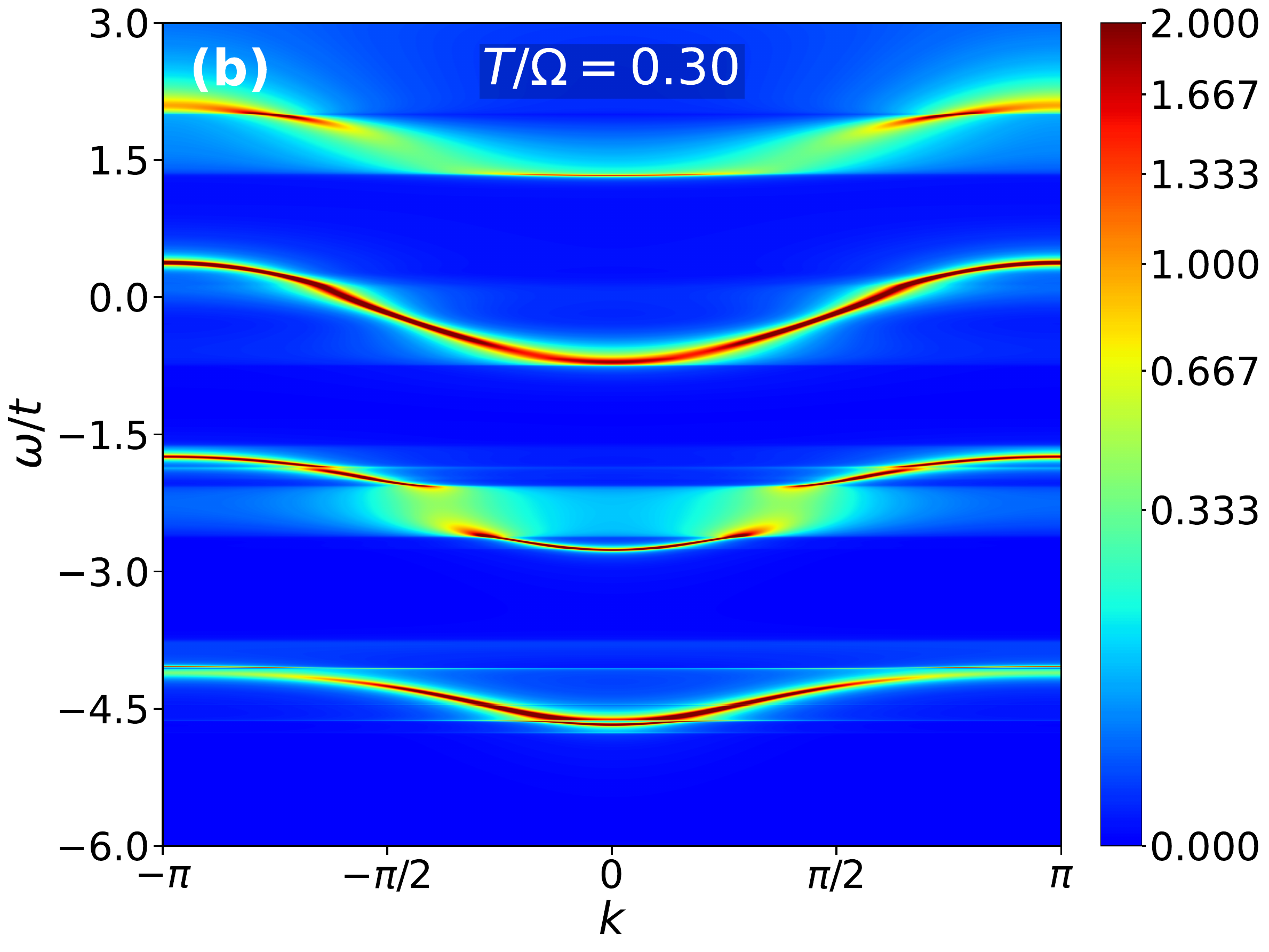}

    \includegraphics[width=0.75\columnwidth]{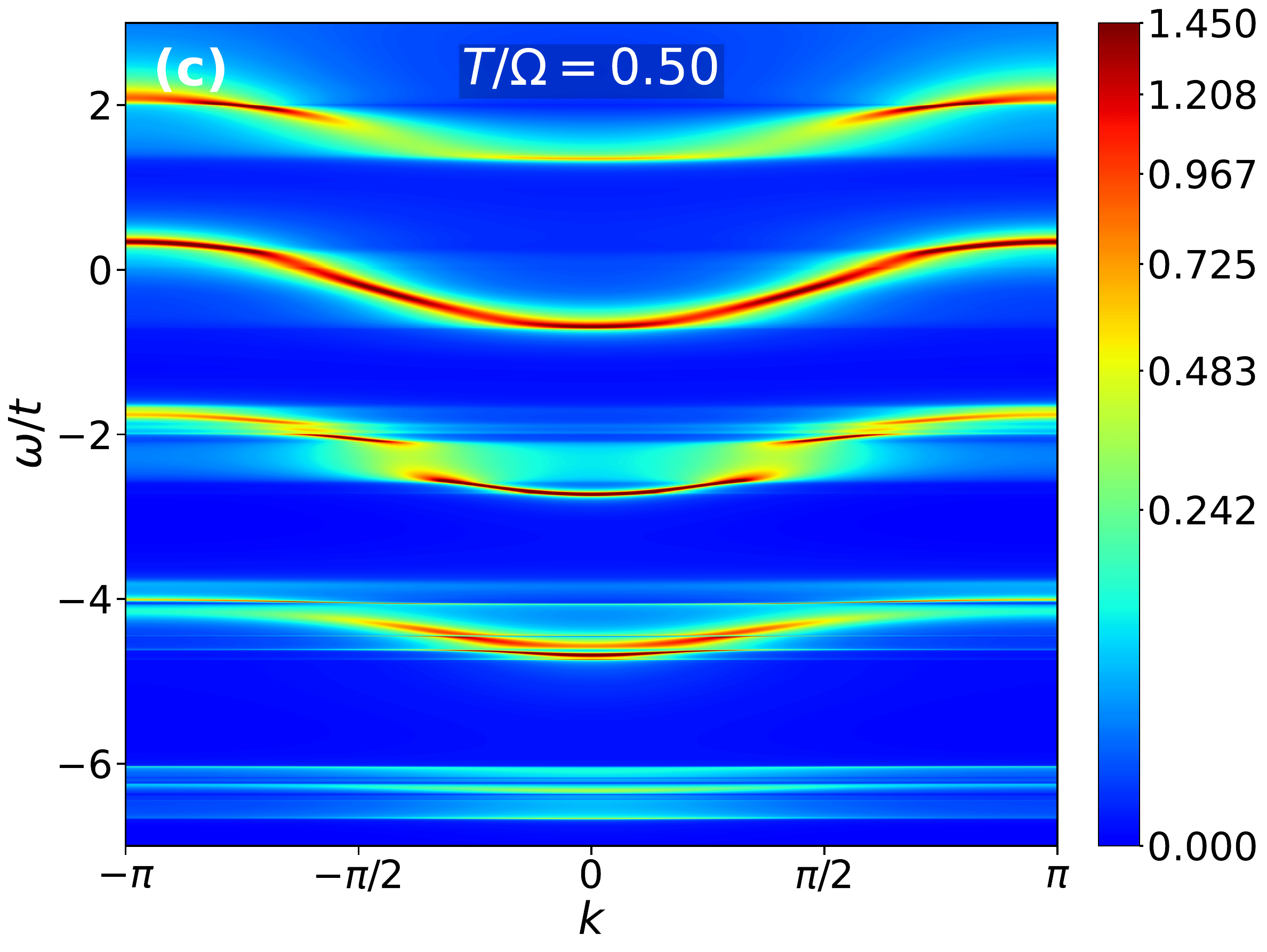}

        \includegraphics[width=0.75\columnwidth]{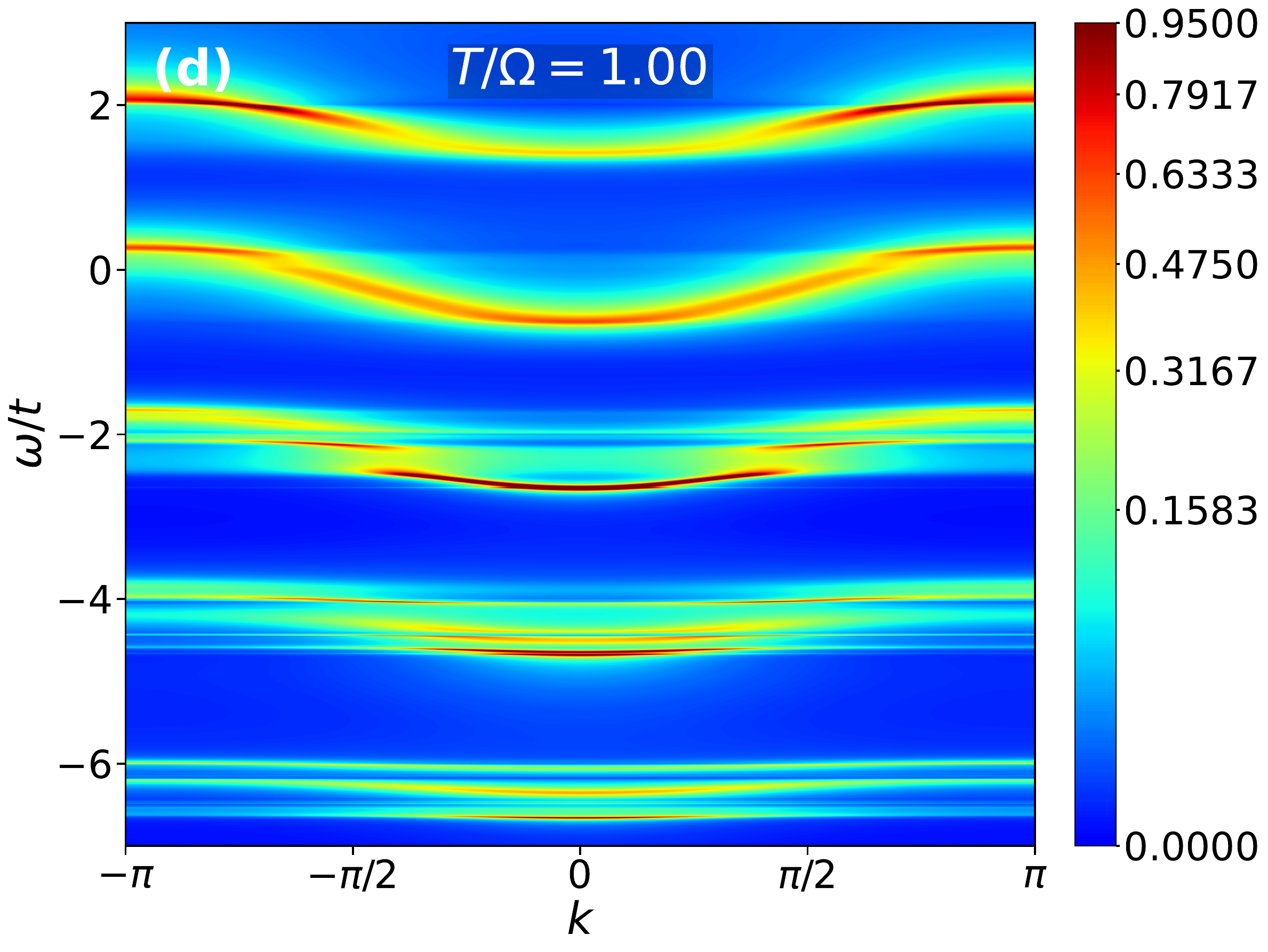}
        
 \caption{(color online) Evolution of the MA$^{(1)}$ $A(\omega,k)$ heat-map with  temperature for the 1D HP with parameters: $\Omega/t=2.0$, coupling strength $\lambda = 2.0$, $t=1$ and $\eta_0=0.001$.}

   \label{fig:heat_map_om_2.0}
\end{figure}

Figs.~\ref{fig:heat_map_om_0.5} and \ref{fig:heat_map_om_2.0} show $A(\omega,k)$ heat-maps across the first Brillouin zone (FBZ), extending results first reported in Ref.~\cite{Bonca2019} to the weakly coupled adiabatic (Fig.~\ref{fig:heat_map_om_0.5}) and strongly coupled anti-adiabatic (Fig.~\ref{fig:heat_map_om_2.0}) regimes. At $T=0$ the \textit{qp} (polaron) band is clearly the lowest, most intense, and sharpest band in both cases; a detailed $T=0$ analysis exists elsewhere (Refs.~\cite{Fehske1997,Fehske2000,Kornilovitch1998,Prokofev1998,Cataudella2007,Linden2005}), so here we simply note the strongly renormalized bandwidth relative to the free-electron band.

With increasing $T$ in Fig.~\ref{fig:heat_map_om_0.5}, two effects dominate: (i) the band loses sharpness and its spectral weight redistributes, particularly near the band center; (ii) the \textit{qp} band is intersected by dispersionless thermal features (as expected from Fig.~5 of the main text) of $T$-dependent width, alongside a faint flat band forming roughly $\Omega$ below it. By $T/\Omega \sim 1.0$ (Fig.~\ref{fig:heat_map_om_0.5}(d)) all $T=0$ features have merged into a continuous, free-electron-like dispersion with a larger bandwidth, echoing observations in Ref.~\cite{Bonca2019} for $\Omega=0.5, \lambda = 1.0$.

Fig.~\ref{fig:heat_map_om_2.0} shows the same qualitative trends (i)--(ii), but with a substantial thermal band forming roughly $\Omega$ below the \textit{qp} band and gaining comparable spectral weight by $T/\Omega\sim0.5$ (panel (c)). Unlike the adiabatic case, the coherent and incoherent parts of $A(\omega,k)$ remain distinguishable, spaced by roughly $\Omega$, even at elevated $T$.

\subsection{Additional finite-$T$ effective mass and broadening predictions}

Complementing the anti-adiabatic regime of Fig.~1 in the main text, Figs.~\ref{fig:eff_mass_lifetime_om_1.0} and \ref{fig:eff_mass_lifetime_om_0.75} report $m^\ast(T)$ and $\eta^\ast(T)$ in regimes where the Lorentzian fit is harder to extract. For $\Omega=1.0t$ (Fig.~\ref{fig:eff_mass_lifetime_om_1.0}) results are limited to weak and intermediate coupling for $D\ge2$, since the \textit{qp} peak carries very low spectral weight even as $T\to0$, complicating reliable Lorentzian fitting; the same restriction applies to the adiabatic case $\Omega=0.75t$ (Fig.~\ref{fig:eff_mass_lifetime_om_0.75}).

Where the Lorentzian form is well defined, $m^\ast(T)$ and $\eta^\ast(T)$ follow the same trends as Fig.~1 of the main text and remain strictly monotonic across all cases analyzed.

\subsection{\texorpdfstring{Fitting of temperature dependence of $m^\ast$ and $\eta^\ast$}{Fitting of temperature dependence of m* and eta*}}

\begin{figure}[t]
    \centering
\includegraphics[width=0.99\columnwidth]{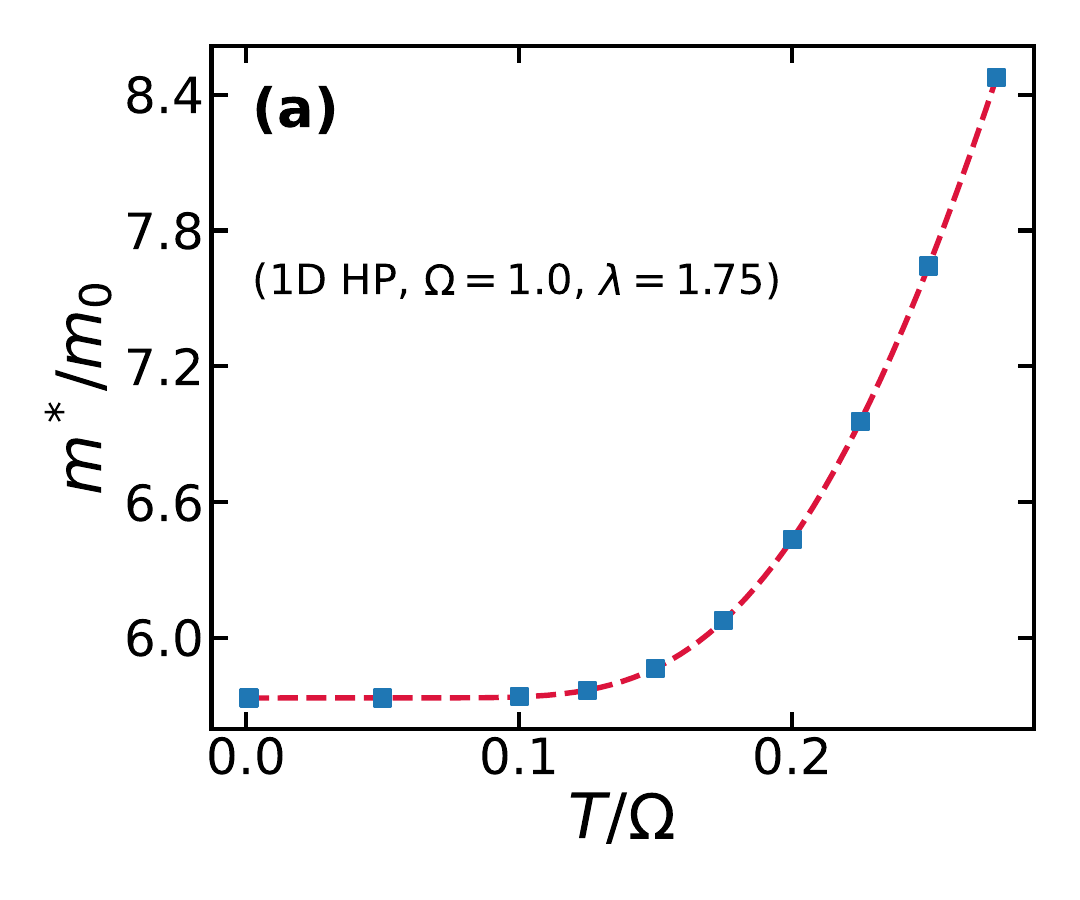}
\includegraphics[width=0.99\columnwidth]{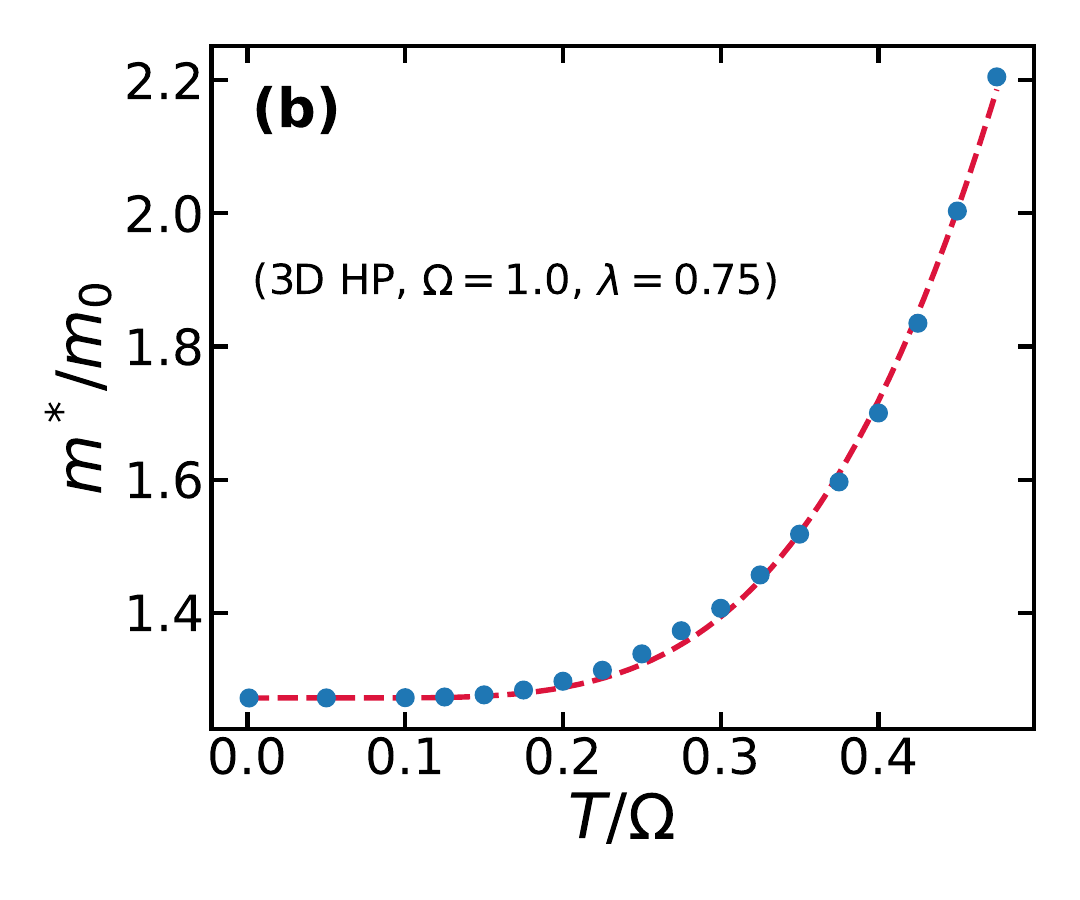}
\includegraphics[width=0.99\columnwidth]{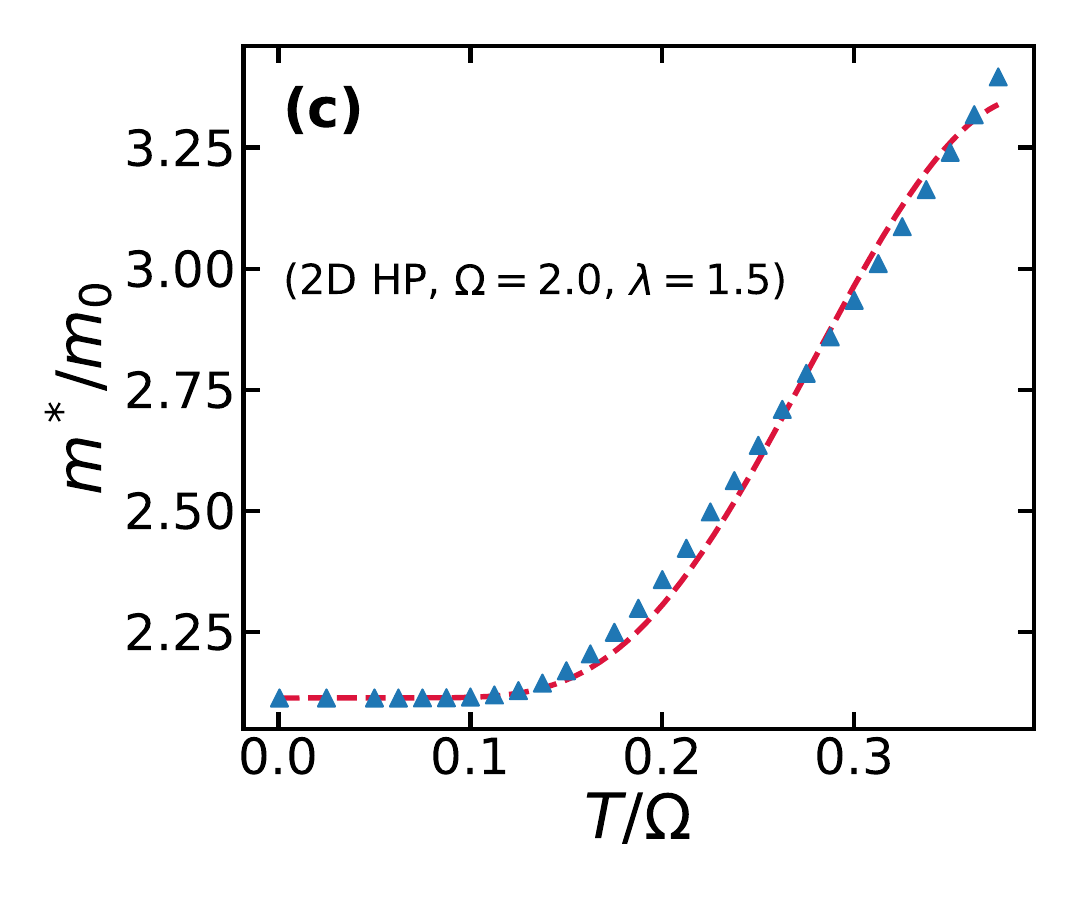}
\caption{(color online) Temperature dependence of the $m^\ast$ fitted to the quadratic form:
$m^\ast(T)/m_{0}=a\,n_{B}^{2}(\Omega)+b\,n_{B}(\Omega)+c$.
The fitted parameters are: (a) $a=-95.60$, $b=104.05$, $c=5.73$;
(b) $a=32.37$, $b=2.10$, $c=1.27$;
(c) $a=-175.39$, $b=29.50$, $c=2.11$.  Original spectral data obtained from  MA$^{(1)}$.}
    \label{fig:m*_fitting}
\end{figure}

\begin{figure}[t]
    \centering

\includegraphics[width=0.99\linewidth]{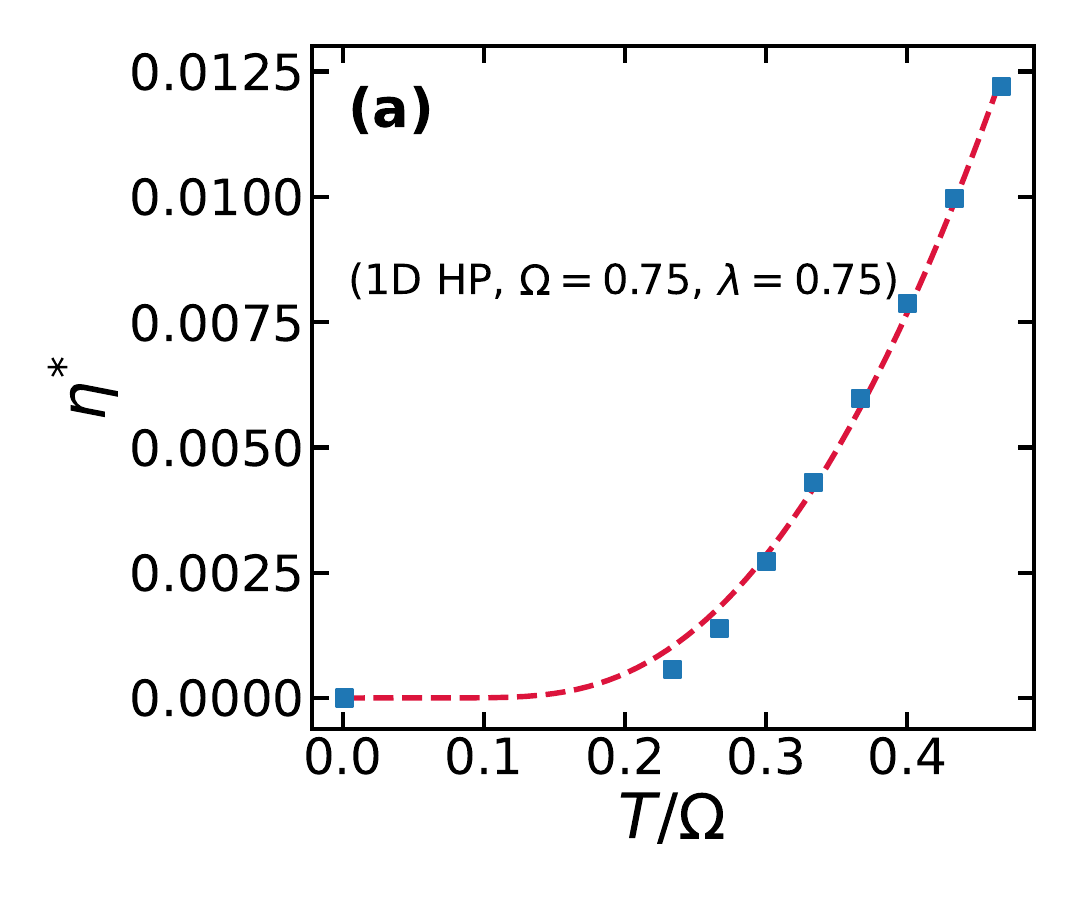}
\includegraphics[width=0.99\linewidth]{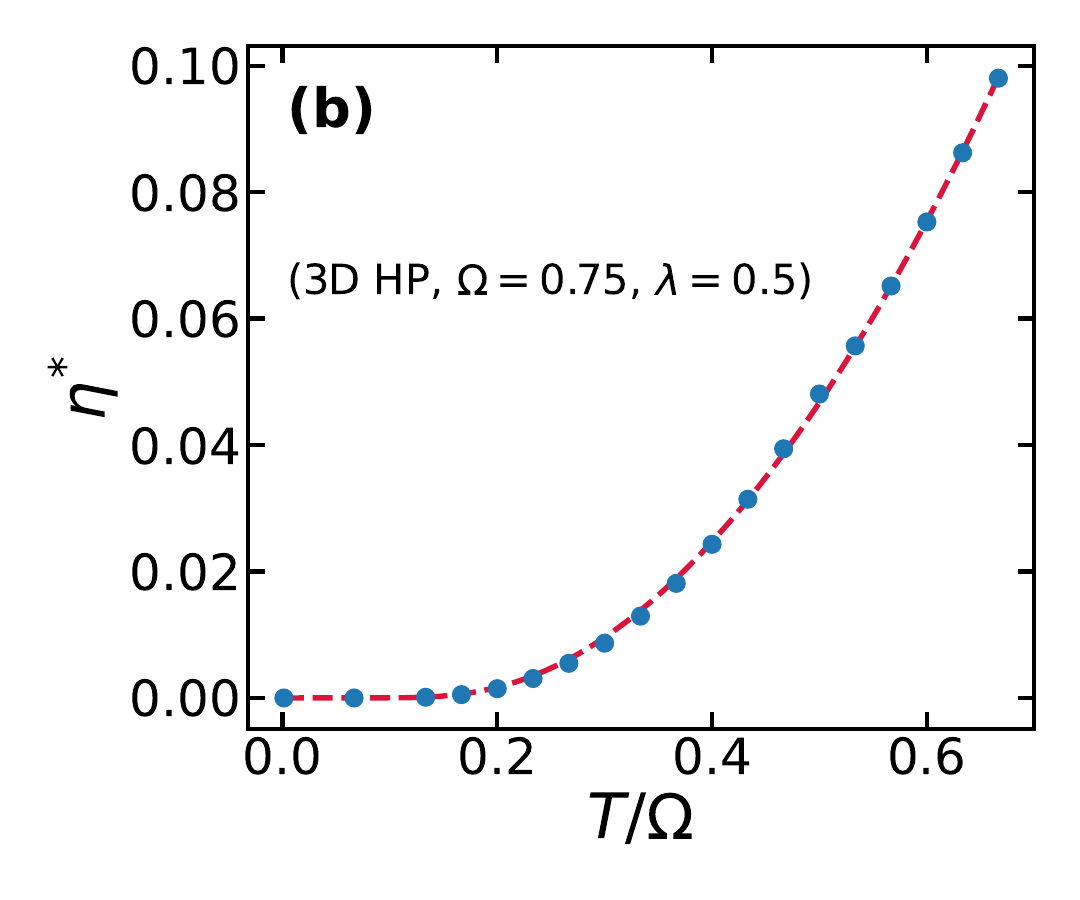}

\includegraphics[width=0.99\linewidth]{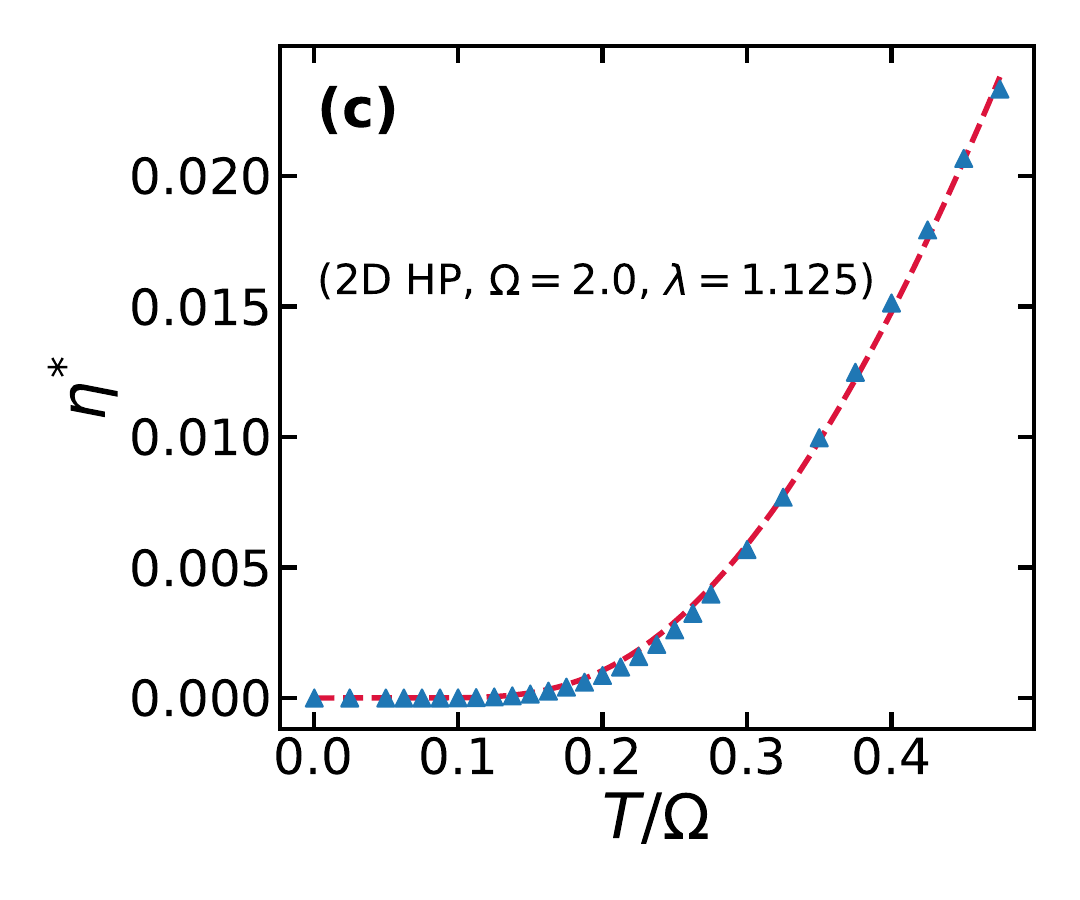}

\caption{(color online) Temperature dependence of the $\eta^\ast$ fitted to:
$\eta^\ast(T)=p\,n_{B}(\Omega)+q\,n_{B}^{2}(\Omega)$.
The fitted parameters are:(a) $p=0.07$, $q=0.17$;
(b) $p=0.25$, $q=0.33$;
(c) $p=0.15$, $q=0.13$. Original spectral data obtained from  MA$^{(1)}$.
}

    \label{fig:tau*_fitting}
\end{figure}

\begin{figure*}[!t]
  \includegraphics[width=0.9\textwidth]{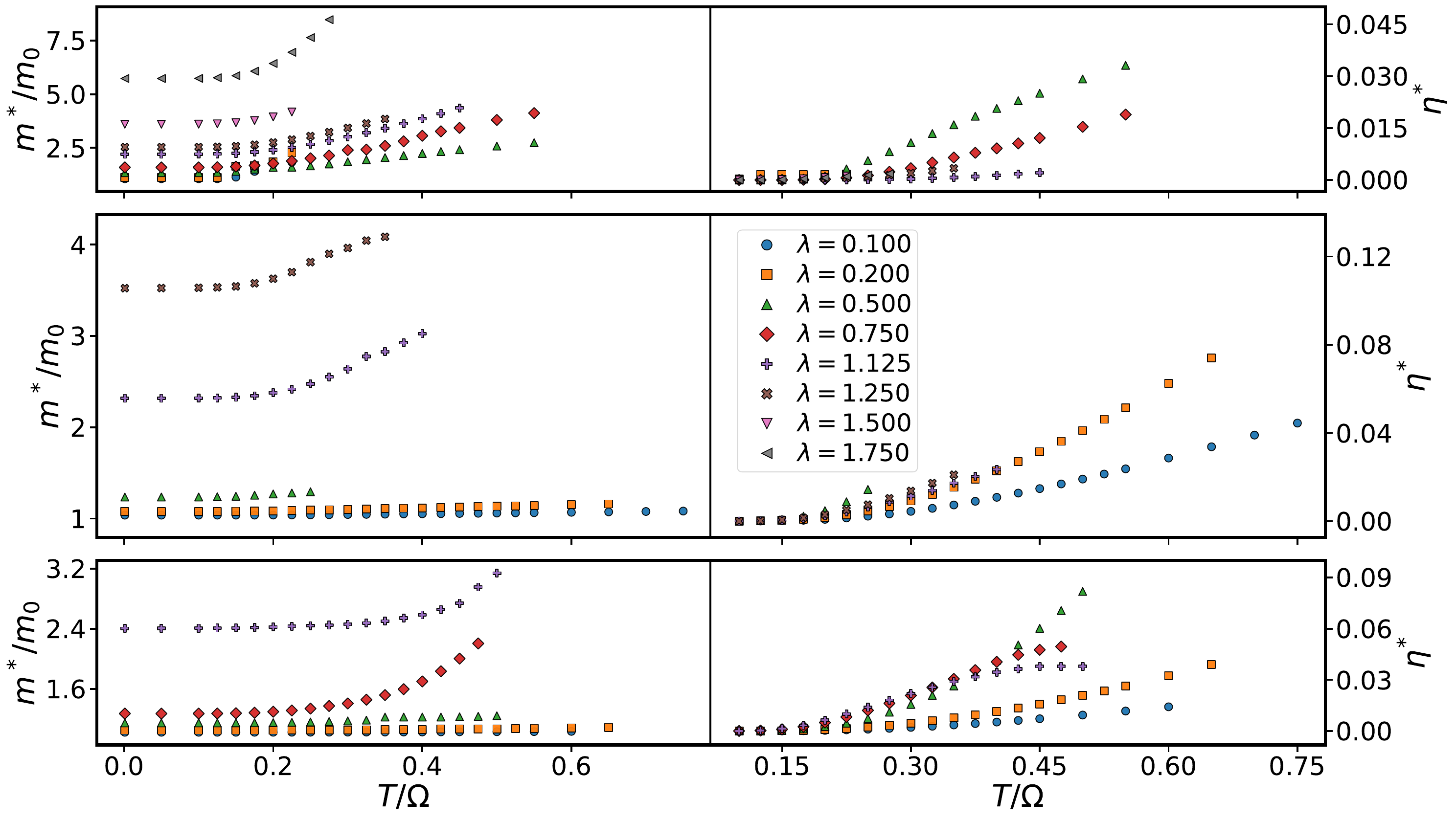}
  \caption{(color online) Temperature dependence up to $T \sim 0.75\Omega$ of  $m^\ast$ and $\eta^\ast$ of the 1D, 2D and 3D HP across coupling regimes, according to MA$^{(1)}$.  Here $t=1, \Omega=1.0$ and $\eta_0=0.001$ and $0.01$ for $D=1$ and $D\ge2$, respectively. The worst-case  fitting standard error was $\sim10^{-2}$ for the \textit{qp}  weight and below  $\sim 10^{-4}$ for $\eta^\ast$.}
   \label{fig:eff_mass_lifetime_om_1.0}
\end{figure*}

\begin{figure*}[!t]
  \includegraphics[width=0.9\textwidth]{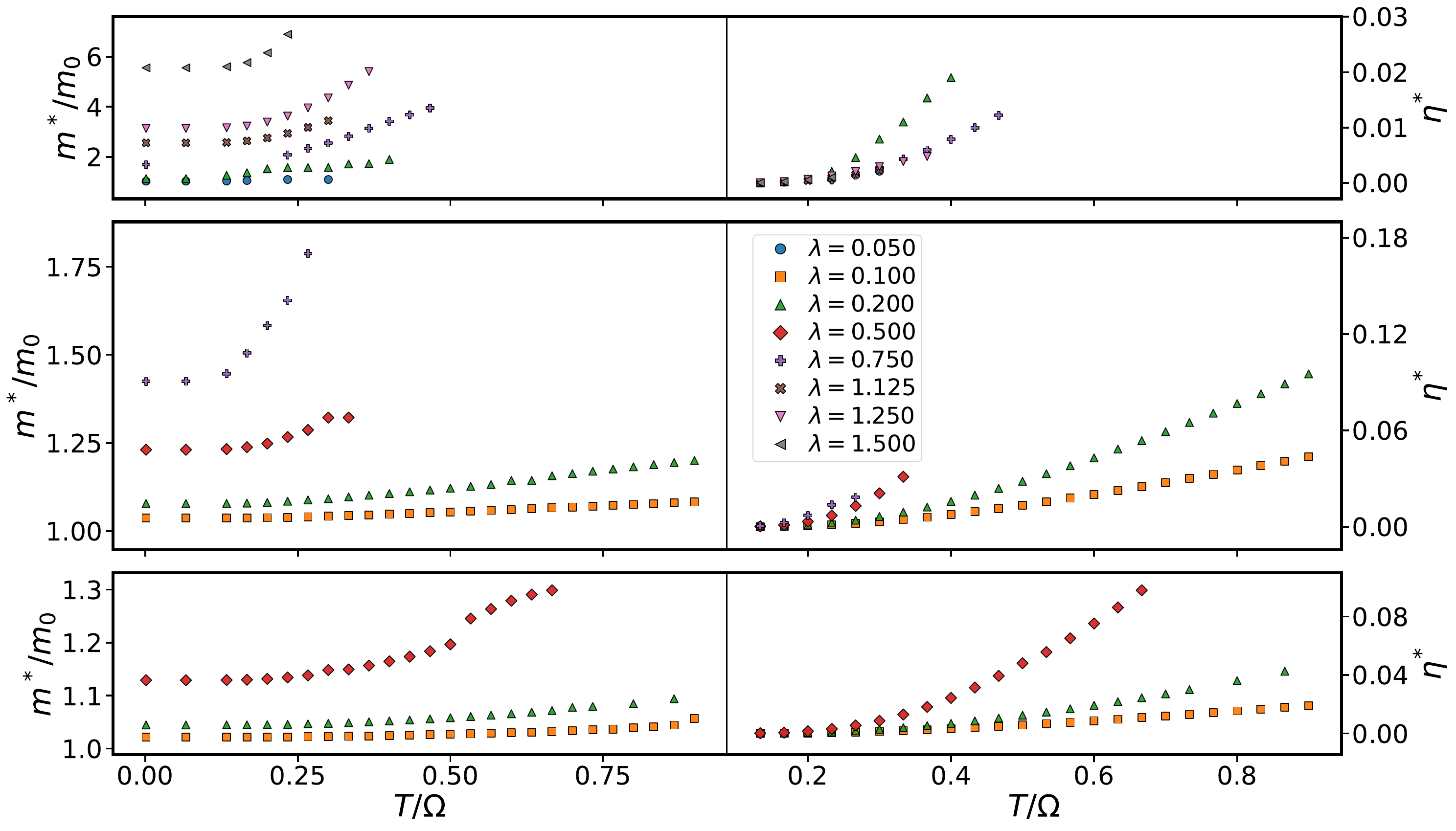}
  \caption{(color online)  Temperature dependence up to $T \sim 0.75\Omega$ of  $m^\ast$ and $\eta^\ast$ of the 1D, 2D and 3D HP across coupling regimes in the weakly adiabatic regime, according to MA$^{(1)}$.  Here $t=1, \Omega=0.75$ and $\eta_0=0.001$ and $0.01$ for $D=1$ and $D\ge2$, respectively. The worst-case  fitting standard error was below $10^{-2}$ for the \textit{qp}  weight and below  $10^{-4}$ for $\eta^\ast$.}
   \label{fig:eff_mass_lifetime_om_0.75}
\end{figure*}

The finite-$T$ BA self energy expression in Eq.~\ref{eqn:SE_BA} gives the following estimate for the temperature dependence of the effective mass $m^\ast(T)$:

\begin{align}
\frac{m^\ast}{m_0}
\approx &
1-g^2\Big[(n_B(\Omega)+1)\,\mathrm{Re} g_0'\!\big(E_P(\textbf{k}=0,T)-\Omega\big)
\nonumber\\ &+n_B(\Omega)\,\mathrm{Re} g_0'\!\big(E_P(\textbf{k}=0,T)+\Omega\big)\Big].
\label{eqn:m*_BA}
\end{align}
where $E_P(\textbf{k}=0,T)\sim-2Dt $ at lowest (BA) order and $g_0'$ is the first derivative of the local propagator. Higher-order diagrams correct the value of $E_P(\textbf{k}=0,T)$ and introduce additional terms proportional to $n_B^2(\Omega)$, $n_B^3(\Omega)$, and so on, so we expect $m^\ast(T)$ to be a power series in $n_B(\Omega)$.  Fitting the MA$^{(1)}$ $m^\ast$ and $\eta^\ast$ data to low-order polynomials in $n_B(\Omega)$, we find that a quadratic form consistently describes the behavior across regimes, as illustrated by the representative examples in Figs.~\ref{fig:m*_fitting} and \ref{fig:tau*_fitting}.

\clearpage

\section{Benchmarking against numerically exact 1D methods}
\label{SM:comparison}

This section benchmarks MA against existing finite-$T$ numerical data \cite{Mitric2022,Bonca2019,Meisner2020} and previously unpublished VED-FTLM results, establishing: (i) the accuracy of MA despite its low computational cost; (ii) the systematic improvement from MA$^{(0)}$ to MA$^{(1)}$; and (iii) the value of MA as an accurate survey tool across the full parameter regime ahead of more resource-intensive numerically exact methods. Throughout, line shapes are colored consistently: DMFT in purple, DMRG in green, VED-FTLM in orange, MA$^{(0)}$ in sky-blue, and MA$^{(1)}$ in red.

\begin{figure}[!htbp]
\includegraphics[width=\columnwidth]{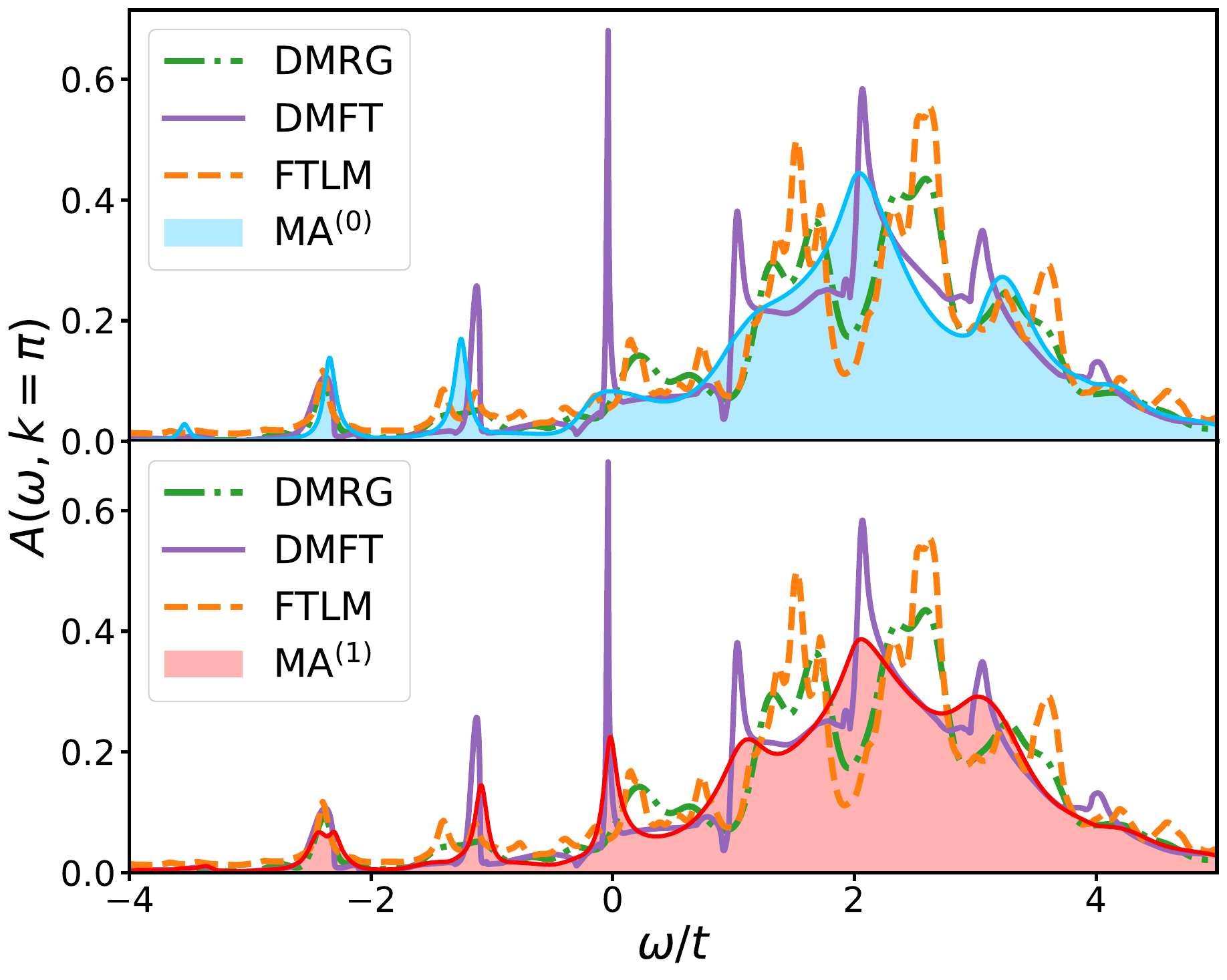}
  \caption{Comparison of MA$^{(0)}$ and MA$^{(1)}$ spectral functions $A(\omega,k=\pi)$ at $T=0.4\,\Omega$ with VED-FTLM \cite{Bonca2019}, finite-$T$ DMRG \cite{Meisner2020}, and DMFT \cite{Mitric2022}. All spectra broadened by $\eta_0=0.05$. DMRG: open chain, $L=101$, quasi-momentum $k=\pi L/(L+1)$; FTLM: six-site ring. Parameters: $\lambda=1$, $\Omega=t=1$.}
   \label{fig:Overlap_pi}
\end{figure}

Fig.~\ref{fig:Overlap_pi} extends the comparison of Fig.~2 in the main text to the Brillouin-zone edge $k=\pi$, at the same parameters ($\lambda=\Omega=t=1$, $T=0.4\,\Omega$). The polaron ground-state peak agrees well across all methods, and DMFT \cite{Mitric2022} and MA$^{(1)}$ — both formulated in the thermodynamic limit — closely agree on the polaron$+$1ph and higher-order continua. The fine structure in VED-FTLM \cite{Bonca2019} and finite-$T$ DMRG \cite{Meisner2020} is attributed to finite-size effects \cite{ReichmannCEI2022}: the DMRG's open boundary conditions give the band center and edge quasi-momenta $k=\pi/(L+1)$ and $k=\pi L/(L+1)$, respectively. None of these methods was designed to capture higher-energy band-edge features, yet all produce qualitatively consistent spectra, confirming MA's reliability there at finite $T$.

\subsection{Comparison with available DMFT data}

MA and DMFT share superficial similarities: both are formulated in the thermodynamic limit with a momentum-independent self-energy for the HP. However, they differ fundamentally. In the single-site DMFT formalism of Ref.~\cite{Mitric2022}, the lattice problem maps to an effective single-site problem whose self-energy is always momentum independent because non-local correlations are neglected entirely \cite{Barisic2007}; cluster extensions such as the dynamical cluster approximation (DCA) restore momentum dependence but have so far been limited to zero temperature \cite{Hague2003}. The Holstein MA self-energy $\Sigma_T(\omega)$ is likewise momentum independent up to MA$^{(1)}$, yet even the lowest order, MA$^{(0)}$, captures momentum dependence for less trivial electron-phonon couplings (breathing mode, Peierls, etc.) \cite{Glenn2008}, and all orders of MA correctly include the non-local correlation function $\chi(i-j)=\langle\psi|c_i^\dagger c_i(b_j+b_j^\dagger)|\psi\rangle$ for $i\ne j$, which vanishes identically in DMFT since the electron and phonons always sit on the same site in that ansatz. Another important  computational difference is that MA is a one-shot calculation, whereas DMFT requires iterations until self-consistency is reached. It is nonetheless instructive to compare the two methods directly.

\begin{figure*}[!ht]
    \centering

    \includegraphics[width=0.24\textwidth]{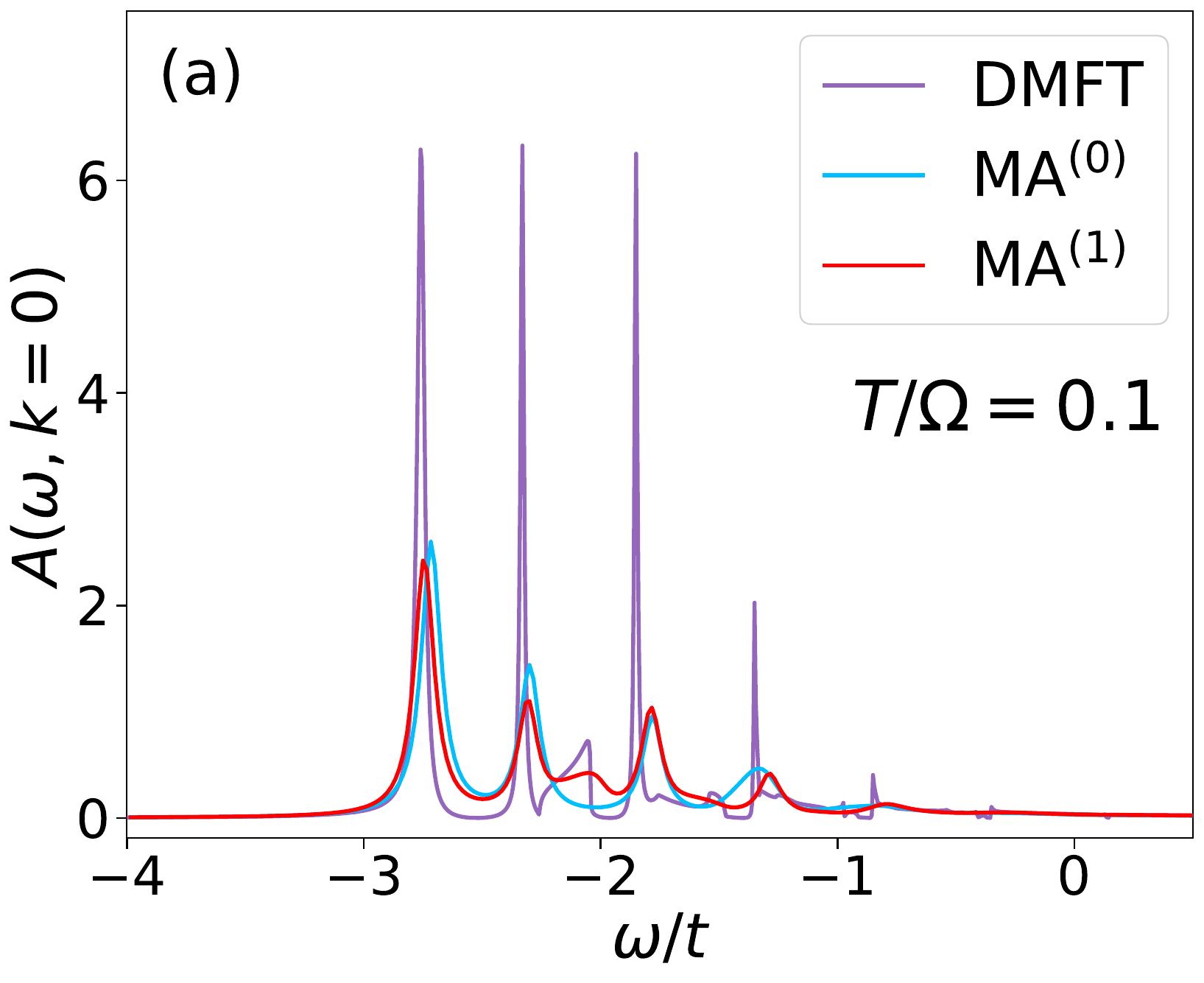}
    \includegraphics[width=0.24\textwidth]{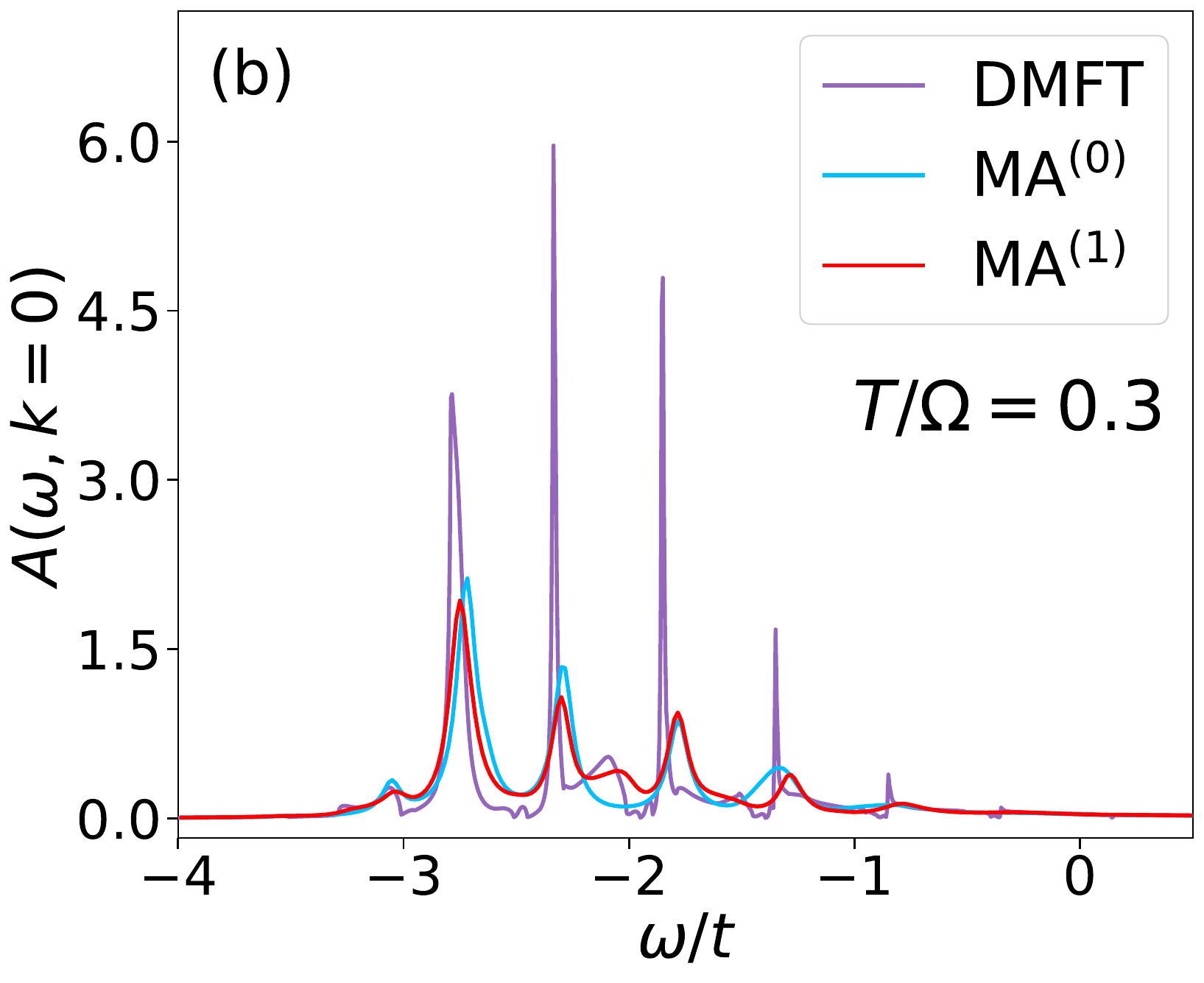}
    \includegraphics[width=0.24\textwidth]{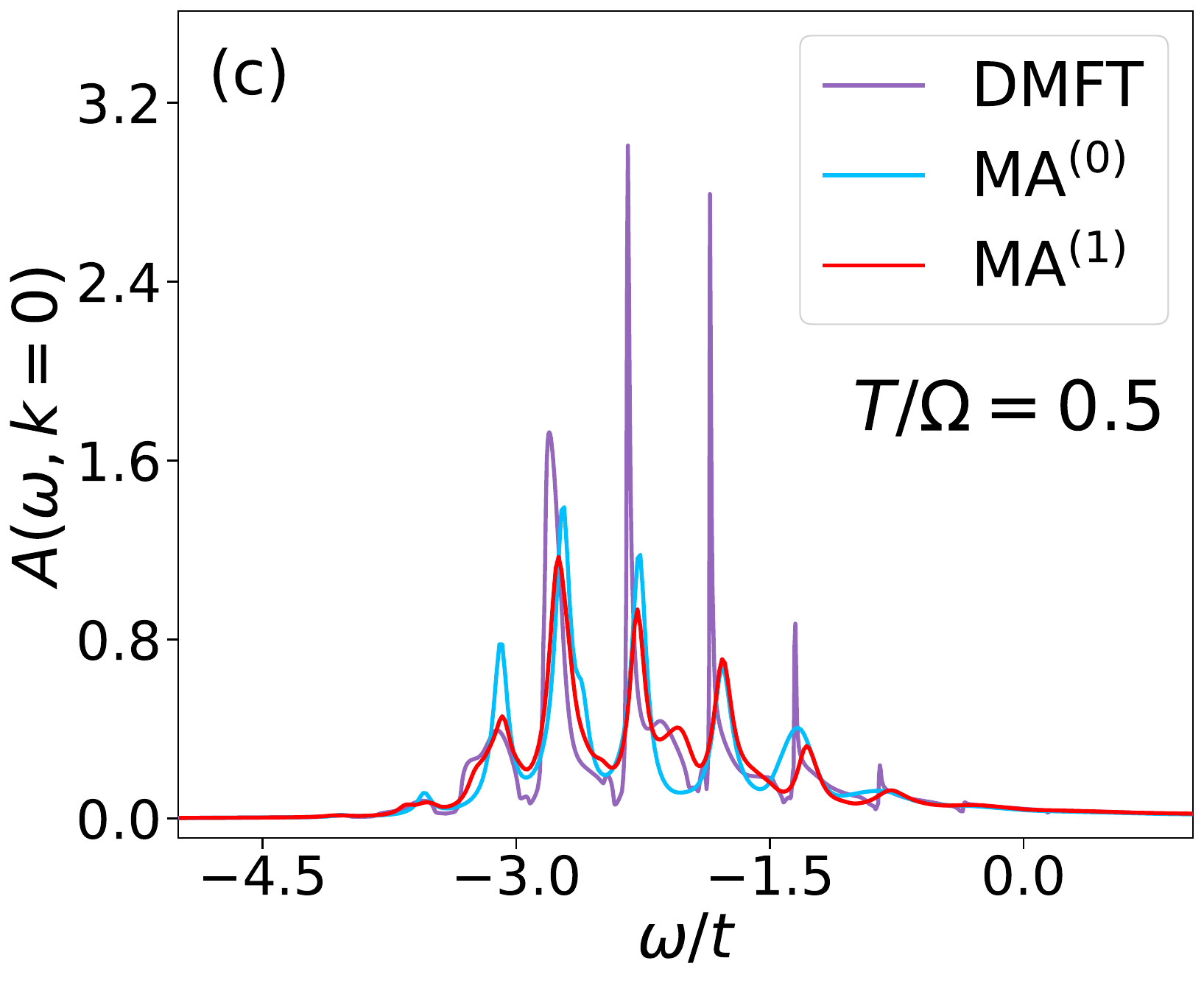}
    \includegraphics[width=0.24\textwidth]{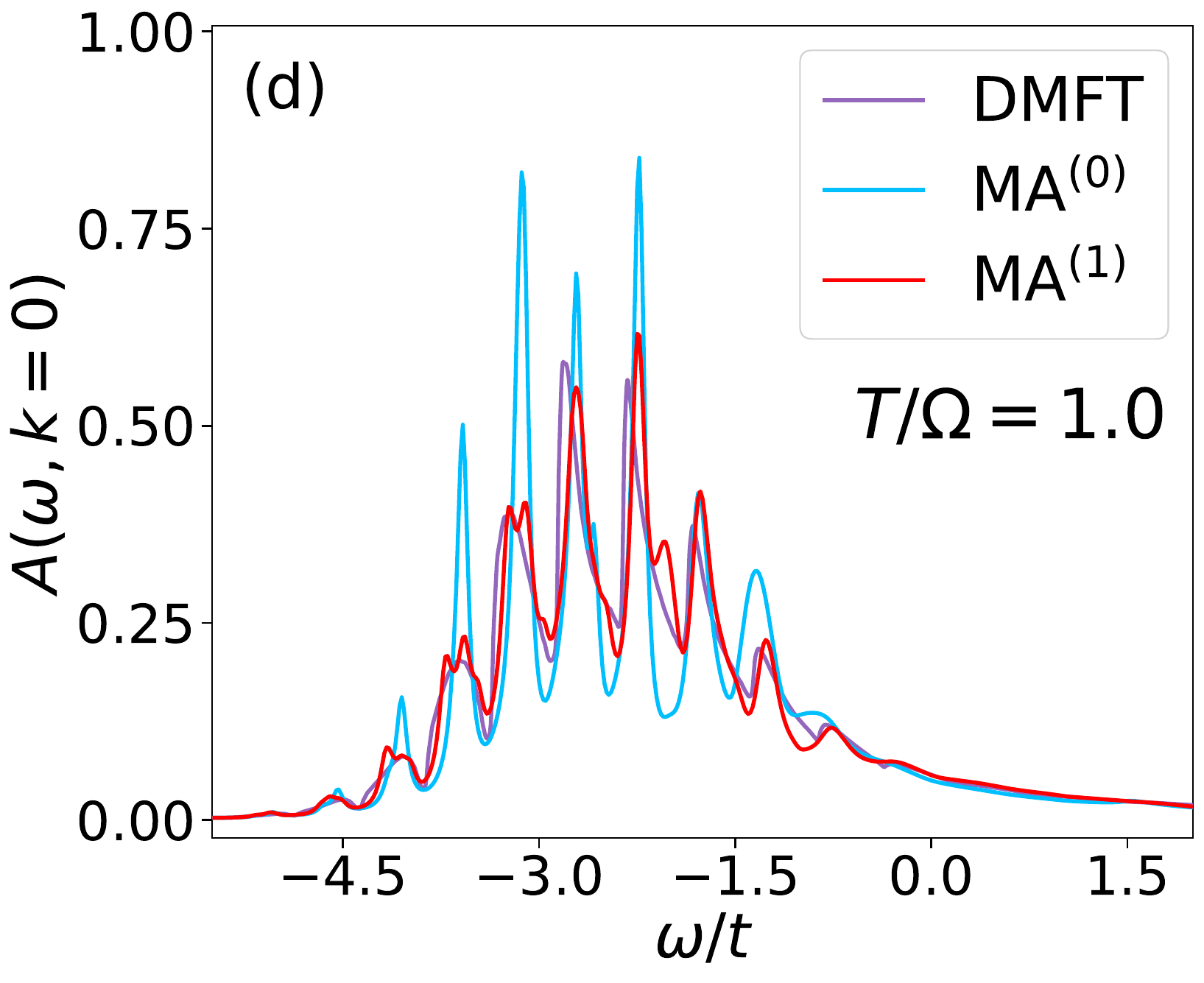}

    \vspace{0.5em}

    \includegraphics[width=0.24\textwidth]{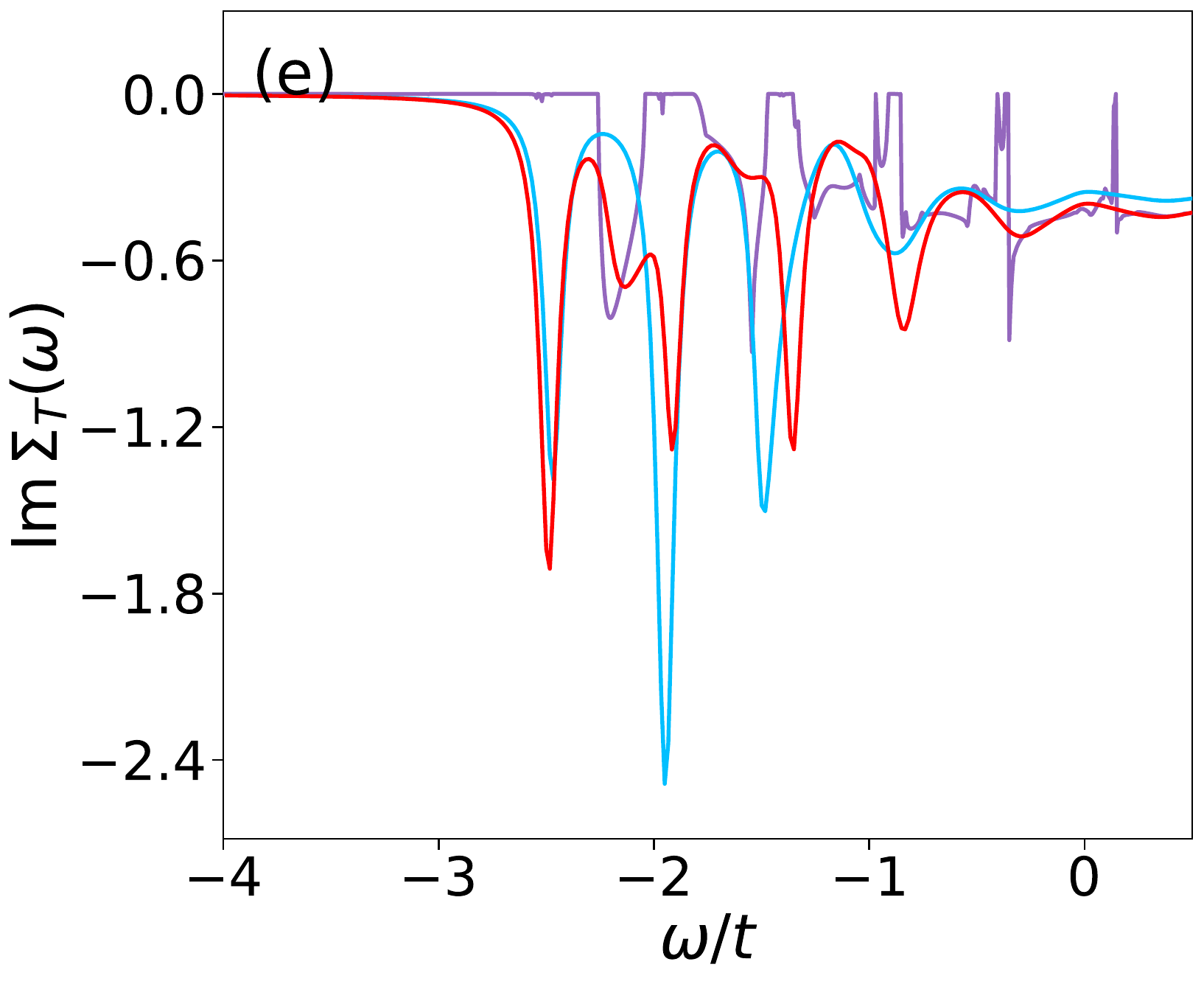}
    \includegraphics[width=0.24\textwidth]{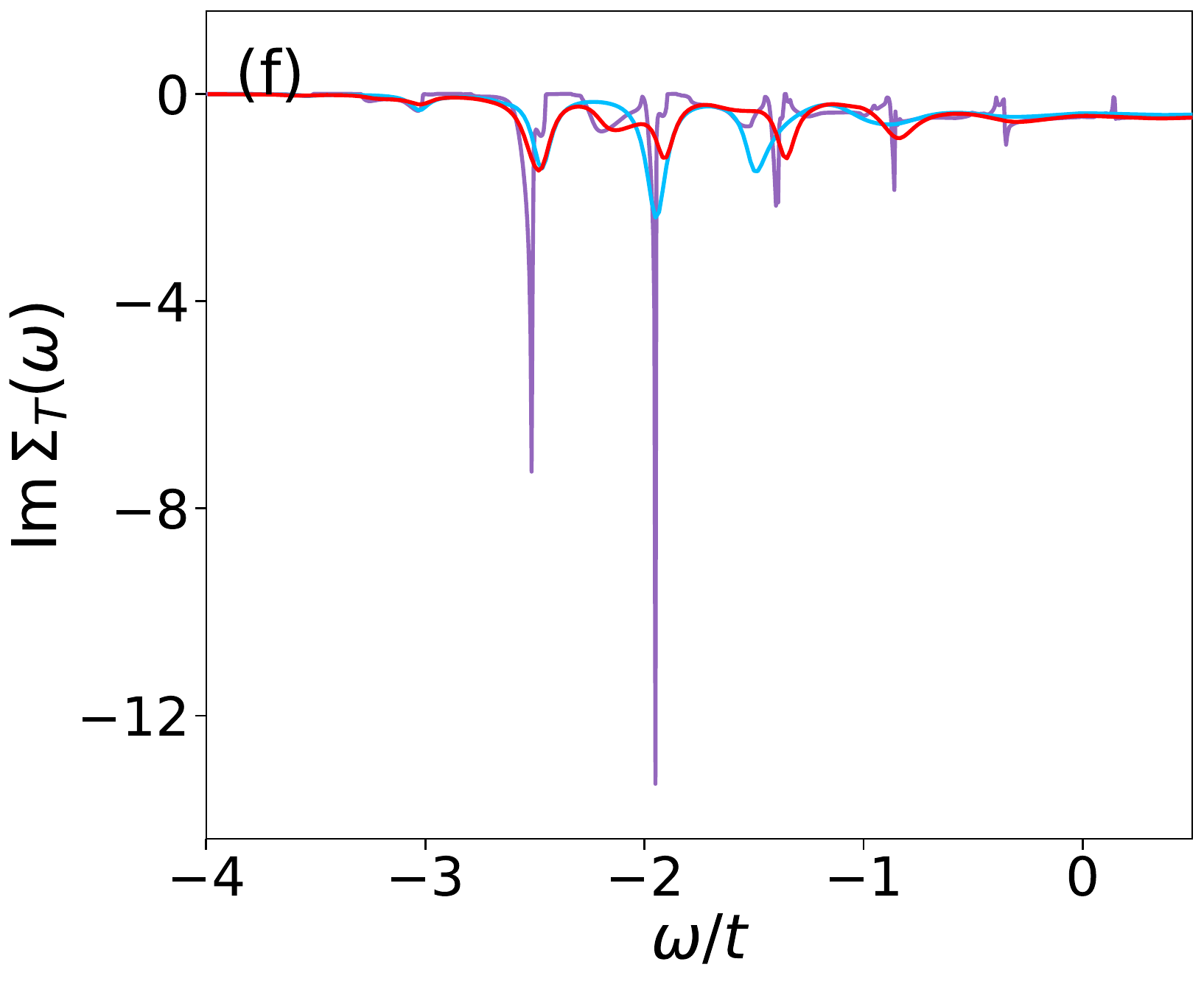}
    \includegraphics[width=0.24\textwidth]{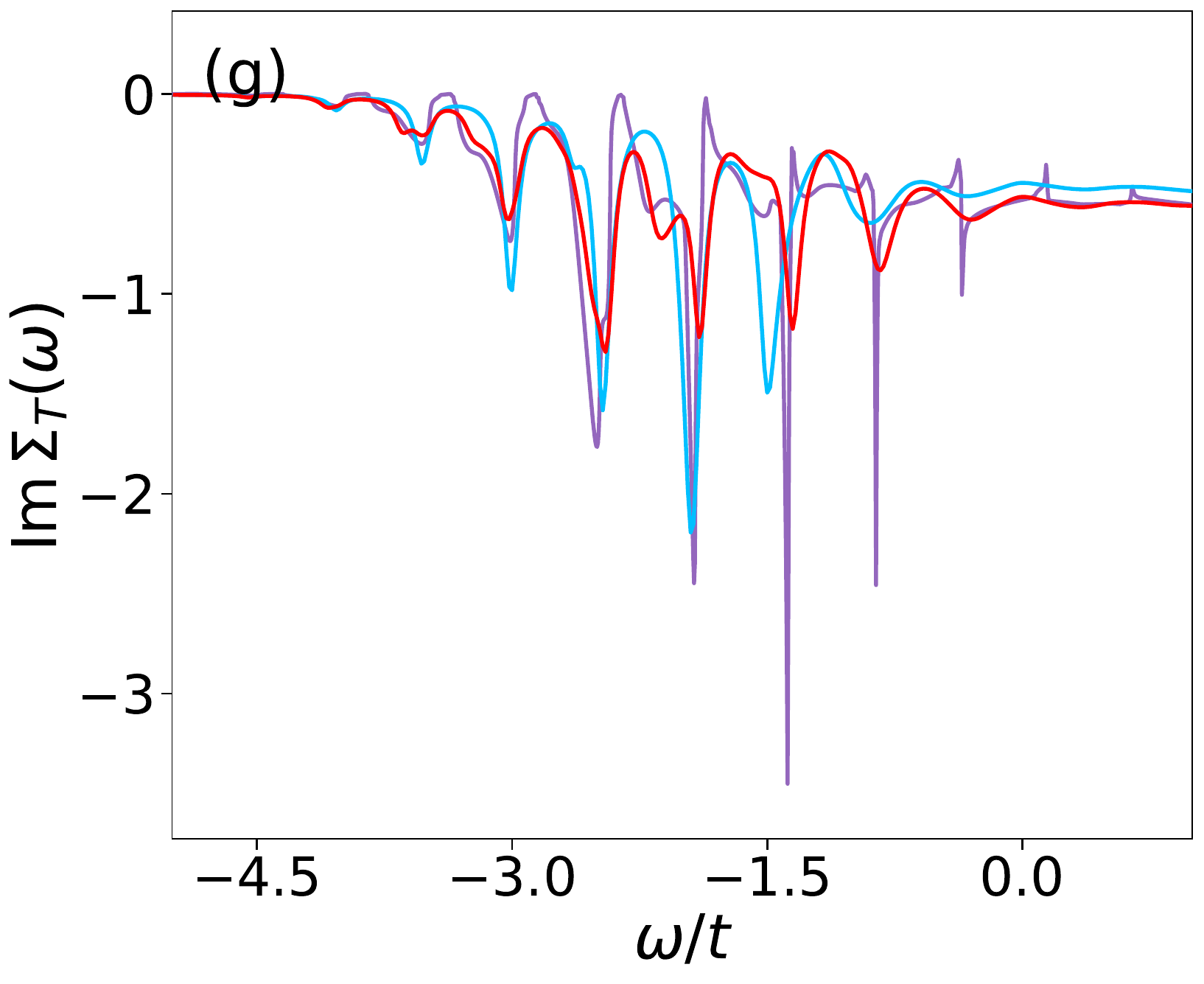}
    \includegraphics[width=0.24\textwidth]{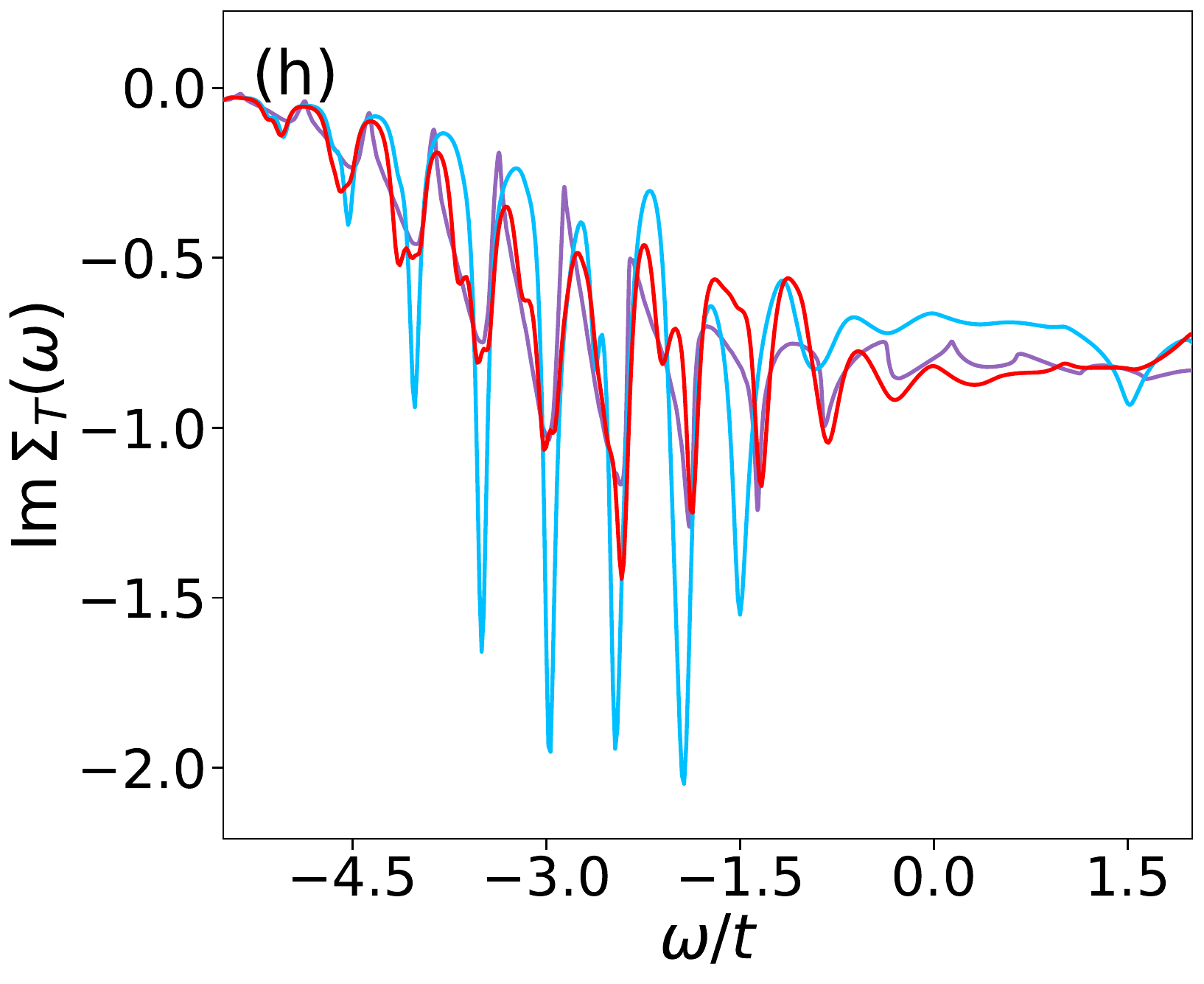}

    \caption{(color online) Comparison of the spectral function $A(\omega,k=0)$ and the imaginary part of the self-energy $\Sigma^{"}_T(\omega)$ from MA$^{(0)}$, MA$^{(1)}$ and DMFT \cite{Mitric2022}, plotted for parameters: $\lambda=1.0$, $\Omega=0.5t, t=1$ and $\eta_0=0.05$ for the MA data. Panels (a)--(d) capture the temperature dependent evolution ($T/\Omega=0.1,0.3,0.5,1$) of $A(\omega,k=0)$ from the three methods;  (e)--(h) showcase the corresponding imaginary part of the self-energy.}
    \label{fig:MA_DMFT_self_energy_comparison}
\end{figure*}

\begin{figure*}[t]
\centering
\begin{minipage}[t]{0.49\textwidth}
\centering
    \includegraphics[width=0.49\linewidth]{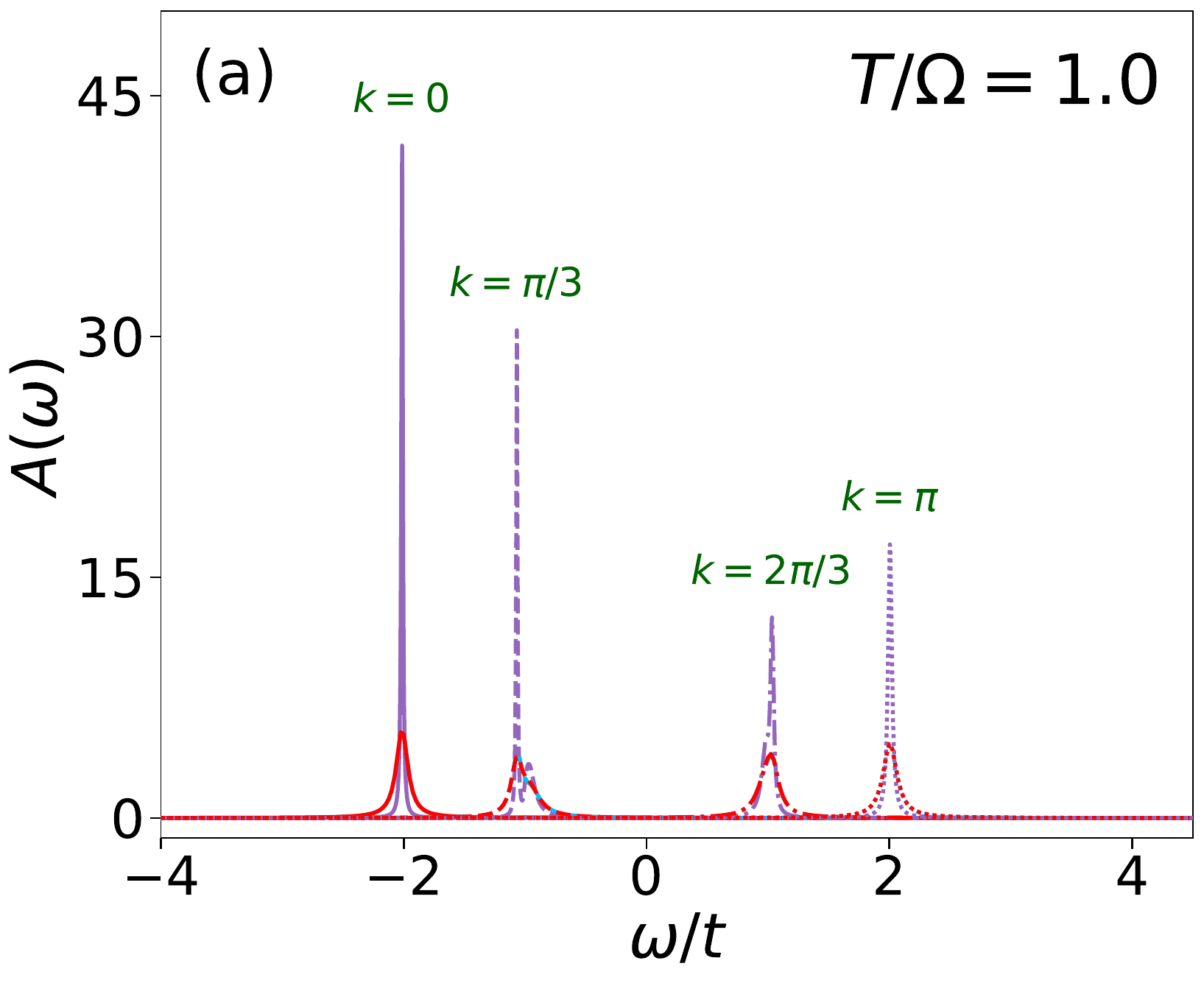}
    \includegraphics[width=0.49\linewidth]{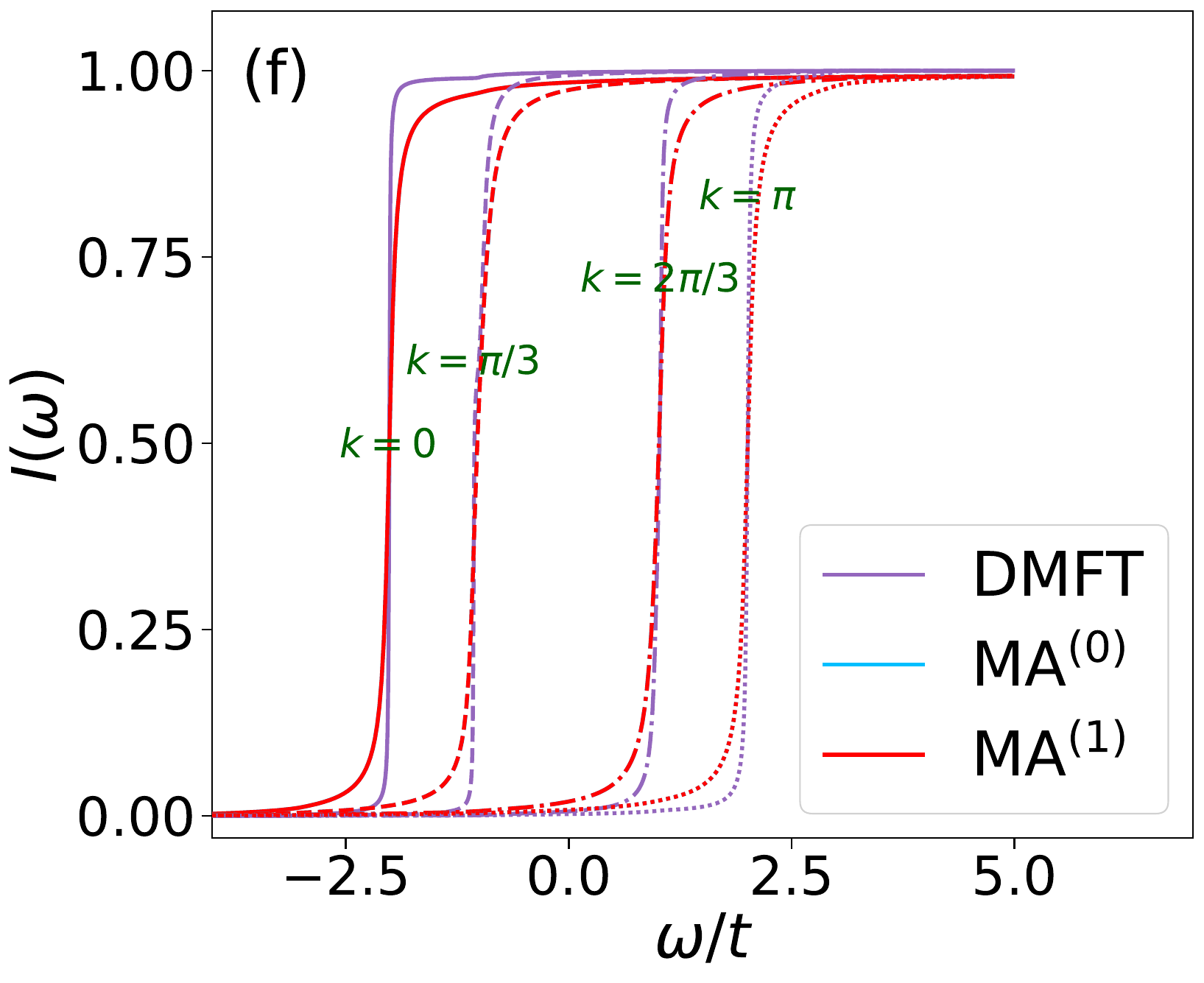}\\[2pt]
    \includegraphics[width=0.49\linewidth]{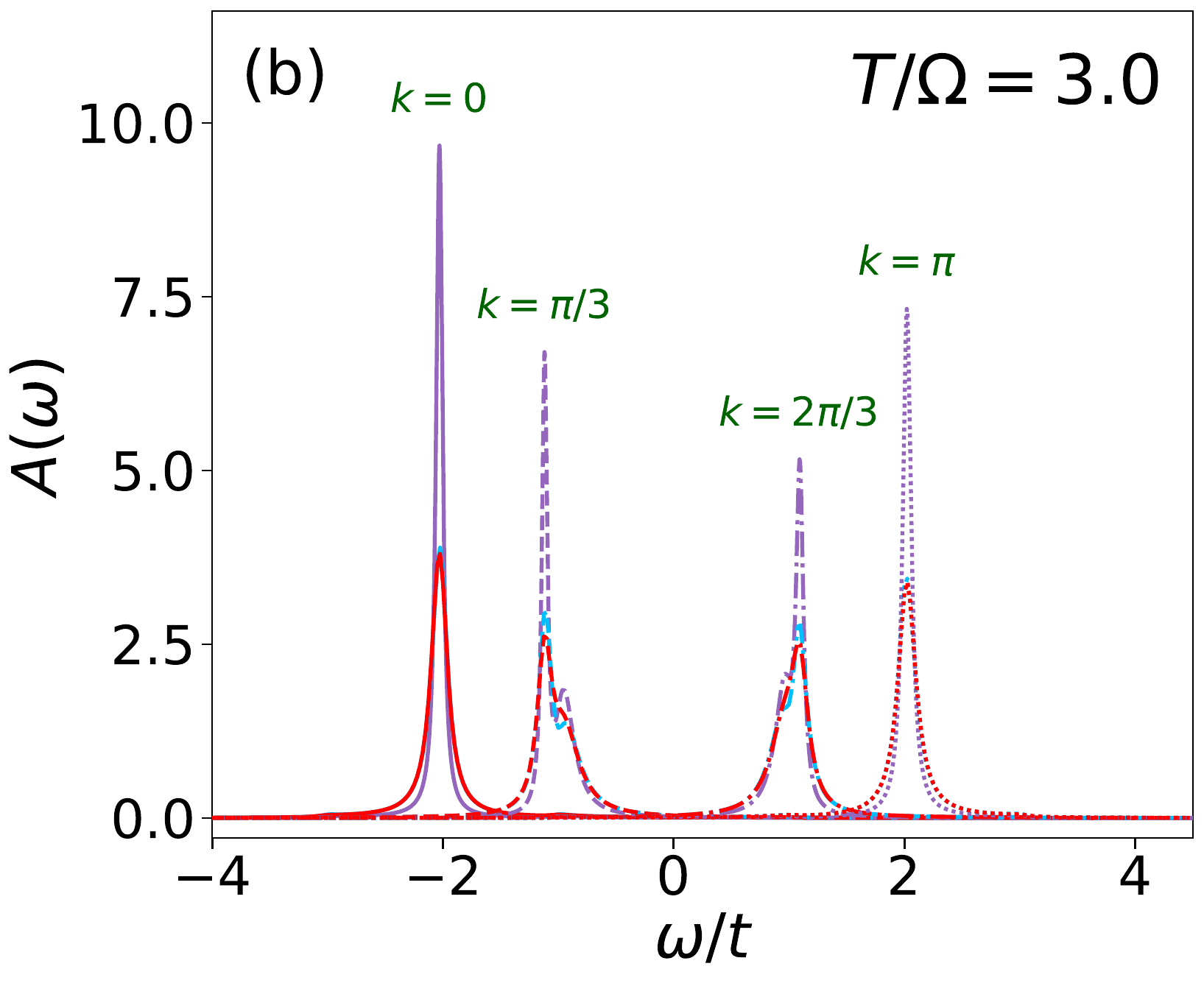}
    \includegraphics[width=0.49\linewidth]{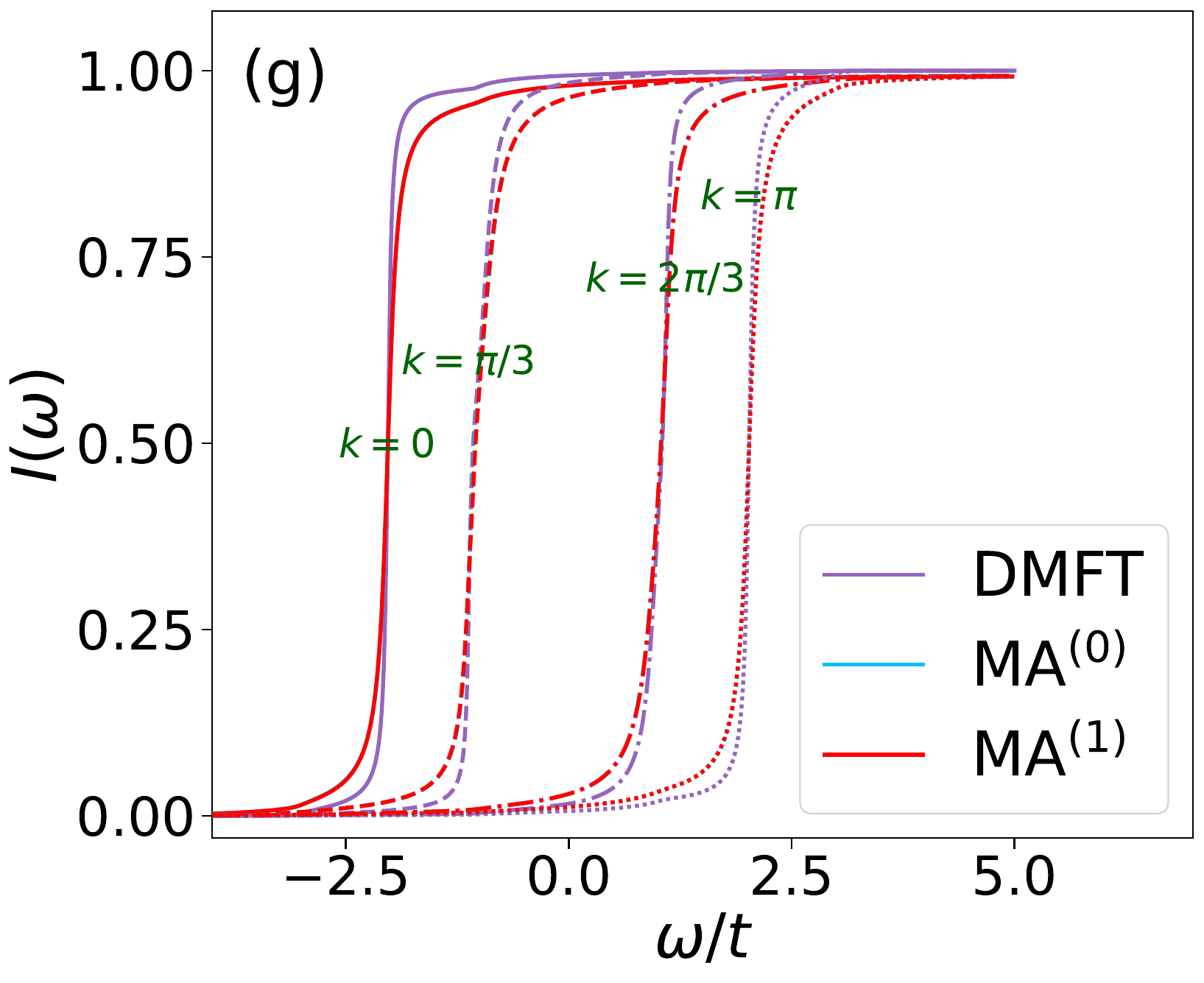}\\[2pt]
    \includegraphics[width=0.49\linewidth]{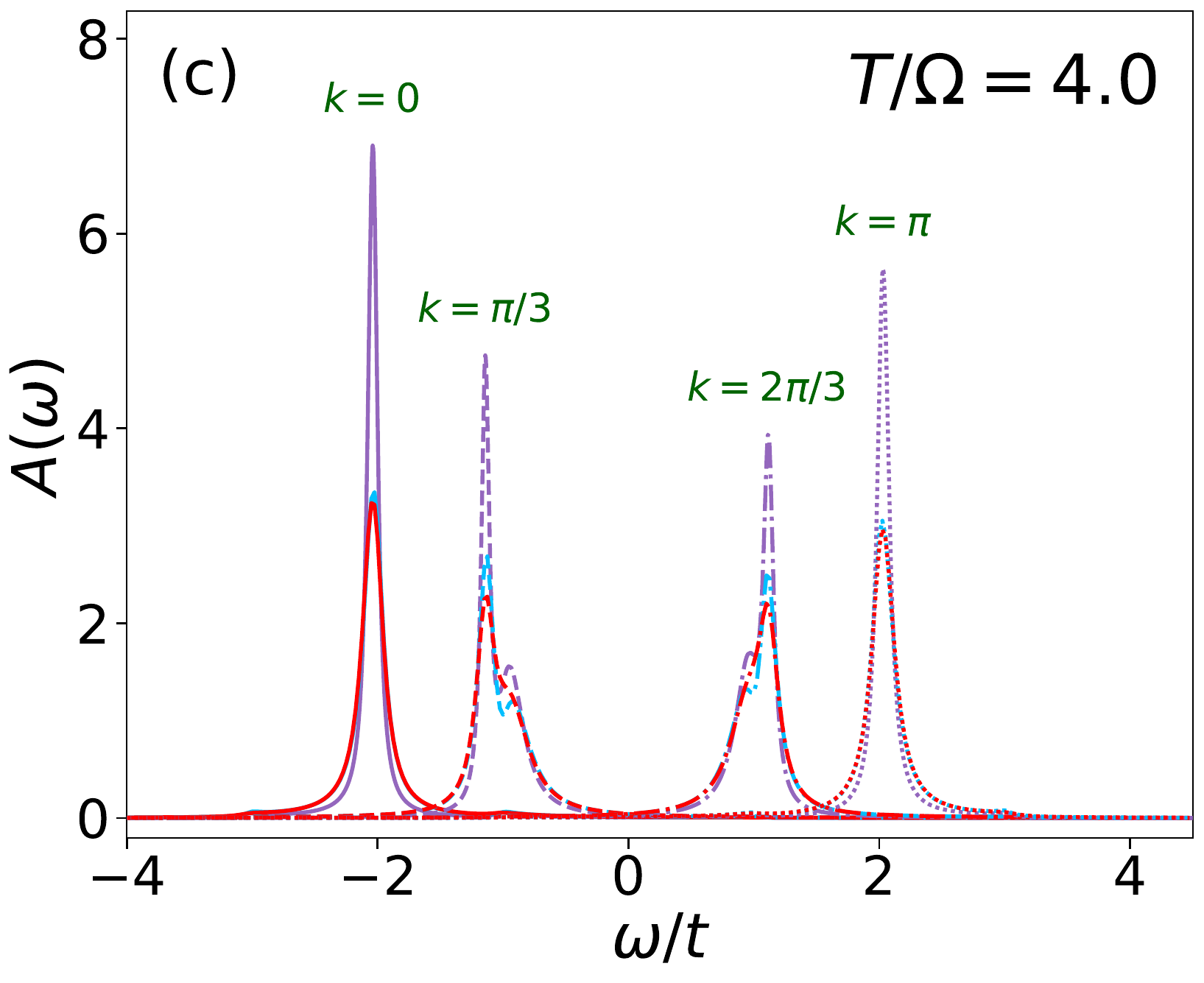}
    \includegraphics[width=0.49\linewidth]{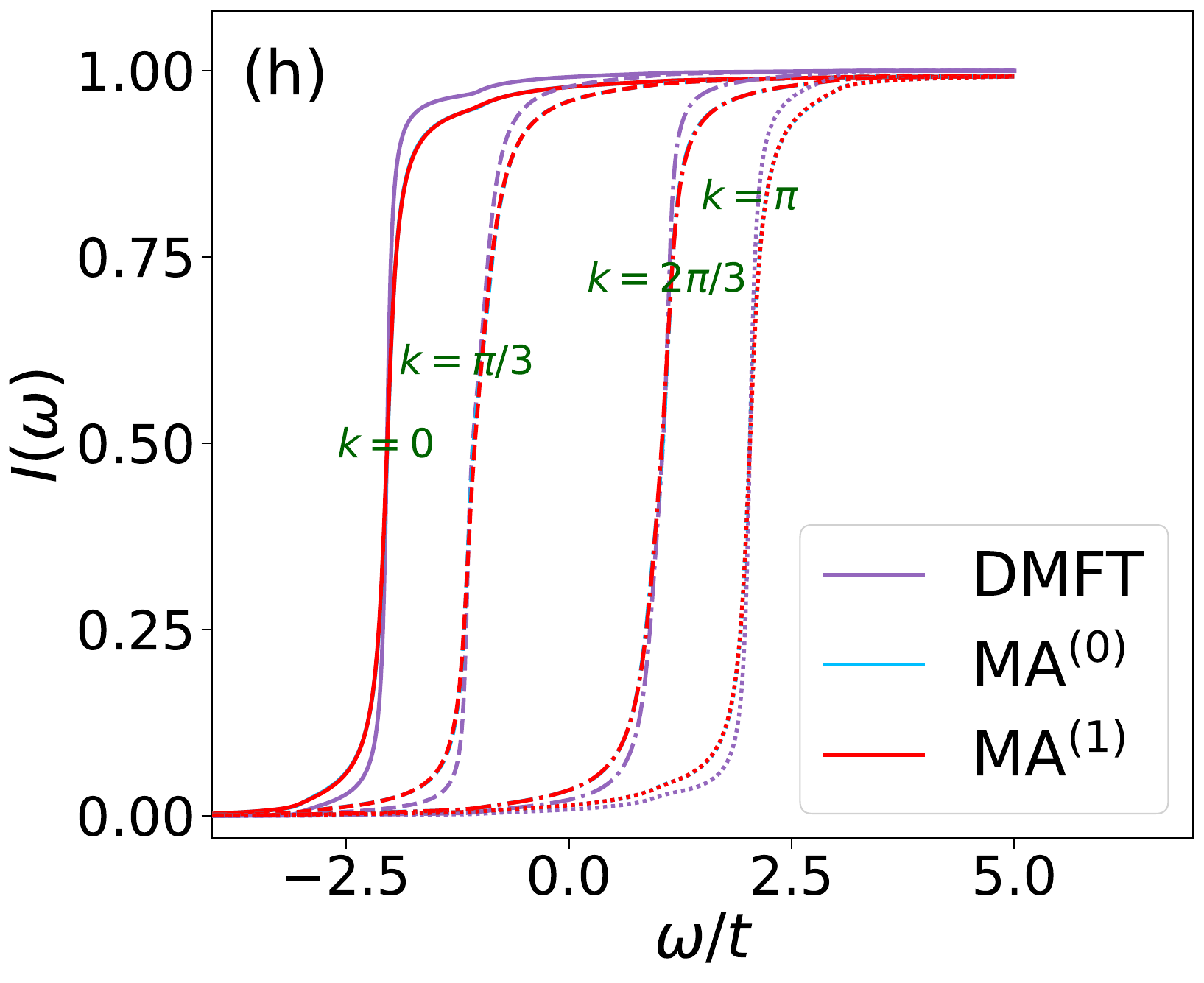}\\[2pt]
    \includegraphics[width=0.49\linewidth]{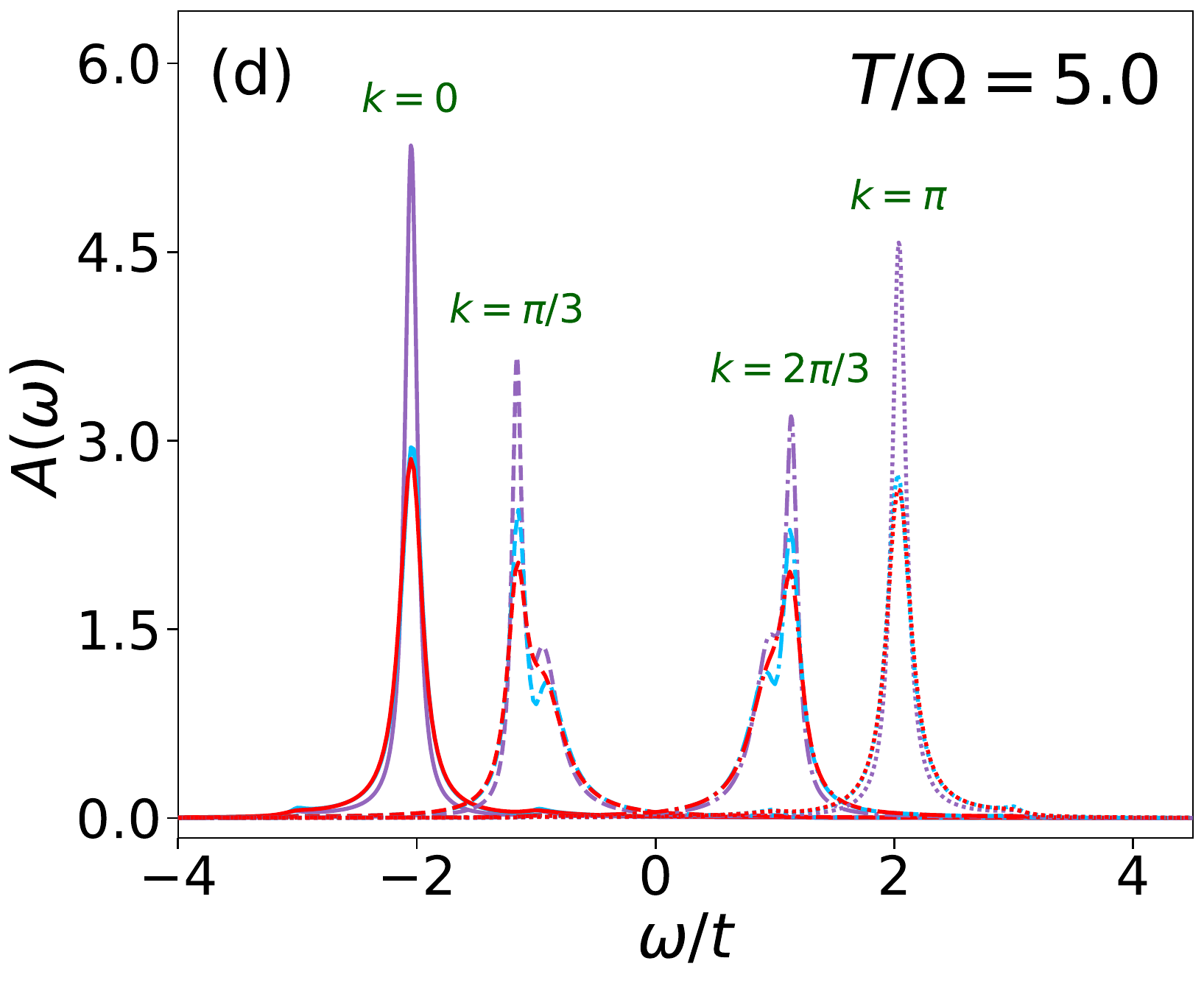}
    \includegraphics[width=0.49\linewidth]{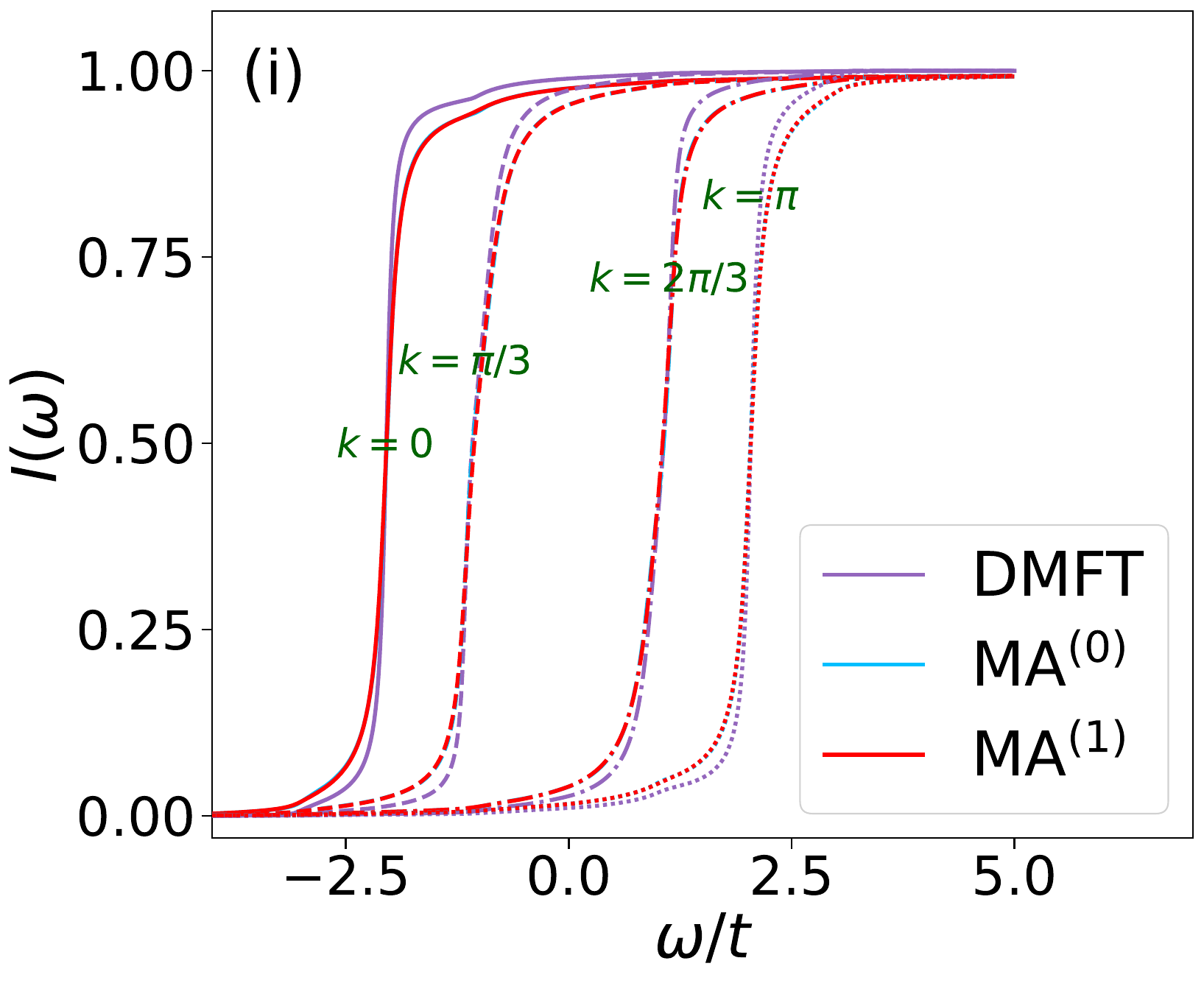}\\[2pt]
    \includegraphics[width=0.49\linewidth]{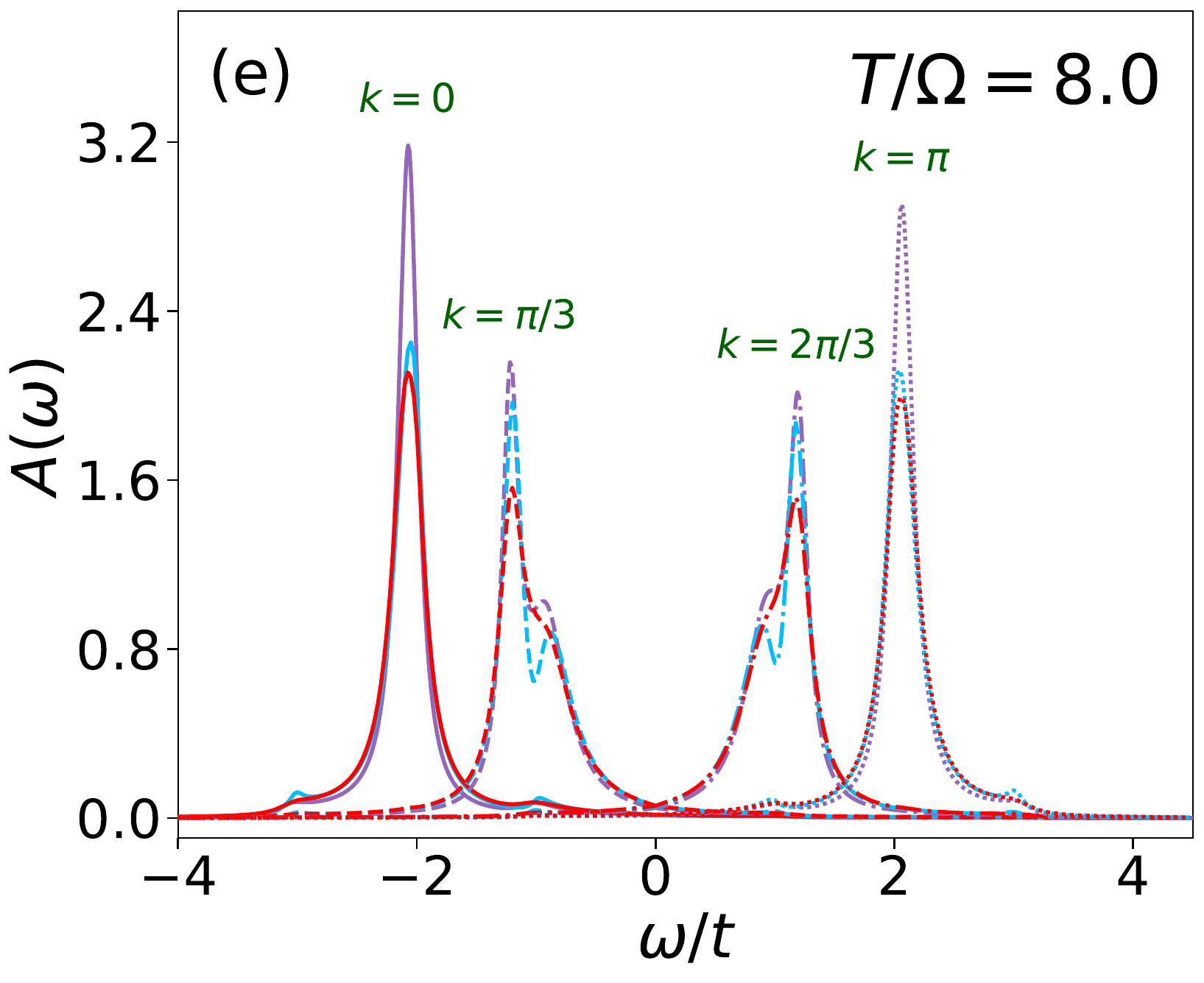}
    \includegraphics[width=0.49\linewidth]{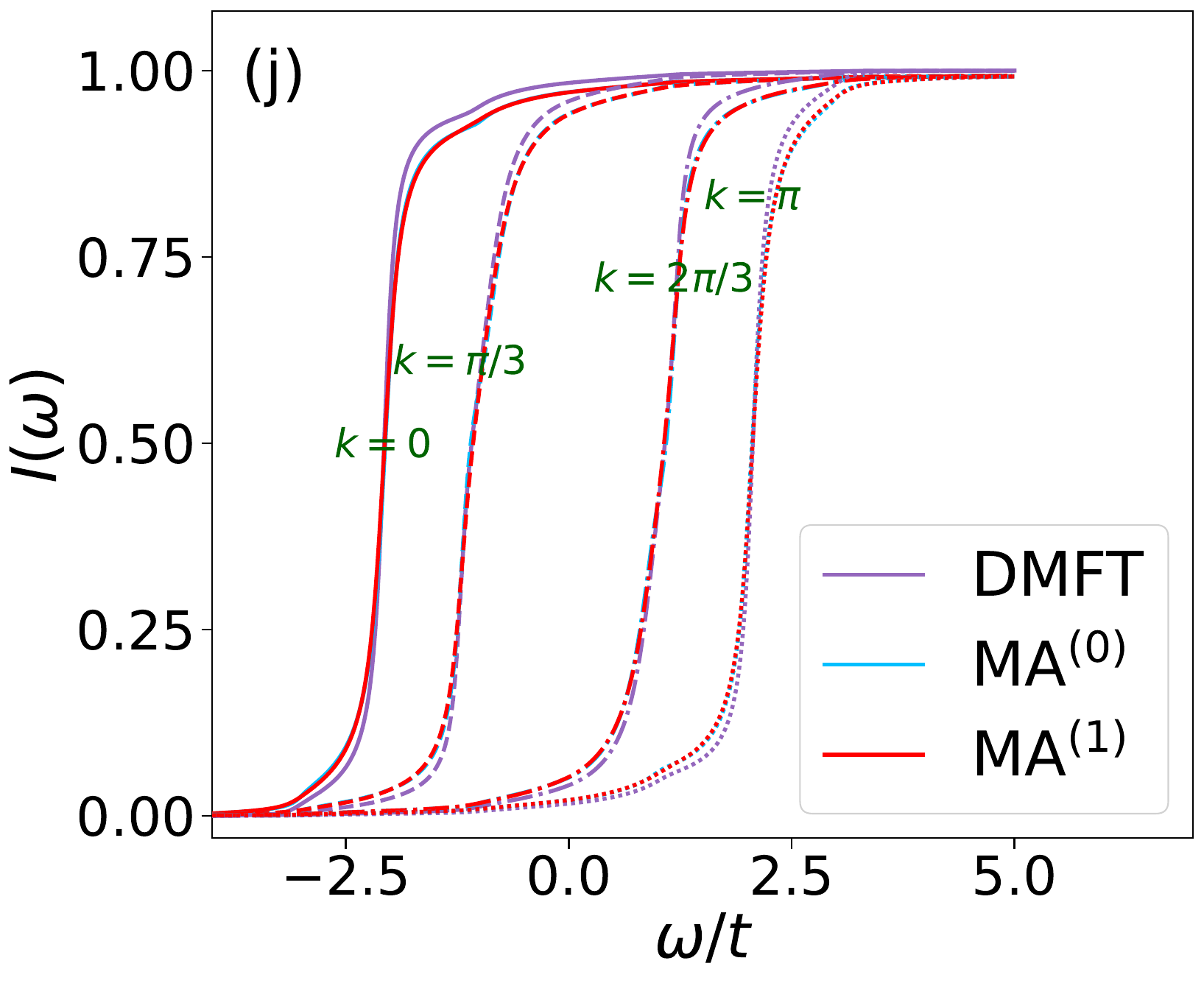}
  \caption{(color online)  Comparison of the spectral function $A(\omega)$ (left panels) and corresponding integrated spectral weight $I(\omega)$ (right panels) from MA$^{(0)}$, MA$^{(1)}$ and DMFT \cite{Mitric2022} for parameters: $\lambda=0.01$, $\Omega=1.0t,\, t = 1$ in the high temperature regime for $k=0,\pi/3,2\pi/3,\pi$ (denoted with different line-styles). Panels (a)--(e) and (f)--(j) represent  $T/\Omega=1.0,\,3.0,\,4.0,\,5.0,\,8.0$ respectively.}
  \label{fig:MA_DMFT_diff_k_comparison}
\end{minipage}\hfill
\begin{minipage}[t]{0.49\textwidth}
\centering
    \includegraphics[width=0.49\linewidth]{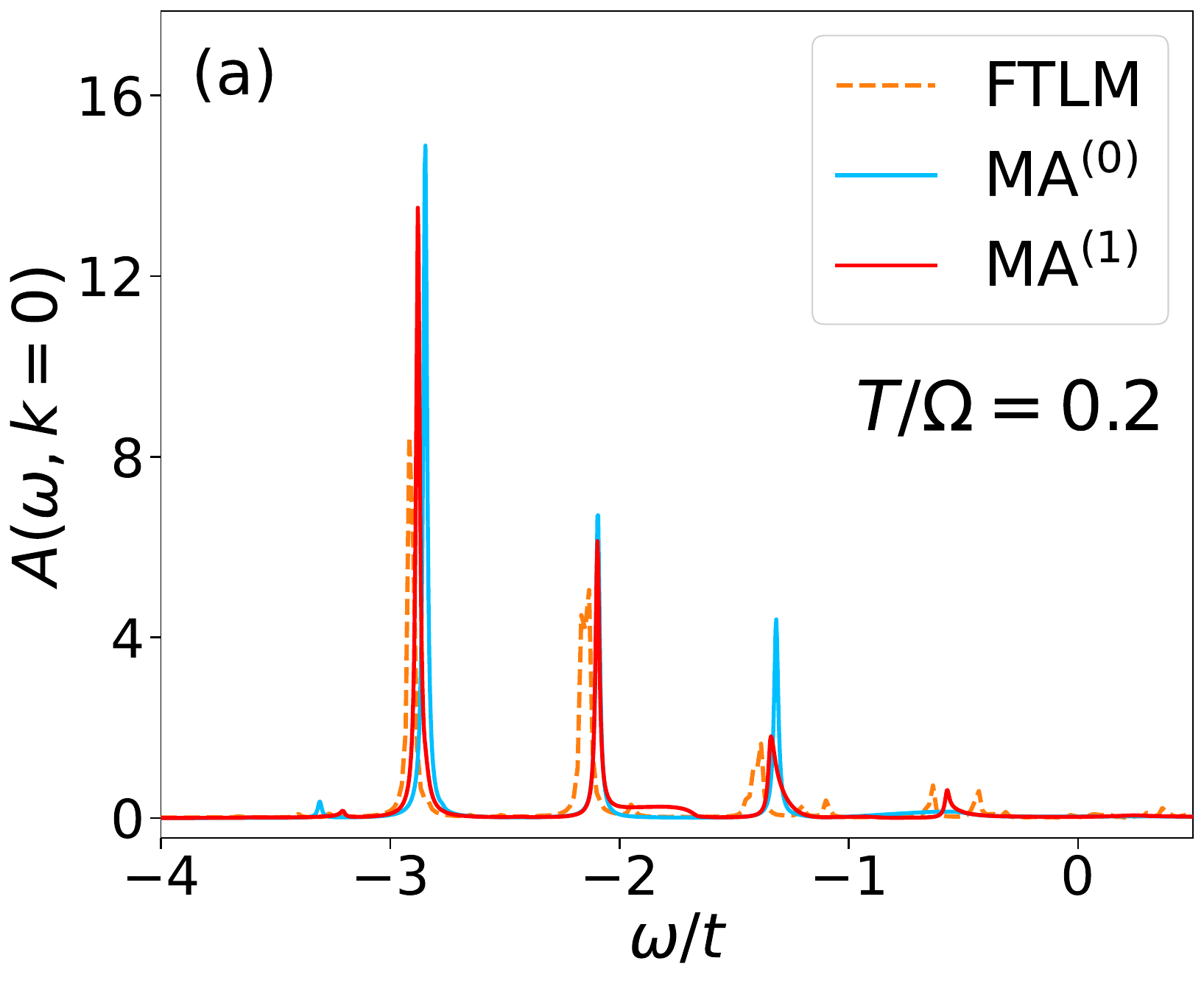}
    \includegraphics[width=0.49\linewidth]{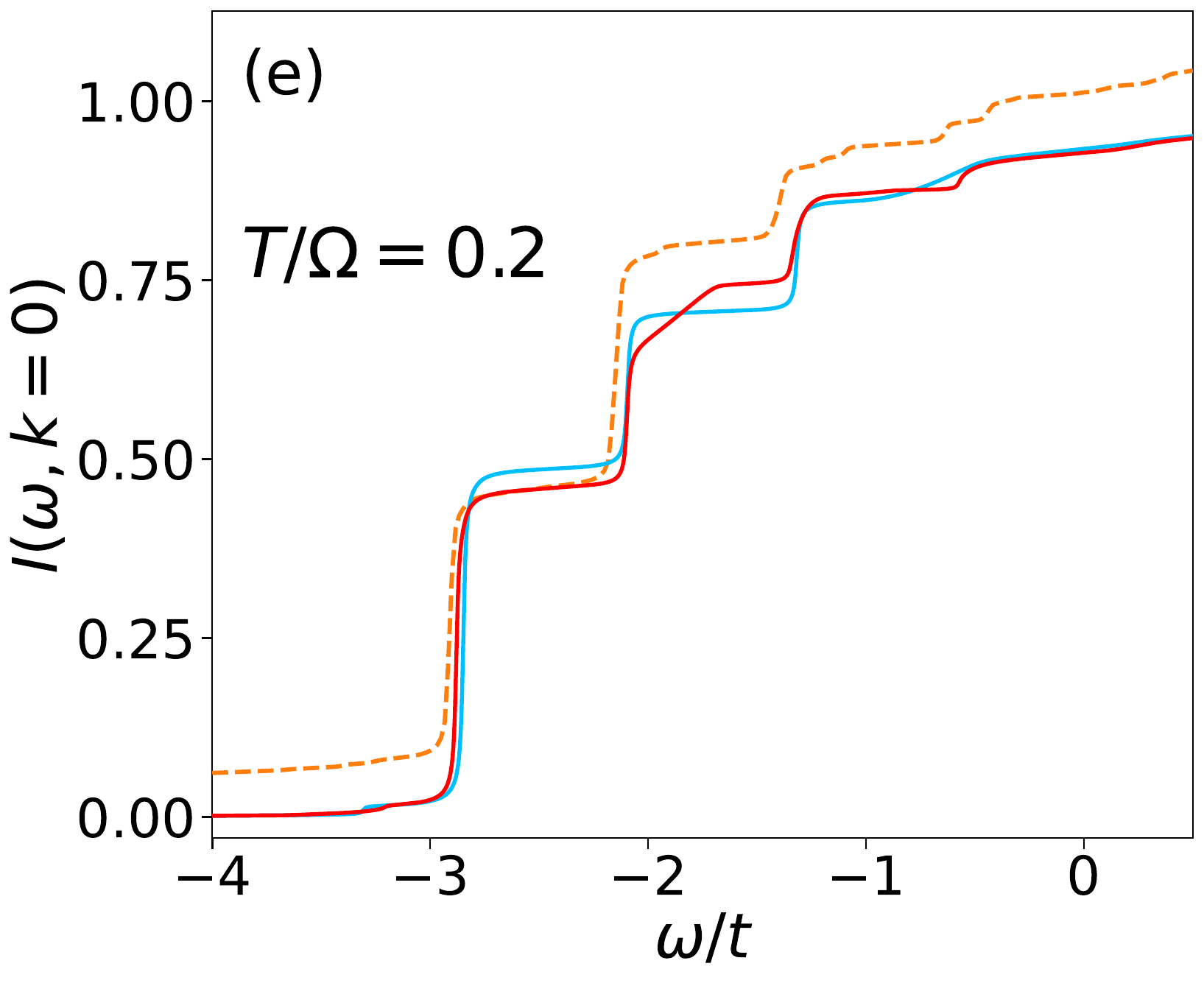}\\[2pt]
    \includegraphics[width=0.49\linewidth]{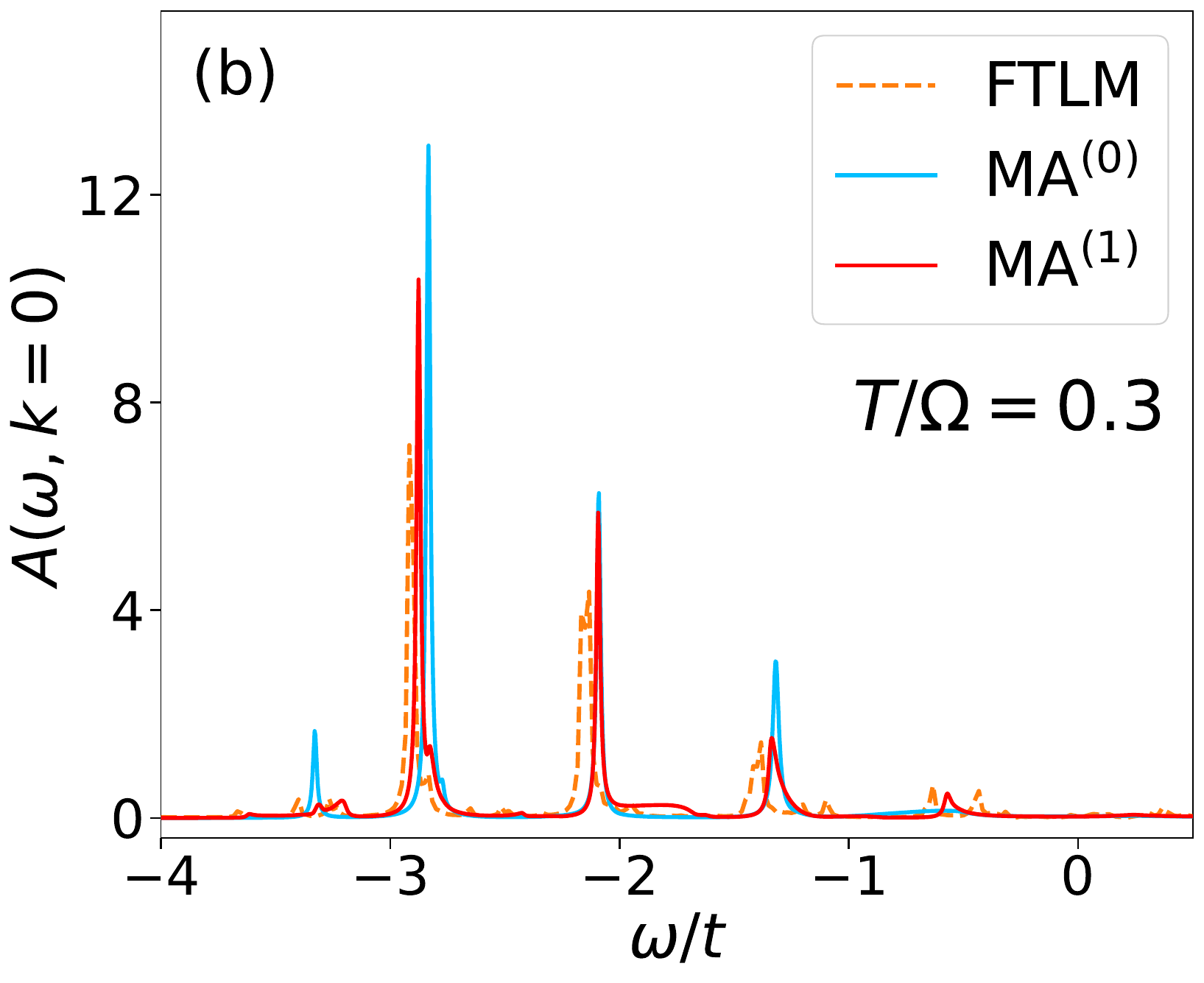}
    \includegraphics[width=0.49\linewidth]{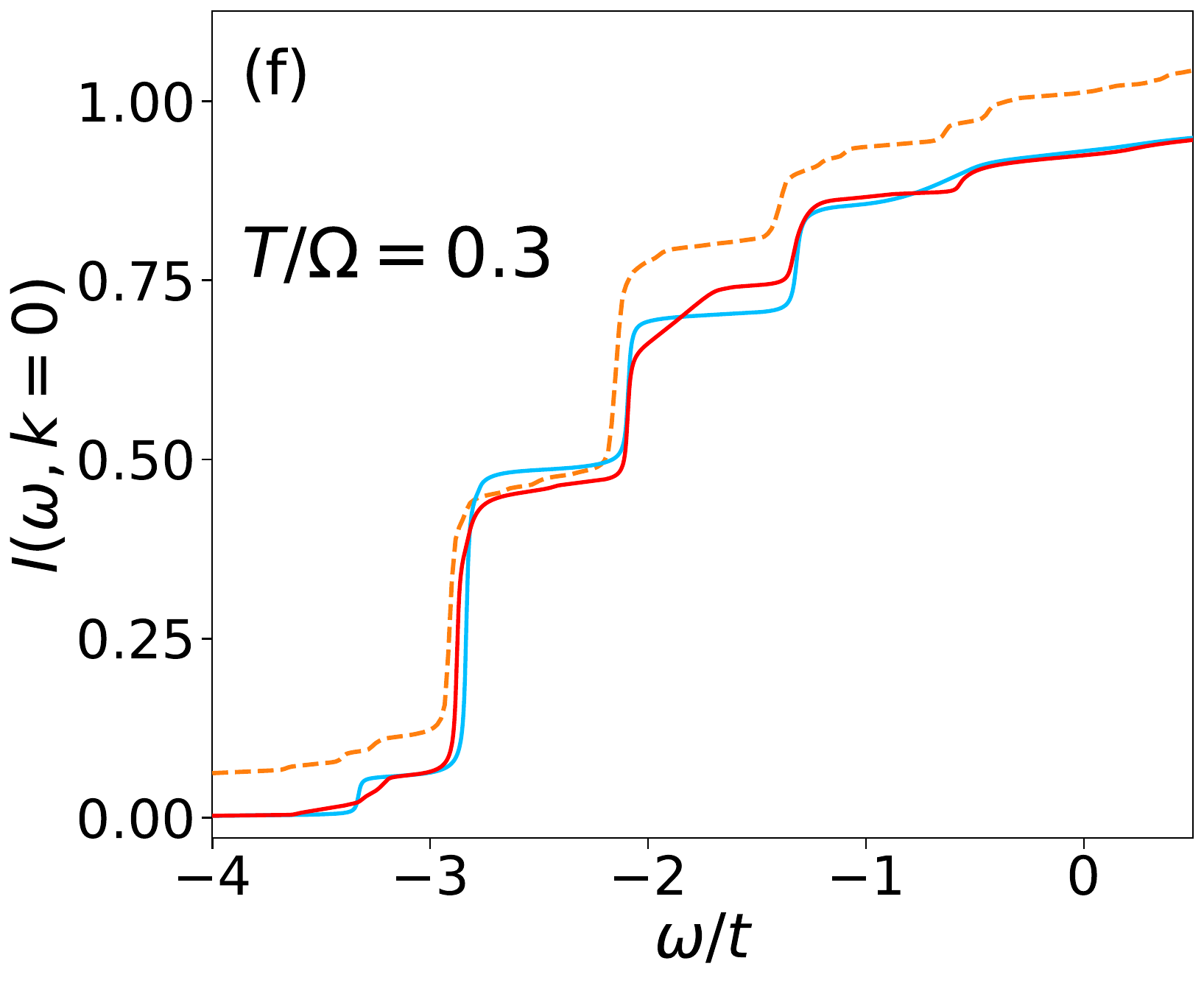}\\[2pt]
    \includegraphics[width=0.49\linewidth]{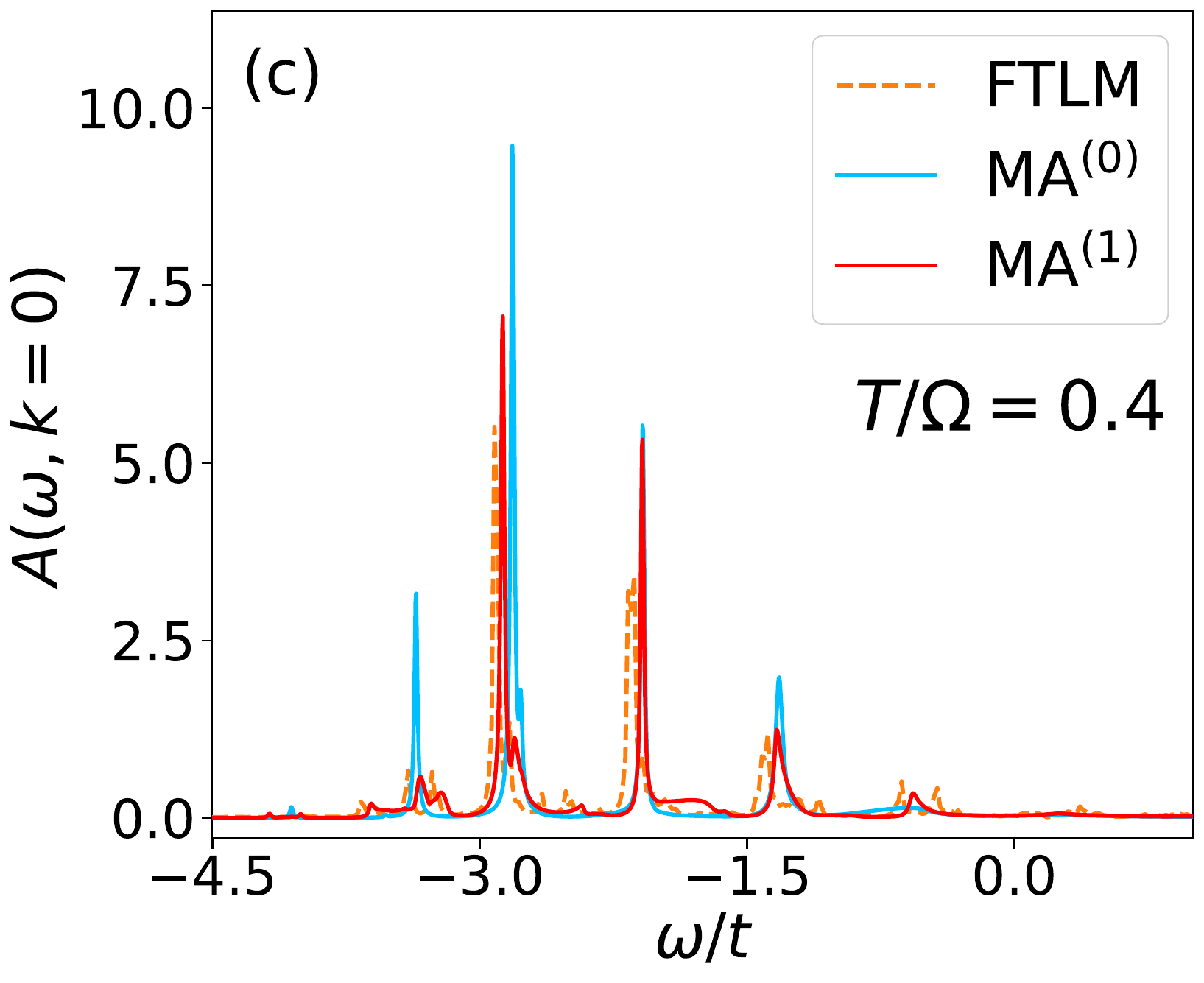}
    \includegraphics[width=0.49\linewidth]{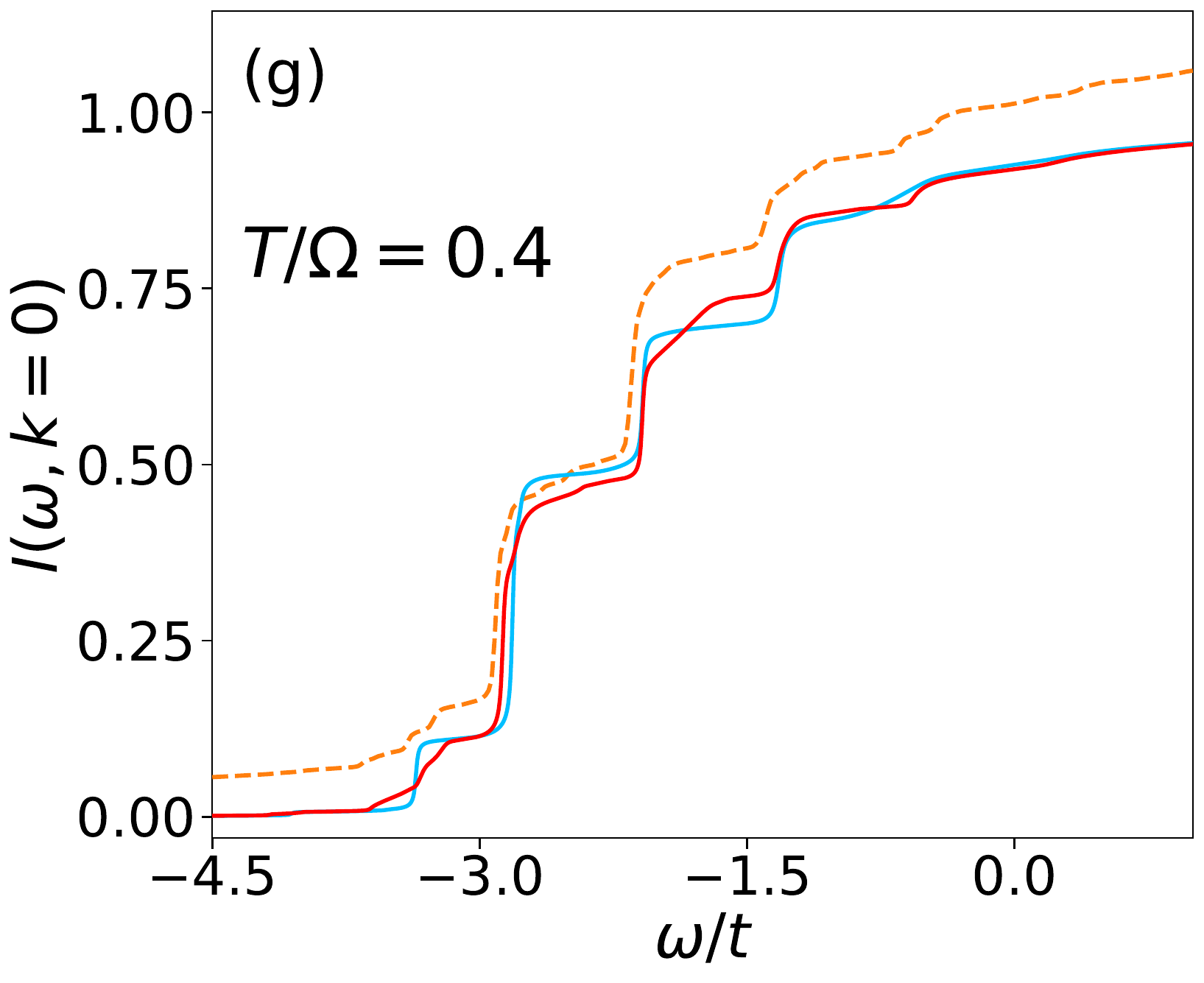}\\[2pt]
    \includegraphics[width=0.49\linewidth]{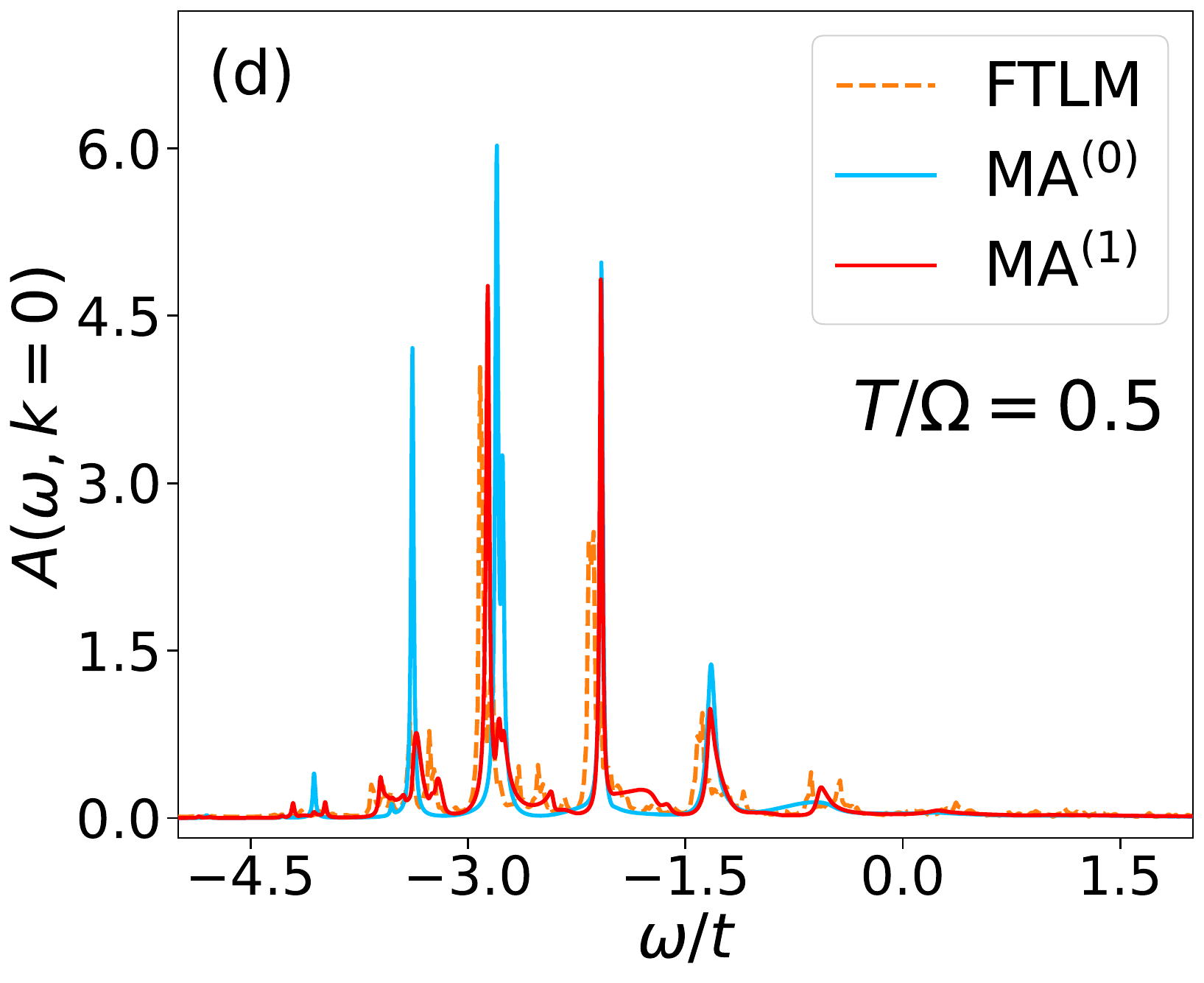}
    \includegraphics[width=0.49\linewidth]{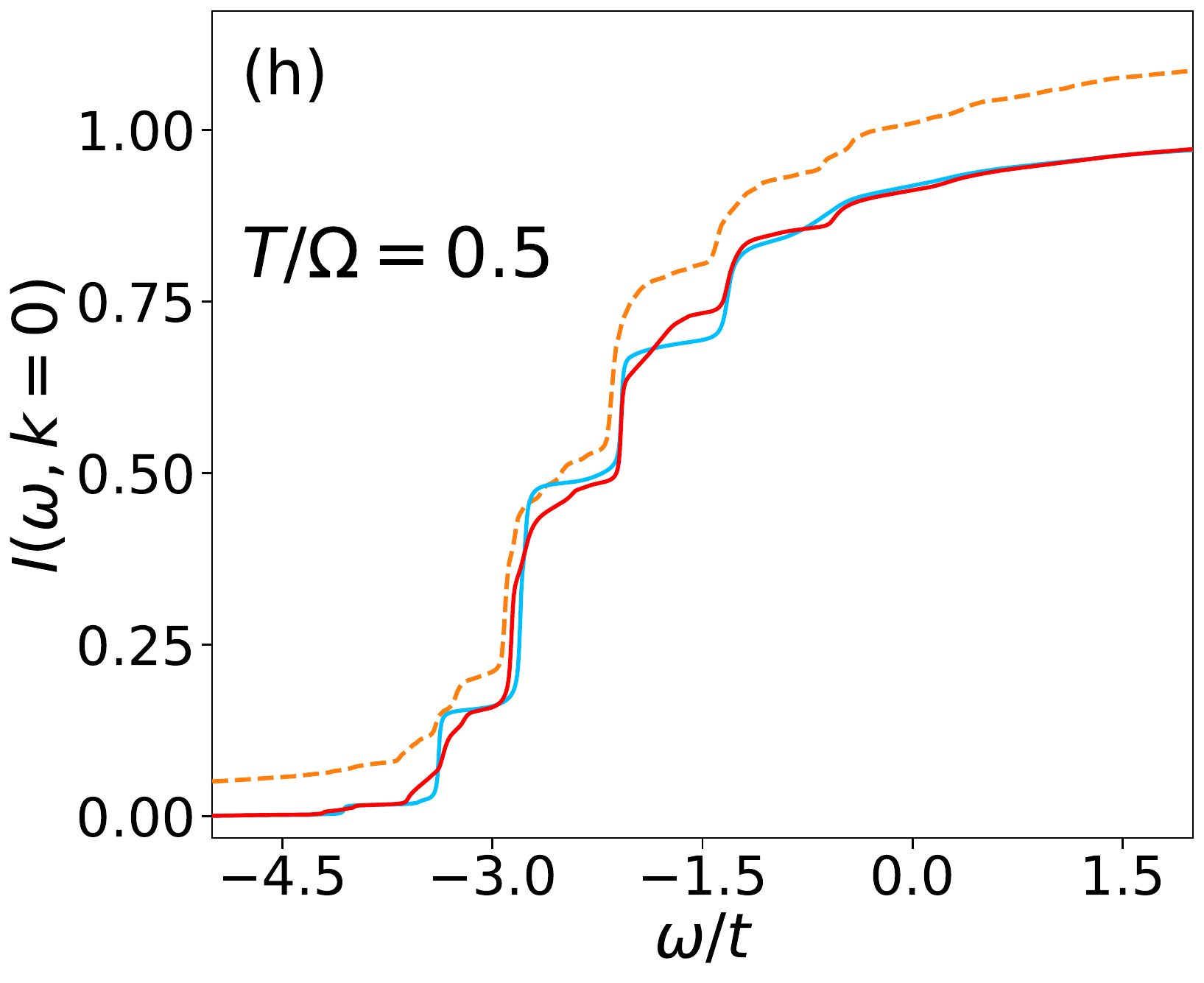}
    \caption{(color online) Comparison of MA$^{(0)}$, MA$^{(1)}$ and VED-FTLM \cite{Bonca2019} through
    (a)--(d) $A(\omega,k=0)$ (left panels) and (e)--(h) $I(\omega,k{=}0)$ (right panels) for $\lambda=1$,
    $\Omega=0.75t,\,t = 1$, $\eta_0=0.01$, at $T/\Omega=0.2,\,0.3,\,0.4,\,0.5$.}
    \label{fig:MA_VED_comparison}
\end{minipage}
\end{figure*}

Since momentum dependence enters DMFT solely through the bare electronic dispersion $\epsilon_k$ in the propagator, the overlap between the MA and DMFT self-energies (Fig.~\ref{fig:MA_DMFT_self_energy_comparison}) directly measures their agreement. Where DMFT self-energies were unavailable, we instead compare the integrated spectral weight $I(\omega)=\int_{-\infty}^{\omega}A(\omega')\,d\omega'$ and the spectral-function overlap across momenta $k$ (Fig.~\ref{fig:MA_DMFT_diff_k_comparison}). $I(\omega)$ tracks where spectral weight concentrates and shows its shift to lower frequencies at higher temperatures; by the zeroth-order sum rule it should tend to $1$ for $|\omega|\gg 2t$.

\renewcommand{\dbltopfraction}{0.95}
\renewcommand{\dblfloatpagefraction}{0.5}
\renewcommand{\textfraction}{0.05}

Figs.~\ref{fig:MA_DMFT_self_energy_comparison} and \ref{fig:MA_DMFT_diff_k_comparison} show that at higher temperatures both the self-energy and spectral functions from DMFT overlap substantially with MA$^{(1)}$, as expected since thermal fluctuations increasingly suppress non-local correlations \cite{Mitric2022}, improving the DMFT ansatz's accuracy.

\subsection{Comparison with previously unpublished VED-FTLM data}

We compare MA against numerically exact VED-FTLM results \cite{Bonca2019} (six-site chain, twisted boundary conditions) via the overlap in spectral functions and integrated spectral weight. Fig.~\ref{fig:MA_VED_comparison} shows excellent agreement between MA$^{(1)}$ and VED-FTLM for both the polaron ground state and the thermal continua across temperatures, holding even in the adiabatic regime. The numerous small peaks that appear in the FTLM line-shape at higher temperature, and the corresponding excess of $I(\omega)$ above $1.0$, are finite-size effects explained elsewhere \cite{ReichmannCEI2022}.

\begin{figure}[t]
  \centering
    \includegraphics[width=0.49\linewidth]{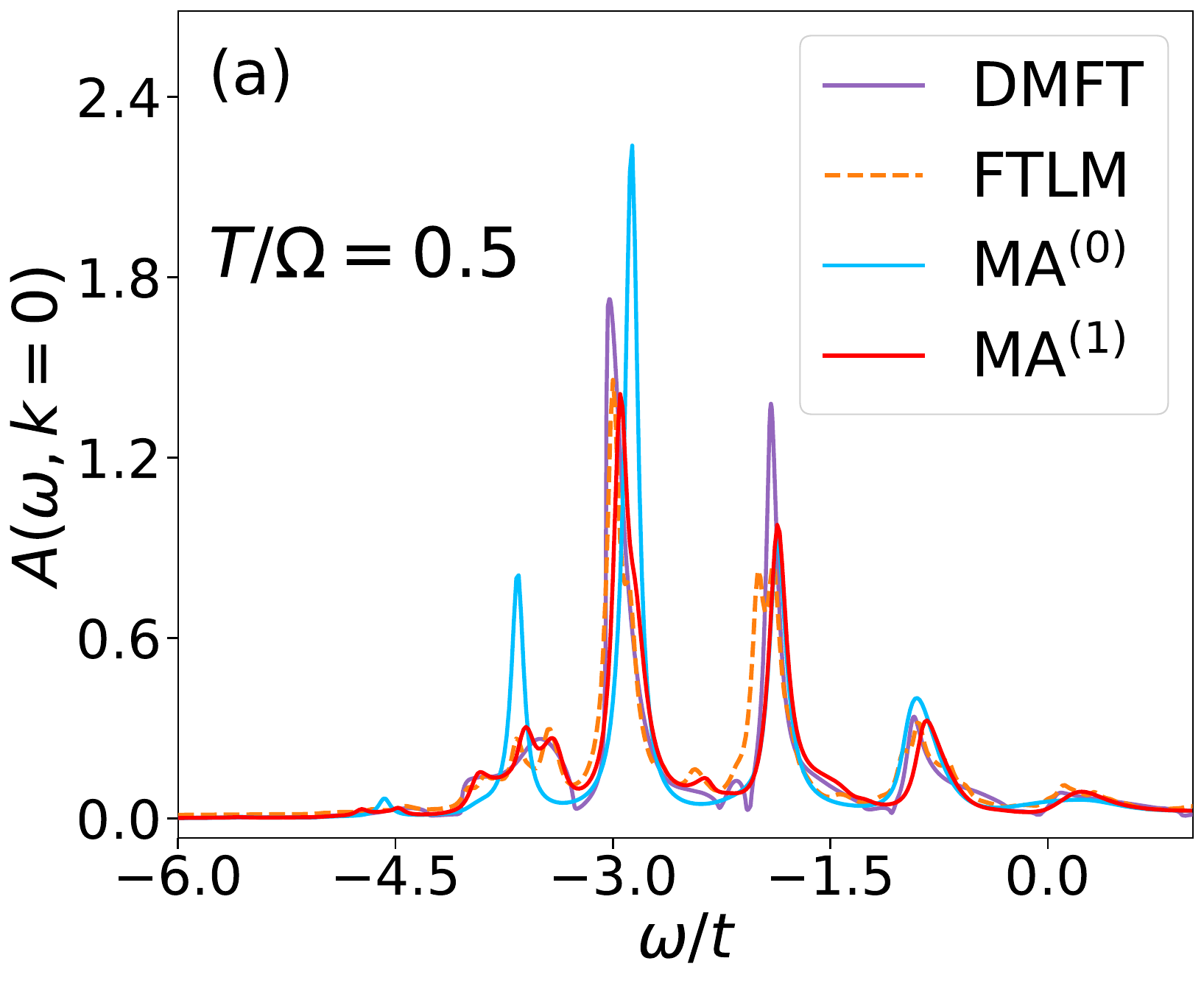}
  \includegraphics[width=0.49\linewidth]{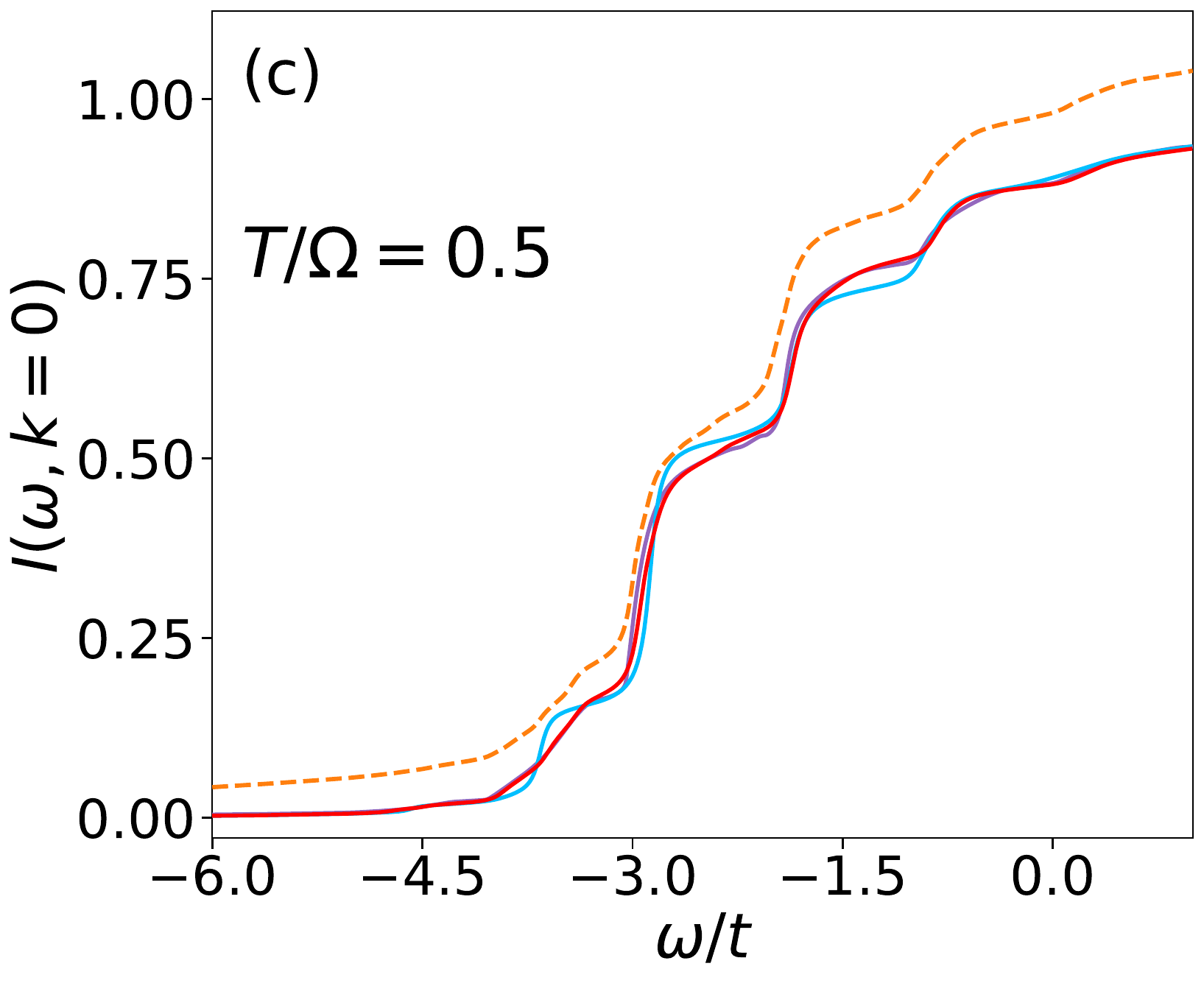}\\[1pt]
  \includegraphics[width=0.49\linewidth]{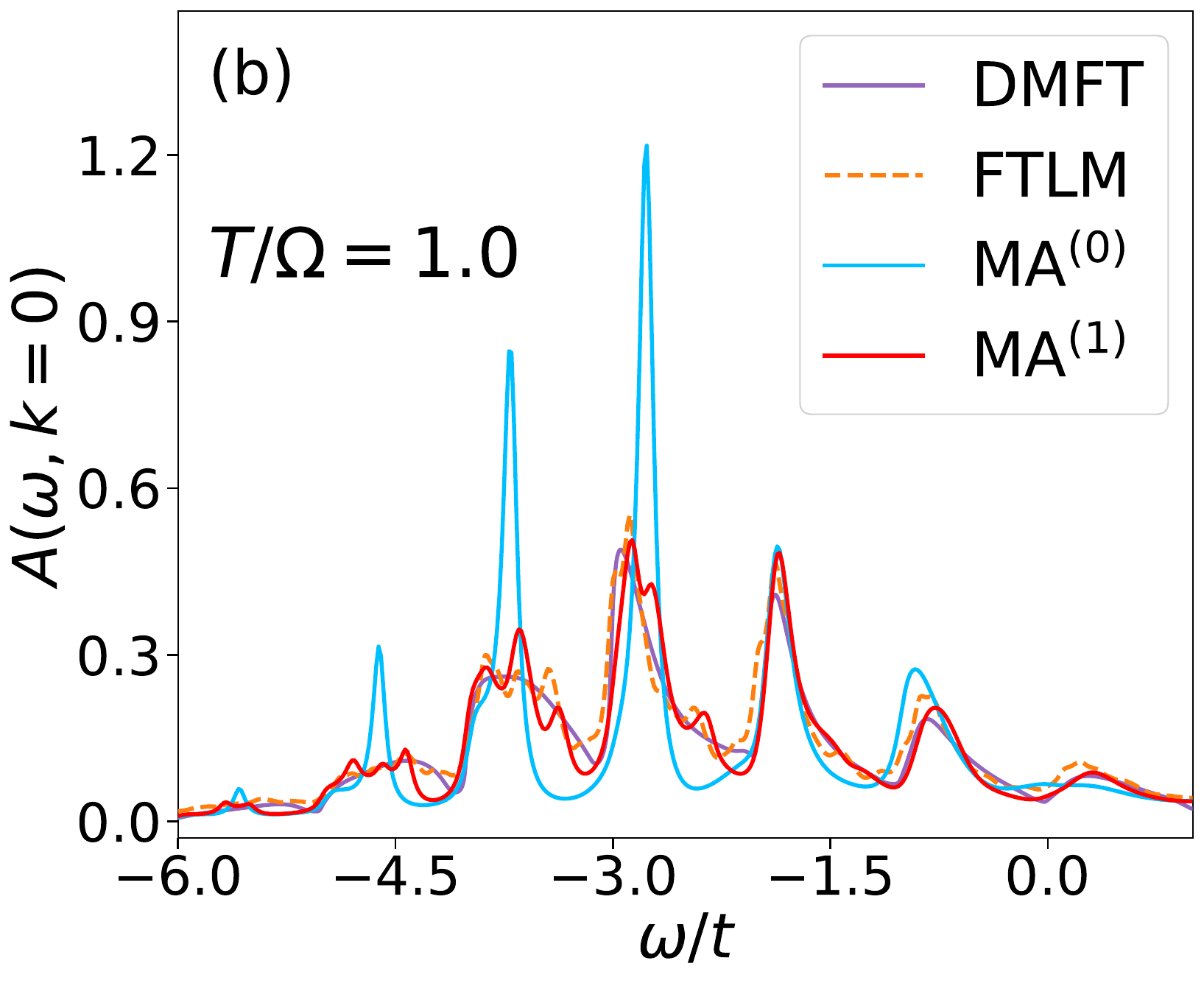}
  \includegraphics[width=0.49\linewidth]{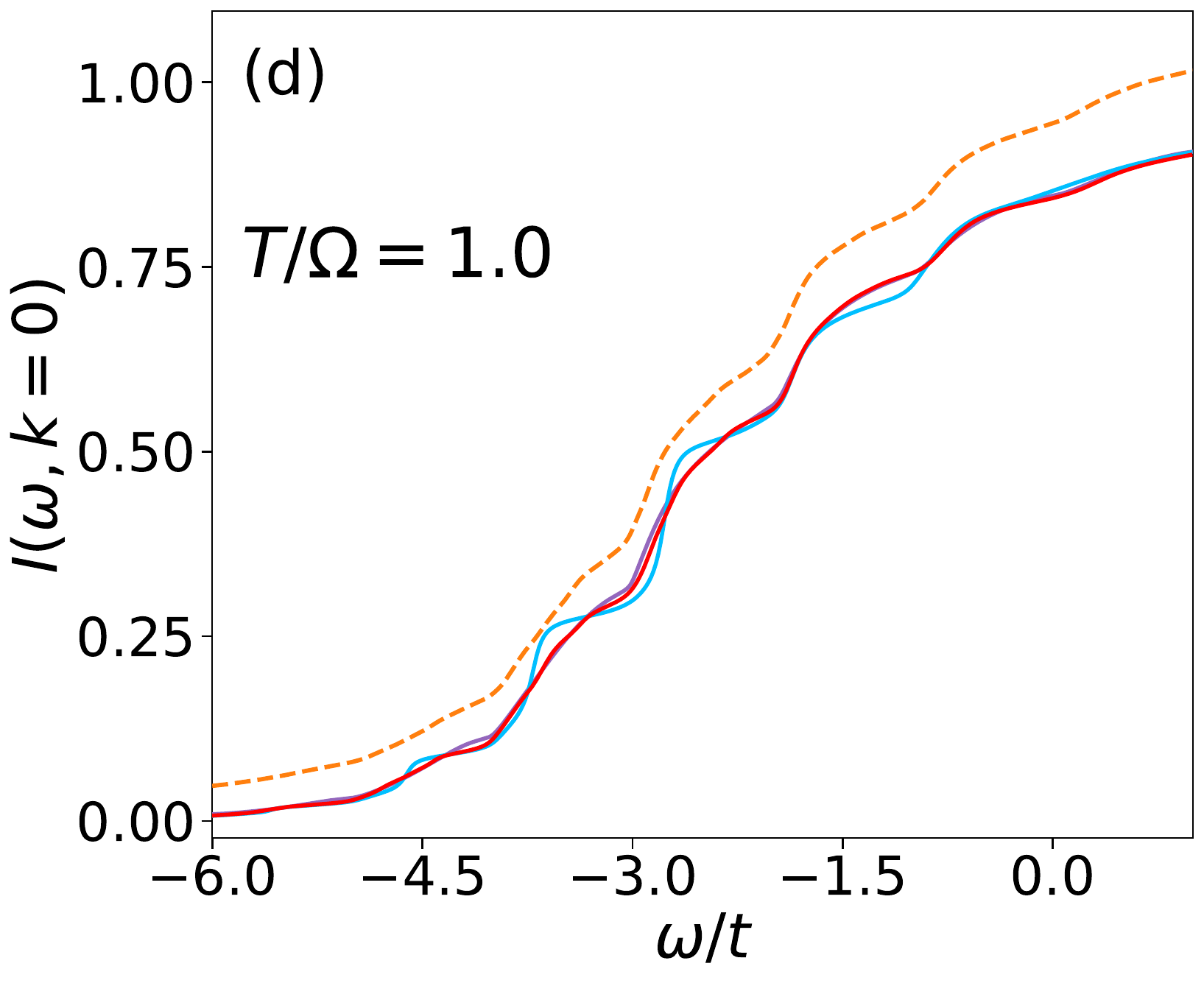}
    \caption{(color online) Comparison of MA$^{(0)}$, MA$^{(1)}$, VED-FTLM \cite{Bonca2019}, DMFT \cite{Mitric2022} $A(\omega,k=0)$ (left panels) and $I(\omega,k{=}0)$ (right panels) in the adiabatic regime for parameters: $\lambda=1$, $\Omega=0.5t,\,t = 1$,  $\eta_0=0.05$. (a),(c) represent $T/\Omega=0.5$ and (b),(d)  represent $T/\Omega=1.0$.}
  \label{fig:VED_DMFT_MA}
\end{figure}

Fig.~\ref{fig:VED_DMFT_MA} plots MA, DMFT, and VED-FTLM together for a common parameter set in the adiabatic regime, as a final validation of MA's accuracy across the full parameter range.

\bibliographystyle{apsrev4-2}
\bibliography{ref}